\documentclass[useAMS]{mn2e}

\usepackage{times}
\usepackage{graphicx}
\usepackage{amsmath}
\usepackage{amssymb}
\usepackage{color}
\newcommand{\beq}{\begin{equation}}
\newcommand{\bea}{\begin{eqnarray}}
\newcommand{\eeq}{\end{equation}}
\newcommand{\eea}{\end{eqnarray}}

\title[Synchrotron signature of a relativistic blast wave]{Synchrotron signature of a relativistic blast wave with
  decaying microturbulence}

\author[M. Lemoine]
{Martin Lemoine\thanks{e-mail:{\tt lemoine@iap.fr}} \\
  Institut d'Astrophysique de Paris, \\
  CNRS, UPMC,
  98 bis boulevard Arago, F-75014 Paris, France\\
}

\begin{document}

\date{}

\pubyear{2008}

\maketitle

\label{firstpage}

\begin{abstract}
  Microphysics of weakly magnetized relativistic collisionless shock
  waves, corroborated by recent high performance numerical
  simulations, indicate the presence of a microturbulent layer of
  large magnetic field strength behind the shock front, which must
  decay beyond some hundreds of skin depths. The present paper
  discusses the dynamics of such microturbulence, borrowing from these
  same numerical simulations, and calculates the synchrotron signature
  of a powerlaw of shock accelerated particles. The decaying
  microturbulent layer is found to leave distinct signatures in the
  spectro-temporal evolution of the spectrum $F_\nu \propto
  t^{-\alpha}\nu^{-\beta}$ of a decelerating blast wave, which are
  potentially visible in early multi-wavelength follow-up observations
  of gamma-ray bursts. This paper also discusses the influence of the
  evolving microturbulence on the acceleration process, with
  particular emphasis on the maximal energy of synchrotron afterglow
  photons, which falls in the GeV range for standard gamma-ray burst
  parameters. Finally, this paper argues that the evolving
  microturbulence plays a key role in shaping the spectra of recently
  observed gamma-ray bursts with extended GeV emission, such as
  GRB090510.
\end{abstract}


\section{Introduction}\label{sec:introd}
The acceleration of particles at a decelerating relativistic
collisionless shock front constitutes a key building block of the
afterglow model of gamma-ray bursts (GRB, M\'esz\'aros \& Rees
1997). The standard phenomenology models the accelerated electron
population as powerlaw ${\rm d}N_e/{\rm d}\gamma_e\propto
\gamma_e^{-p}$, which radiates powerlaw photon spectra of the form
$F_\nu \propto t^{-\alpha}\nu^{-\beta}$, with a temporal decay index
$\alpha$ and a frequency index $\beta$ that are direct functions of
$p$, see e.g. Piran (2005) for a review, or e.g. Sari et al. (1998),
Panaitescu \& Kumar (2000) for detailed formulae. From both
microscopic and observational points of view, the situation however
appears more complex, in spite of several remarkable results of the
past decade.

On the microscopic level, for instance, one understands the formation
of a relativistic collisionless shock front in a weakly magnetized
medium -- such as the interstellar medium (ISM) -- through the
self-generation of intense small scale electromagnetic fields that act
as the mediating agents for the transition from the far upstream
unshocked state to the far downstream shocked state. The accelerated
particles, as forerunners of the shock front, play a central role in
triggering the microinstabilities that build the self-generated
field. In turn, this self-generated microturbulence controls the
scattering of these particles and it therefore directs the
acceleration process, which becomes intricately non-linear. This
general scheme has been validated so far in high performance
particle-in-cell (PIC) simulations (e.g. Spitkovsky 2008, Keshet et
al. 2009, Sironi \& Spitkovsky 2009, 2011, Martins et al. 2009) and
understood on the basis of analytical arguments at the linear level
(e.g. Medvedev \& Loeb 1999, Gruzinov \& Waxman 1999, Lyubarsky \&
Eichler 2006, Lemoine \& Pelletier 2010, 2011a, Rabinak et
al. 2011). The situation becomes more complex when one tries to bridge
the gap between the limited simulation timescales and the much longer
timescales probed by the observations.  On such timescales, one would
indeed expect that the microturbulence has died away (e.g. Gruzinov \&
Waxman 1999), yet GRB afterglow modelling has generally pointed to a
field strength close to a percent of equipartition permeating the
blast, on day timescales. The origin of this field and its relation
with the microturbulence behind the shock front has remained a nagging
issue for many years.

On the observational level, the recent era of rapid follow-up
observations in the X-ray and GeV domain has brought its wealth of
surprises. The Swift satellite has revealed X-ray afterglow light
curves that differ appreciably in the $10^2-10^4\,$s domain from the
canonical afterglow model (Nousek et al.  2006, O'Brien et
al. 2006). Of more direct interest to the present work, the {\it
  Fermi}-LAT telescope has reported the discovery of long-lived
($\sim100-1000\,$s) GeV emission in a fraction of observed bursts. In
one case (GRB090510), this emission has been measured almost
contemporarily to emission in the X-ray and optical domains as early
as $100\,$s (Ackermann et al. 2010, de Pasquale et al. 2010). This
long-lived high energy emission has been shown to fit nicely the
predictions of a model in which the electrons cool slowly through
synchrotron radiation in a background shock compressed magnetic field,
without any need for microturbulence (Barniol-Duran \& Kumar 2009,
2010, 2011a; see also Gao et al. 2009; de Pasquale et al. 2010; Corsi
et al. 2010; Ghirlanda et al., 2010; He et al 2011; and see Ghisellini
et al. 2010, Razzaque 2010 and Panaitescu 2011 for alternative points
of view).  Given the past history in GRB afterglow modelling, such a low
magnetization of the blast may come as a surprise, but it may also
point out that after all, the microturbulence does decay away as
theoretically expected, and that the high level of turbulence seen on
day timescales has been seeded through some other
instability\footnote{De Pasquale et al. 2010 and Corsi et al. 2010
  have shown that the afterglow emission could be modelled with a more
  traditional estimate $\epsilon_B\sim10^{-2}-10^{-3}$, but this comes
  at the price of an extraordinarily low external density $n\lesssim
  10^{-6}\,$cm$^{-3}$. This interpretation is not considered here, see
  also Sec.~\ref{sec:fcons} for further discussion.}.

Depending on how fast and how far from the shock this microturbulent
layer decays, it is likely to influence the particle energy gains from
Fermi acceleration, and losses through synchrotron radiation. The
microturbulent layer must actually ensure the scattering of
accelerated particles, because in the absence of microturbulence,
these particles would be advected away with the transverse magnetic
field lines to which they are tied and acceleration would not take
place (e.g. Begelman \& Kirk 1990, Lemoine et al. 2006, Niemiec et
al. 2006, Pelletier et al. 2009)\footnote{see also
  Sec.~\ref{sec:fcons}.}. Now, scattering in small scale turbulence is
so slow that producing GeV photons at an external blast wave of
Lorentz factor of a few hundreds represents a challenge, see e.g. Kirk
\& Reville (2010), Lemoine \& Pelletier (2011c), and see also Piran \&
Nakar (2010), Sagi \& Nakar (2012). It would be about impossible if
the particles were to scatter in the microturbulent layer then radiate
in the much weaker shock compressed background field. Therefore, the
interpretation of Barniol-Duran \& Kumar (2009, 2010, 2011a) actually
suggests that the microturbulence also plays a role in the radiation
of GeV photons, at the very least, if not in shaping the synchrotron
spectra over the broad spectral range. In other words, the observation
of extended GeV emission and its early follow-up in other wavebands
may be opening a rare window on the dynamics of the microturbulence in
weakly magnetized relativistic collisionless shocks.

In this context, the present paper proposes to discuss the synchrotron
spectra and more generally the afterglow spectrum $F_\nu\propto
t^{\-\alpha}\nu^{-\beta}$ of a decelerating relativistic blast wave,
accounting for the time evolution of the microturbulence behind the
shock front. While the initial motivation of this work was to provide
a concrete basis for the scenario of Barniol-Duran \& Kumar (2009,
2010, 2011a), in which particles scatter in a time decaying
microturbulence but radiate in a region devoid of microturbulence, it
has become apparent that the possibility of decaying microturbulence
opens a quite rich diversity of phenomena, which deserve to be
discussed in detail. This problem has been tackled by Rossi \& Rees
(2003), who considered the simplified case of a homogeneous
microturbulent layer that dies instantaneously beyond some distance,
and by Derishev (2007), who showed that a particle radiating in a time
evolving magnetic field can lead to spectra quite different from the
standard one-particle synchrotron spectra. Borrowing from the latest
PIC simulations, the present paper establishes a model of the
microturbulence strength and coherence length evolving as powerlaws in
time beyond some distance, until the decay saturates down to the
background shock compressed magnetic field; the present study then
calculates the synchrotron spectra in this structure for various
typical configurations (slow cooling, fast cooling, with and without
inverse Compton losses) and it discusses the problem of particle
scattering, acceleration timescale and maximum photon energy in this
setting. As such, it generalizes and encompasses these former studies,
in the spirit of providing new tools with which one can analyse
existing and forthcoming data. A brief comparison to present early
afterglow observations is provided.

The detailed spectra and the spectro-temporal indices $\alpha$,
$\beta$ of $F_\nu\,\propto\,t^{-\alpha}\nu^{-\beta}$ are provided in
Appendix~\ref{sec:appFnu}, while Section~\ref{sec:synchspec} details
the model for the evolution of microturbulence and provides the
general characteristics of the afterglow light curves and spectral
energy distributions in various configurations. Section~\ref{sec:disc}
discusses the scattering process and the maximal acceleration energy,
and it confronts the above models to existing data. The results are
summarized in Sec.~\ref{sec:concl}.  Throughout, this paper adopts the
standard notation $Q_{x}\equiv Q/10^{x}$ with $Q$ a generic quantity
in cgs units. Fiducial values used for numerical applications
correspond to those derived in Barniol-Duran \& Kumar (2009, 2010,
2011a), and He et al. (2011), e.g. an external density $n\sim
10^{-3}\,$cm$^{-3}$, a blast Lorentz factor $\gamma_{\rm b}\sim 300$
at $100\,$s and it is assumed that the blast has entered the
deceleration regime. One must distinguish the time $t_{\rm obs}$ in
the observer frame, also written $t_{\rm obs}=100\,t_2\,$s in
numerical applications, from the time experienced by a particle since
shock entry; this difference is manifest everywhere. For convenience,
the notation $z_{+,0.3}\,\equiv\,(1+z)/2$ is introduced, $z$ denoting
the redshift of the GRB.

\section{Synchrotron spectra with time decaying
  microturbulence}\label{sec:synchspec}

The self-generation of microturbulence in the precursor of a
relativistic collisionless shock front propagating in a very weakly
magnetized medium appears both guaranteed and necessary to the
maintenance of the shock. It is necessary because in the absence of a
background magnetic field, self-magnetization is required to build up
a magnetic barrier that initiates the shock transition. It is
guaranteed because the development of microinstabilities follows
naturally from the penetration of the unshocked plasma by the
anisotropic beam of supra-thermal particles moving ahead of the shock
(e.g. Medvedev \& Loeb 1999).

Studies of non-relativistic collisionless magnetospheric shocks have
shown that dissipation ahead of the shock is initiated by the
reflection of ambient particles (i.e. from the unshocked plasma)
on the shock front in the compressed magnetic field (e.g. Leroy et
al. 1982). Particle-in-cell simulations indicate that a similar
phenomenon takes place at a relativistic unmagnetized shock, although
the particles now reflect on the small scale electromagnetic fields
self-generated by the microinstabilities (e.g. Spitkovsky 2008). The
reflected and accelerated particle populations merge together and
trigger microinstabilities such as the Weibel (filamentation)
instability or oblique electrostatic instabilities, provided the
precursor extends far enough for these modes to grow on the precursor
crossing timescale (Lemoine \& Pelletier 2010, 2011a). In a very
weakly magnetized shock wave, with magnetization typical of the ISM
and blast Lorentz factor $\gamma_{\rm b}\lesssim 10^3$, this appears
guaranteed.

As seen from the shock frame (in which the shock front lies at rest)
the incoming kinetic energy is carried by the protons, the electrons
carrying only a fraction $m_e/m_p$ of the incoming flow kinetic
energy. Energy transfer between the two species in the microturbulence
leads to heating of the electron population, close to equipartition by
the time it reaches the shock front, as observed in current PIC
simulations (Sironi \& Spitkovsky 2011), see also the discussion in
Lemoine \& Pelletier (2011a). Equipartition means that the incoming
electrons carry Lorentz factor $\gamma_e\sim \gamma_{\rm b}m_p/m_e$,
hence their skin depth scale (downstream frame) $c/\left[4\pi
  \gamma_{\rm b}n_{\rm u}/\left(\gamma_e m_e\right)\right]^{1/2}\sim
c/\omega_{\rm pi}$.  The natural length scale of the electromagnetic
structures produced by these microinstabilities is therefore the ion
skin depth scale $c/\omega_{\rm pi}$ of the upstream plasma.

\subsection{Input from Particle-in-cell simulations}
Particle-in-cell simulations not only validate the above general
scheme, they also provide interesting constraints on the shape and
evolution of microturbulence ahead and behind the shock front.  Two
most recent and most detailed studies are of direct interest to the
present work.

Chang et al. (2008) have performed long simulations of the evolution
of micro-turbulence behind a relativistic shock front. The simulations
have been computed for an unmagnetized pair plasma with relative
Lorentz factor $\gamma_{\rm b}=15$ between upstream and downstream. In
weakly magnetized relativistic shock waves, the preheating of
electrons in an electron-ion shock of similar configuration appears so
efficient that for all practical matters, the downstream plasma
behaves as a relativistic pair plasma; hence the results of Chang et
al. (2008) can be transposed to an electron-ion shock.  These
simulations show that the microturbulence remains mostly static in the
downstream rest frame, and that it is composed of an intermittent
magnetic field structure that can be roughly described as a collection
of magnetic loops and islands on typical length scales $\sim 10-30\,
c/\omega_{\rm pi}$. One clear observation made in this work is that
the small scale structures dissipate first, leaving the large scale
clumps unaffected over the timescale of the simulation. Chang et
al. (2008) interpret this gradual erosion as collisionless damping,
with a damping frequency $\Im\omega \propto \lambda^{-3}$ ($\lambda$
denoting the spatial scale). If the magnetic turbulence is described in
Fourier space as a power law spectrum with most of magnetic power on
small length scales, this implies a decay of the magnetic field
strength accompanied by an evolution of the coherence scale
$\lambda_{\delta B}\propto t^{1/3}$. This is  made explicit
further below.

The longest PIC simulation so far for a relativistic shock is that of
Keshet et al. (2009), which extends to about $10^4 \omega_{\rm
  pi}^{-1} \,\sim\,240\,n_{-3}^{-1/2}\,$s (comoving time). For values
envisaged by Barniol-Duran \& Kumar (2009, 2010, 2011a) and He et
al. (2011) to describe the GeV extended emission, i.e. an ejecta of
energy $E\sim 10^{53}\,$erg, launched into a medium of density $n\sim
10^{-3}\,$cm$^{-3}$ with initial Lorentz factor $\gamma_{\rm ej}\sim
10^3$, the above timescale represents close to $1\,$\% of a dynamical
timescale at an observer time of a hundred seconds. Keshet et
al. (2009) provide a detailed study of the magnetic field power
spectrum of the turbulence and its evolution. They confirm most of the
findings of Chang et al. (2008); in particular, they show that the
magnetic field does decay behind the shock wave, but on rather long
length scales compared to a skin depth $c/\omega_{\rm pi}$. More
importantly, they find that the presence of shock accelerated
particles influences the decay timescale of the magnetic field with a
general trend being that higher energy particles ensure a longer
lifetime for the downstream microturbulence. Given that the
simulations of Keshet et al. (2009) extend for a time that is much
smaller than the dynamical time, it has not had time to produce very
high energy particles and to probe their impact. Such high energy
particles would tend to populate the magnetic perturbation spectrum
with longer wavelength modes, which would then decay on longer
timescales when downstream. In any event, this should not call into
question what has been said above, since low energy particles carry
most of the energy of a shock accelerated population with index
$p>2$. Nevertheless, to probe how far the perturbation spectrum may be
populated, one can conduct the following exercise. The maximal size of
the precursor is given in a reasonable approximation by $r_{{\rm
    L,max},i,0}/\gamma_{\rm sh}^3$, where $r_{{\rm L,max},i,0}$
represents the gyration radius of the highest energy ions in the
background upstream magnetic field, and $\gamma_{\rm
  sh}=\sqrt{2}\gamma_{\rm b}$ the shock Lorentz factor as measured
upstream. This result is discussed in detail in Plotnikov et
al. (2012) but it can be understood as follows. The highest energy
ions are those that travel the furthest away from the shock, since
electrons are generally accelerated to a smaller energy due to
synchrotron losses; furthermore, the particles gyrate by an angle
$1/\gamma_{\rm sh}$ over a timescale $t_{\rm res,u}\simeq c^{-1}
r_{{\rm L,max},i,0}/\gamma_{\rm sh}$ before being caught up by the
shock front (Achterberg et al. 2001); finally, the typical distance
between the shock front and the particle is $c\,t_{\rm
  res,u}\left(1-\beta_{\rm sh}\right)\sim c\,t_{\rm
  res,u}/(2\gamma_{\rm sh}^2)$. Assuming that the ions are accelerated
on a timescale $t_{\rm res,u}$ (see also Sec.~\ref{sec:accel}) and
comparing the time available for acceleration with the age of the
shock wave $r/c$, one finds the precursor size $\sim r/\gamma_{\rm
  sh}^2 \sim 2\times 10^4 \left(c/\omega_{\rm pi}\right)
n_{-3}^{1/2}t_2 z_{+,0.3}^{-1}$. This indicates that the perturbation
may well extend on several decades. In the absence of ions, the
precursor size would be set by the highest energy electrons; balancing
acceleration at a Bohm rate (as experienced downstream) to synchrotron
losses, one would find a precursor size about 20 times smaller for the
adopted fiducial parameters, nevertheless much larger than the typical
size of the fluctuations.

Finally, Keshet et al. (2009) suggest a damping frequency $\Im\omega
\propto \lambda^{-2}$, which implies, when combined with the above
result that the decay of the microturbulence might extend over quite
long spatial scales.

Following Chang et al. (2008), the magnetic field power spectrum is
described as a time decaying powerlaw form in the downstream
(comoving) frame, with
\begin{equation}
  \langle\delta B(t)^2\rangle\, = \,a_B\,\delta B_\mu^2
  \,\int_{\lambda_{\mu}}^{\lambda_{\rm max}} 
\frac{{\rm d}\lambda}{\lambda_{\mu}} 
  \left(\frac{\lambda}{\lambda_{\mu}}\right)^{\alpha_B}\exp\left(-\frac{
      t}{\tau_\lambda}\right)\ , \label{eq:Bturb}
\end{equation}
with $t$ denoting the time (downstream frame) since shock entry of the
corresponding plasma element, $\lambda_{\mu}$ (resp. $\lambda_{\rm
  max}$) the minimum (resp. maximum) wavelength scale of the
microturbulence at $t=0$, $\alpha_B<-1$ so that the turbulent power
lies at the smallest scales, $a_B\,\equiv\,\vert 1+\alpha_B\vert$ for
normalization purposes, $\delta B_\mu$ denotes the rms field strength
at $t=0$ and $\tau_{\lambda}=\left\vert\Im\omega\right\vert^{-1}$ the
damping time, which depends on $\lambda$:
\begin{equation}
\tau_\lambda \,\equiv\, \omega_{\rm pi}^{-1}
\left(\omega_{\rm pi}\lambda/c\right)^{\alpha_\lambda}\ .\label{eq:taul}
\end{equation}

Assuming $\lambda_{\rm max}\gg\lambda_{\mu}$ for the moment,
Eq.~(\ref{eq:Bturb}) can be integrated in terms of an incomplete Gamma
function,
\begin{eqnarray}
  \langle\delta B(t)^2\rangle &\,=\,& \delta B_\mu^2 \,\alpha_{\lambda}^{-1}
  \mu(t)^{(1+\alpha_B)/\alpha_\lambda}\nonumber\\
  & & \,\,\times \left\{\Gamma\left(-\frac{1+\alpha_B}{\alpha_\lambda}\right) 
    - \Gamma\left[-\frac{1+\alpha_B}{\alpha_\lambda};\,
      \mu(t)\right]\right\}\ ,\nonumber\\
& & \label{eq:dbt}
\end{eqnarray}
with
\begin{equation}
  \mu(t)\,\equiv\, \omega_{\rm pi}\,t\,\left(\frac{\omega_{\rm pi}\lambda_{\mu}}{c}\right)^{-\alpha_\lambda}\ .\label{eq:ufac}
\end{equation}
For convenience, one may approximate Eq.~(\ref{eq:dbt}) with
respectively the small and large argument limits to describe the
evolution of the magnetic field strength as
\begin{equation}
  \langle\delta B(t)^2\rangle \,\simeq\,\begin{cases}

\delta B_\mu^2  &\text{if}\,\,\mu(t)< 1\ ,\\
\delta B_\mu^2 \Gamma\left(1+\vert\alpha_t\vert\right)
  \mu(t)^{\alpha_t} & \text{if}\,\,\mu(t)\gg1\ ,
\end{cases}\label{eq:Bevol}
\end{equation}
with the following definition
\begin{equation}
\alpha_t \,\equiv\, \frac{1+\alpha_B}{\alpha_\lambda}\,<0\ .\label{eq:alphat}
\end{equation}
In the following, the numerical factor
$\Gamma\left(1+\vert\alpha_t\vert\right)$, of order unity, will be
dropped henceforth.

Equation~\ref{eq:alphat} shows that the temporal decay index of the
magnetic field behind the shock is inherently linked to how power is
distributed on scales larger than the minimum scale $\lambda_\mu$ --
as characterized by $\alpha_B$ -- and to how fast small scale features
are dissipated -- as characterized by $\alpha_\lambda$. The
interpretation for this is clear: as small scales are erased, magnetic
power is removed, but the rate at which the total strength erodes
depends on how much strength is left at longer wavelengths. In fine,
the uncertainty on $\alpha_t$ is related to the sourcing of large
wavelength fluctuations, which are likely related to the dynamics of
high energy particles in the upstream. The above assumption
$\lambda_{\rm max}\gg\lambda_{\mu}$ has been discussed above. Its
robustness depends crucially on the influence of maximal energy
particles upstream of the shock front. In the extreme opposite case
$\lambda_{\rm max}\sim\lambda_{\mu}$, one should observe a roughly
constant magnetic field while $\mu(t)<1$, followed by fast decay once
$\mu(t)>1$. This situation may be accounted for by Eq.~\ref{eq:Bevol}
with a more pronounced value of $\alpha_t$.

The following therefore considers a range of possibilities for
$\alpha_t$, even though the PIC simulations of Chang et al. (2008) and
Keshet et al. (2009) both suggest $-1<\alpha_t<0$. More specifically,
Chang et al. (2008) suggest that the magnetic field Fourier spectrum
(for $\delta B$, not $\delta B^2$) in wavenumber has slope $\simeq
0\rightarrow 1/2$, which corresponds to $1+\alpha_B\sim -2\rightarrow
-1$, and $\alpha_\lambda=3$ leading to $\alpha_t \sim -1/3 \rightarrow
-2/3$. Keshet et al. (2009) show that right behind the shock front,
the magnetic field decays exponentially on short a distance scale to
level off at a strength corresponding to $\epsilon_B\sim 10^{-2}$ for
some hundreds of skin depth\footnote{This result motivates the present
  choice of $\epsilon_B=10^{-2}$ as a fiducial value, even though the
  magnetic energy density reaches $\sim15\,$\% of the incoming energy
  at the shock transition itself (see Chang et al. 2008, Keshet et
  al. 2009).}. A closer inspection of their Fig.3 however reveals that
the initial exponential decay leaves way to a powerlaw decay at late
simulation times (thus meaning far downstream) and by eye, one
estimates $\alpha_t\sim -0.5$. These simulations thus indicate a value
of $\alpha_t$ between $-1$ and $0$; however, given the present
limitations of the PIC simulations, and the above possible caveat
related to the extension of the magnetic perturbation spectrum, one
cannot exclude yet that $\alpha_t<-1$. In this respect, one must point
out that recent simulations of the development and the dynamics of
relativistic Weibel turbulence indeed suggest a value $\alpha_t\simeq
-2$ (Medvedev et al. 2010). Although these simulations do not simulate
the shock itself, but a Weibel turbulence through the interpenetration
of two relativistic beams, these are 3D while the shock simulations of
Chang et al. (2008) and Keshet et al. (2009) are 2D.

In order to account for these different possibilities in the
following, Eq.~\ref{eq:Bevol} is kept in its present form, but the
decay exponent $\alpha_t$ is assumed to take possibly mild or more
pronounced values. Depending on whether $\alpha_t<-1$ or
$-1<\alpha_t<0$, it will be seen that radically different radiative
signatures are to be expected.

In summary, theoretical arguments combined with recent high
performance PIC simulations suggest the following characterization for
the evolution of the microturbulence behind a relativistic (weakly
magnetized) shock front. Immediately behind the shock, the magnetic
field carries strength $\delta B_\mu$ corresponding to an
equipartition parameter $\epsilon_B\equiv \delta
B_\mu^2/\left(32\pi\gamma_{\rm b}^2n m_p c^2\right)$ with fiducial
value $\epsilon_{B}\sim 10^{-2}$, while $\lambda_{\mu}\sim 10-30
c/\omega_{\rm pi}$ represents the fiducial value for the coherence
scale at that same location. The magnetic field strength decays as
$t^{\alpha_t}$ after a time
\begin{equation}
  t_{\mu+}\,\equiv\,\omega_{\rm
    pi}^{-1}\left(\omega_{\rm
      pi}\lambda_{\mu}/c\right)^{\alpha_\lambda}\ ,
\end{equation}
defined through $\mu(t_{\mu+})\,\equiv\,1$, of the order of hundreds
to thousands of inverse plasma times, until it eventually settles at
the shock compressed value $B_{\rm d}=4\gamma_{\rm b}B_{\rm u}$.  In
the following, this timescale is rewritten in units of inverse plasma
times as
\begin{equation}
\Delta_\mu\,\equiv\,\omega_{\rm pi} t_{\mu+}\ ,\label{eq:dmu}
\end{equation}
meaning also that the undecayed part of the microturbulence extends
for $\Delta_\mu$ skin depths. Note that $\Delta_\mu\,\gg\,1$ according
to the above simulations.  Finally, the coherence length of the
microturbulence evolves as $t^{1/\alpha_\lambda}$, with fiducial value
$\alpha_\lambda\sim 2-3$.

\subsection{Radiation in time decaying microturbulence}\label{sec:radmu}
As a particle gets Fermi accelerated, it interacts with the turbulent
layer within a scattering length scale $l_{\rm scatt}$ of the shock
front. This scattering length scale controls the residence time hence
the acceleration timescale hence the maximal energy that can be
reached; it is discussed in more detail in
Sec.~\ref{sec:accel}. For the time being, it suffices to note that the
Larmor radius of the bulk of the electrons, with minimum Lorentz
factor $\gamma_{\rm m}=\vert p-1\vert^{-1}\vert p-2\vert\epsilon_e
\gamma_{\rm bl}m_p/m_e$ is so small that these electrons can only
explore the undecayed part of the turbulence:
\begin{equation}
  r_{\rm L}\left(\gamma_{\rm m}\right)
  \,\approx\,\epsilon_{e,-0.3}\epsilon_{B,-2}^{-1/2}
  \frac{c}{\omega_{\rm pi}}\,\ll\,\lambda_\mu\ .
\end{equation}
The microturbulence thus controls the acceleration of the bulk of
electrons, independently of how fast this microturbulence decays or
how large the blast Lorentz factor may be.  At Lorentz factors
$\gg\gamma_{\rm m}$ of interest for high energy radiation, the
electrons may start to explore the region $\mu(t)>1$. Then the
transport becomes non trivial; its impact on acceleration is discussed
in Sec.~\ref{sec:accel}.

During the acceleration stage, the particle moves in a near ballistic
manner and diffusive effects can be neglected, given that the return
probability decreases fast with the number of steps of length $l_{\rm
  scatt}$ taken. Therefore, as a particle moves away from the shock on
a distance scale $x$ along the shock normal during a time $t_{\rm
  p}\sim x/(c\cos\theta)$, with $\theta$ the angle to the shock
normal, it explores a microturbulence that has decayed according to
the laws given above with $t\simeq3\cos\theta\,t_{\rm p}$. This factor
of $3$ of course results from the convective velocity $c/3$ of the
downstream plasma.

On length scales much larger than $l_{\rm scatt}$, a particle diffuses
in the microturbulence and in a first approximation, one can describe
its transport by advection with the downstream plasma. For such
particles, $t\simeq t_{\rm p}$. The above slight difference between
$t$ and $t_{\rm p}$ does not impact the results given further below
and can be neglected in view of the uncertainties related to the time
evolution of the turbulence. In the following, $t$ and $t_{\rm p}$
are thus be used interchangeably.

The cooling history of an electron of Lorentz factor $\gamma_e$ obeys
the standard law
\begin{equation}
\frac{{\rm d}\gamma_e}{{\rm d}t} \,=\, -\frac{1}{6\pi} \sigma_{\rm
  T}\frac{\delta B(t)^2(1+Y)}{m_e c}\gamma_e^2\ ,\label{eq:cool1}
\end{equation}
with $Y$ the Compton cooling factor. For the time being, one considers
the simple case $Y\ll1$; the influence of inverse Compton losses is
discussed in detail in App.~\ref{sec:appFnu} and further below. One
then defines a Lorentz factor $\gamma_{\mu+}$ such that, if
$\gamma_e>\gamma_{\mu+}$, the particle cools in the undecayed
microturbulence where $\mu(t)<1$ (at time $t<t_{\mu+}$), while if
$\gamma_e<\gamma_{\mu+}$, the particle does not cool in that layer but
further on. Writing the synchrotron cooling time for particles of
Lorentz factor $\gamma_e$ in a (constant) magnetic field of strength
$\delta B$ as $t_{\rm syn}\left[\gamma_e;\delta B\right]$, the Lorentz
factor $\gamma_{\mu+}$ can also be defined as the solution of
\begin{equation}
\gamma_{\mu+}:\quad \mu\left[t_{\rm
    syn}\left(\gamma_{\mu+};\delta B_{\mu}\right)\right]\equiv 1\ .\label{eq:gmu+}
\end{equation}
In terms of the fiducial values of interest here,
\begin{equation}
  \gamma_{\mu+}\,\simeq\, 3 \times 10^{9}\, t_{2}^{3/4}
  E_{53}^{-1/4}n_{-3}^{-1/4}\epsilon_{B,-2}^{-1}\Delta_{\mu,2}^{-1} \ ,\label{eq:gmu+v}
\end{equation}
with $\Delta_{\mu,2}=\Delta_{\mu}/100$. This value of $\gamma_{\mu+}$
generally exceeds the maximal Lorentz factors that can be achieved
through shock acceleration, therefore particles cool outside this
undecayed turbulent layer, unless $\Delta_{\mu}$ is larger than
expected, as discussed in Sec.~\ref{sec:accel}. The latter may well
happen, if for instance $\alpha_\lambda>2$ and/or
$\lambda_\mu\gg10c/\omega_{\rm pi}$.

If a particle exits the acceleration process with a Lorentz factor
$\gamma_{e,0}$, then at time $t$,
\begin{equation}
  \gamma_e \,\simeq\,\begin{cases}
    \displaystyle{\frac{\gamma_{e,0}}{1+\gamma_{e,0}/\gamma_{\mu+}}} &
    (\mu(t)\ll1)\\  
    \displaystyle{\frac{\gamma_{e,0}}{1+(1+\alpha_t)^{-1}
        \left[\mu(t)^{1+\alpha_t}+\alpha_t\right]\gamma_{e,0}/\gamma_{\mu+}}}&
    (\mu(t)\gg1)\ .
\end{cases}
\label{eq:cool2}
\end{equation}
If $\gamma_{e0}>\gamma_{\mu+}$, the particle cools down to
$\gamma_{\mu+}$ within the layer where $\mu(t)<1$ (i.e. $t<t_{\mu+}$)
and subsequently, it continues cooling if $-1<\alpha_t<0$ or stops its
cooling if $\alpha_t<-1$. If $\gamma_{e0}<\gamma_{\mu+}$ on the
contrary, the particle either cools later in the decaying
microturbulence if $-1<\alpha_t<0$, or not if $\alpha_t<-1$. In any
case, the particle  of course eventually cools in the background
shock compressed field (notwithstanding issues related to the
available hydrodynamical timescale). Whichever occurs influences the
afterglow light curve and spectral energy distribution.

If $-1<\alpha_t<0$, it is convenient to define a second Lorentz
factor, $\gamma_{\mu-}$, as the Lorentz factor for which cooling
occurs on a timescale $t_{\mu-}$ such that $\delta
B\left(t_{\mu-}\right)=B_{\rm d}$, i.e. at the time at which the
turbulence field has relaxed to the background shock compressed value
$B_{\rm d}=4\gamma_{\rm bl}B_{\rm u}$. For the time being, no
consideration is made of the hydrodynamical timescale of the
blast. Then, if $\gamma_{\rm m}>\gamma_{\mu-}$ (and $\alpha_t>-1$),
most particles cool in the decaying microturbulent layer. One finds
\begin{equation}
\frac{\gamma_{\mu-}}{\gamma_{\mu+}}\,\simeq\, \left(\frac{\sigma_{\rm
      u}}{\epsilon_{B}}\right)^{-(1+\alpha_t)/\alpha_t}\ ,
\end{equation}
in terms of the upstream magnetization parameter $\sigma_{\rm u}
\,\equiv\, B_{\rm u}^2/(4\pi n m_p c^2)$. The Lorentz factor
$\gamma_{\mu-}$ depends exponentially on $\alpha_t$; it may therefore
take very different values.

To calculate the radiative signature, one integrates over the cooling
history of the electron population, as in Gruzinov \& Waxman (1999),
although for simplicity, the calculation is done in a one-dimensional
quasi-steady state approximation, meaning that the secular
hydrodynamical evolution of the blast is neglected in the course of
this integration over the blast width. This method corresponds to the
steady state approximation of Sari et al. (1998), when calculating the
stationary electron distribution in a homogeneous shell. The spectral
power density of the blast can then be written in the downstream frame
in terms of an integral over a particle history, up to a (comoving)
hydrodynamical timescale $t_{\rm dyn}$ (see Eq.~\ref{eq:tdyn})
\begin{equation}
  P'_{\nu} \,=\, \int_{\gamma_{\rm m}}^{\gamma_{\rm max}}\, \frac{{\rm
      d}\dot N_e}{{\rm d}\gamma_{e,0}}\,{\rm d}\gamma_{e,0}\,\,\int_{0}^{t_{\rm dyn}} {\rm d}t\,\frac{{\rm d}E_{\rm
      syn}(\gamma_{e,0})}{{\rm d}\nu{\rm d}t}\ ,
\end{equation}
with ${\rm d}E_{\rm syn}/{\rm d}\nu{\rm d}t$ the spectral power
density radiated by an electron at time $t$, of initial Lorentz factor
$\gamma_{e,0}$ and of cooling history given by Eq.~(\ref{eq:cool2});
see also Eq.~\ref{eq:appdFnu}. The above expression is folded over the
injection distribution ${\rm d}\dot N_e/{\rm d}\gamma_{e,0}$, which is
assumed to take a power law form between $\gamma_{\rm m}$ and
$\gamma_{\rm max}\gg\gamma_{\rm m}$:
\begin{equation}
{\rm d}\dot N_e \,=\, \dot N_e \, \frac{\vert1-p\vert}{\gamma_{\rm
  m}}\left(\frac{\gamma_{e,0}}{\gamma_{\rm m}}\right)^{-p}\,{\rm
  d}\gamma_{e,0}\ ,\label{eq:dNedg}
\end{equation}
with
\begin{equation}
  \dot N_e \,=\, \gamma_{\rm b}\left(\beta_{\rm b}+\frac{1}{3}\right)4\pi n r^2 c\label{eq:dNedt}
\end{equation}
the number of electrons swept and shock accelerated by the shock wave
per unit time, as measured in the downstream frame.

The detailed calculation of the spectral power $P_{\nu}$ is carried
out in Appendix~\ref{sec:appFnu} for different relevant cases. In
particular, two distinctions have to be made: whether the
microturbulence decays rapidly beyond $t_{\mu+}$ or not and whether
inverse Compton losses contribute significantly to the cooling of
electrons. These cases are examined in turn in the next subsections,
in parallel to the discussion of App.~\ref{sec:appFnu}.

\subsection{Gradual decay, no inverse Compton losses}\label{sec:gradnoIC}
As discussed in App.~\ref{sec:appFnu}, the gradual evolution of the
microturbulence behind the shock affects the spectro-temporal flux
$F_\nu$ in various ways. For one, particles of different Lorentz
factors cool in different magnetic fields, with particles of lower
energy experiencing lower magnetic fields at cooling. This implies
that the characteristic synchrotron frequencies are modified and more
specifically, that ratios of characteristic frequencies are stretched
with respect to the standard case of a homogeneous turbulent
layer. Regarding the cooling frequency $\nu_{\rm c}$, the modification
is non-trivial because the cooling Lorentz factor $\gamma_{\rm c}$
itself depends in a non-trivial way on the temporal decay index of the
turbulence, see Eq.~\ref{eq:gammac2}. The characteristic frequency
$\nu_{\rm m}$ associated to particles of Lorentz $\gamma_{\rm m}$ can
also be modified in a non-trivial way, since $\nu_{\rm m}\,=\,\nu_{\rm
  p}\left[\gamma_{\rm m}; \delta B_{\gamma_{\rm m}}\right]$, with
$\nu_{\rm p}$ defined as the synchrotron peak frequency of particles
of Lorentz factor $\gamma_{\rm m}$ in a magnetic field of strength
$\delta B_{\gamma_{\rm m}}$, see Eq.~\ref{eq:nusyn}. The magnetic
field $\delta B_{\gamma_{\rm m}}$ in which the particles of Lorentz
factor $\gamma_{\rm m}$ radiate most of their energy takes the shock
compressed value $B_{\rm d}$ if both $\gamma_{\rm m}<\gamma_{\mu-}$
and $t_{\rm dyn}>t_{\mu-}$, but $\delta B_\mu (t_{\rm
  dyn}/t_{\mu+})^{\alpha_t/2}$ if $t_{\rm dyn}<t_{\mu-}$ and
$\gamma_{\rm m}<\gamma_{\rm c}$, or $\delta B_\mu (\gamma_{\rm
  m}/\gamma_{\mu+})^{-\delta_t}$, with
$\delta_t=\alpha_t/[2(1+\alpha_t)]$ defined in Eq.~\ref{eq:delta}.
The ratio $t_{\rm dyn}/t_{\mu-}$ determines whether the turbulence has
relaxed to $B_{\rm d}$ by the back of the blast or not, and this value
depends exponentially on $\alpha_t$:
\begin{eqnarray}
  \frac{t_{\rm dyn}}{t_{\mu-}}&\,\simeq\,& 1.6\times 10^4
  e^{6.85/\alpha_t} 
  B_{-5}^{-2/\alpha_t} n_{-3}^{3/8+1/\alpha_t}\nonumber\\
& &\,\,\,\times E_{53}^{1/8}\epsilon_{B,-2}^{1/\alpha_t}
  \Delta_{\mu,2}^{-1}t_2^{5/8}z_{+,0.3}^{-5/8}  \ .
\end{eqnarray}
In practice, it can take small or large values at different times,
even for $\alpha_t=-0.5$ for which the numerical prefactor becomes
$0.018$.

In direct consequence of the above, the deceleration of the blast
implies a non-trivial temporal evolution of the characteristic
frequencies. This modifies the standard temporal evolution of $F_\nu$.
Furthermore, as a particle gets advected away from the shock with the
microturbulence, it radiates at decreasing frequencies, whether it
cools efficiently or not. Consequently, the changing magnetic field
also modifies the spectral slope of the flux $F_\nu$.

Appendix~\ref{sec:appFnu} provides a detailed discussion of the
possible cases, with detailed expressions for the characteristic
frequencies $\nu_{\rm m}$, $\nu_{\rm c}$, $\nu_{\mu+}$ and
$\nu_{\mu-}$, depending on their respective orderings. Following
App.~\ref{sec:appFnu}, the present discussion does not consider the
cases in which either $\nu_{\rm m}>\nu_{\mu+}$ or $\nu_{\rm
  c}>\nu_{\mu+}$, because such cases appear rather extreme in terms of
the parameters characterizing the turbulence, as discussed in the
former Section. Moreover, these cases tend to the standard model of a
synchrotron afterglow in a homogeneous turbulence when both $\nu_{\rm
  m}>\nu_{\mu+}$ and $\nu_{\rm c}>\nu_{\mu+}$ and they can  be
easily recovered from App.~\ref{sec:appFnu}.

\begin{figure}
\includegraphics[width=0.49\textwidth]{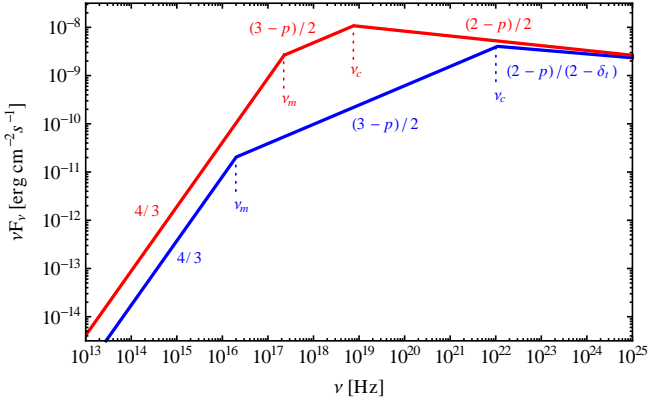}
\caption{Comparison between two synchrotron spectra: in a homogeneous
  turbulence of strength $\epsilon_B=10^{-2}$ (top red curve), and in
  a decaying microturbulence such that $\alpha_t=-0.5$,
  $\Delta_{\mu}=10^{2}$.  For both, the blast parameters are $\gamma_{\rm
    b}=245$, $n_{-3}=1$ and the injection distribution index
  $p=2.2$. The regime is slow cooling, and $t_{\rm dyn}<t_{\mu-}$
  (corresponding to case 1 of Fig.~\ref{fig:spec_asupm1} for the
  microturbulent model).
  \label{fig:spec1} }
\end{figure}

Figure~\ref{fig:spec1} presents a concrete example of a spectral
energy distribution, comparing the standard prediction for a
homogeneous turbulence with $\epsilon_B=10^{-2}$ (blue line) to a time
evolving microturbulence with $\alpha_t=-0.5$, also starting at
$\epsilon_B=10^{-2}$ (red line), with $\Delta_{\mu}=10^{2}$. Both models
assume $\gamma_{\rm b}=245$ corresponding to an observer time $t_{\rm
  obs}=100\,$s for a blast at $z=1$ with $E_{53}=1$, $n_{-3}=1$, an
injection slope $p=2.2$ and a circumburst medium of constant
density. Inverse Compton losses are neglected throughout the blast in
this example.  Figure~\ref{fig:spec1} reveals the characteristic
stretch of frequency range, with $\nu_{\rm m}\simeq 2\times
10^{16}\,$Hz for the dynamical microturbulent model (resp. $\nu_{\rm
  m}\simeq 2.3\times 10^{17}\,$Hz in the homogeneous turbulence) and
$\nu_{\rm c}\simeq 1.1\times10^{22}\,$Hz (resp. $\nu_{\rm c}\simeq
7.6\times 10^{18}\,$Hz). The frequency $\nu_{\mu+}=2\times
10^{27}\,$Hz lies outside the range of Fig.~\ref{fig:spec1}.

For the above case of slow cooling, the spectral indices at low and
intermediate frequencies, respectively $\beta=1/3$ and
$\beta=-(1-p)/2$ (with $F_\nu\propto t^{\alpha}\nu^{-\beta}$) remain
unaffected in the presence of decaying microturbulence. However the
temporal index $\alpha$ is modified in these cases, even at low
frequencies, which opens the possibility of testing such cases through
the temporal behavior of an early follow-up in the
optical. Section~\ref{sec:grb} below offers a comparison of such
spectra with the observed light curve of GRB090510. Regarding the fast
cooling part of the electron population, both the spectral and
temporal indices are modified, see Table~\ref{tab:gradnoIC} for their
detailed values. See also Fig.~\ref{fig:spec_asupm1} for an
illustration of possible synchrotron spectra.

Synchrotron self-absorption is negligible at all frequencies shown in
this figure, see Sec.~\ref{sec:synch-self} for details.

\subsection{Rapidly decaying microturbulence (no inverse Compton
  losses)}\label{sec:fastnoIC}
There are two main differences between the synchrotron spectra of
gradually vs rapidly decaying microturbulence. In the former case,
particles may cool in the decaying microturbulence layer, while in the
latter, particles either cool in the undecayed region of short extent,
if their Lorentz is sufficiently large, or in the background magnetic
field beyond the microturbulent layer otherwise, in the absence of
inverse Compton losses that is. This implies in particular that there
is no well defined cooling Lorentz factor for the microturbulence.
Secondly, as the turbulence decays rapidly, most of the synchrotron
power is emitted in the region of largest magnetic power, hence the
flux $\nu F_\nu$ associated to the microturbulent layer peaks at
$\nu_{\mu+}$. At times such that $t_{\rm dyn}>t_{\mu-}$, cooling in
the background shock compressed field leads to the emergence of a
secondary synchrotron component on top of the former, and one may now
define a cooling Lorentz factor in the background compressed field. At
frequency $\nu_{\rm m}$, the ratio between these two components is of
order $t_{\rm dyn}/t_{\mu-}>1$, at the benefit of the latter. At the
exit of the microturbulent layer, the maximal Lorentz factor cannot
exceed $\gamma_{\mu+}$, so that the secondary component cuts-off at
most at $\nu_{\rm p}\left[\gamma_{\mu+};B_{\rm d}\right]$, which falls
short of the GeV range, see the discussion in Sec.~\ref{sec:accel}. In
practice, this suggests that most of the low energy emission in the
optical and X-ray domains result from cooling in the background field,
while the highest energy emission can be attributed to the presence of
the microturbulence.

\begin{figure}
\includegraphics[width=0.49\textwidth]{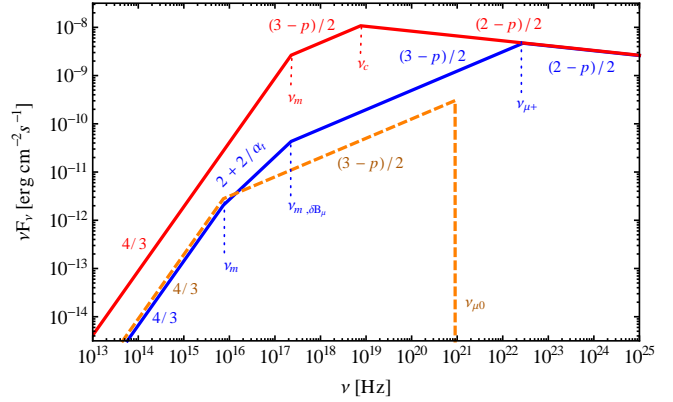}
\caption{Same as Fig.~\ref{fig:spec1}, now comparing the synchrotron
  spectrum in a homogeneous turbulence of strength
  $\epsilon_B=10^{-2}$ (top red curve), with a decaying
  microturbulence such that $\alpha_t=-1.8$, $\Delta_{\mu}=2.7\times
  10^{4}$ (bottom blue curve).  For both, the blast parameters are
  $\gamma_{\rm b}=245$, $n_{-3}=1$ and the injection distribution
  index $p=2.2$. The orange dashed line represents the secondary
  synchrotron component associated to cooling in the background shock
  compressed field (here $B_{\rm u}=10\,\mu$G), which emerges on top of
  the microturbulent component because $t_{\rm dyn}=1.3t_{\mu-}$
  (corresponding to case 2 of Fig.~\ref{fig:spec_ainfm42pp1}). The
  regime is slow cooling.
  \label{fig:spec2} }
\end{figure}

Figure~\ref{fig:spec2} presents a concrete example of a spectral
energy distribution, comparing the standard prediction for a
homogeneous turbulence with $\epsilon_B=10^{-2}$ (top red line) to a
time evolving microturbulence with $\alpha_t=-1.8$, also starting at
$\epsilon_B=10^{-2}$ (bottom blue curve). The microturbulent layer is
such that $\Delta_{\mu}=2.7\times 10^{4}$, corresponding for instance to
$\alpha_\lambda=3$, $\lambda_\mu=30\,c/\omega_{\rm pi}$. The upstream
magnetic field $B_{\rm u}=10\,\mu$G and as in Fig.~\ref{fig:spec1},
$n_{-3}=1$, $\gamma_{\rm b}=245$, $t_{\rm obs}=100\,$s, $z=1$, which
implies in particular that $t_{\rm dyn}=1.3t_{\mu-}$. Therefore, a
secondary synchrotron associated to cooling in $B_{\rm d}$ emerges on
top of the microturbulent component.  The corresponding characteristic
frequencies for the microturbulent component are $\nu_{\rm
  m}\,\simeq\, 5.7\times 10^{15}\,$Hz, $\nu_{{\rm m},\delta
  B_\mu}\,\simeq\, 2.3\times 10^{17}\,$Hz and $\nu_{\mu+}\,\simeq\,
2.8\times 10^{22}\,$Hz; the cut-off frequency of the secondary
synchrotron component $\nu_{\mu0} = 0.9\times 10^{21}\,$Hz. In such a
scenario, the microturbulent component would dominate in the X-ray and
at higher energies, while the secondary component dominates in the
optical; at later times, the secondary component would come to dominate
as well in the X-ray.

Regarding the spectro-temporal evolution of the flux, one recovers
features similar to those discussed above in the case
$\alpha_t>-1$. In particular, the slow cooling slopes $1/3$ and
$(1-p)/2$ remain unaffected, but the temporal index is different when
$t_{\rm dyn}<t_{\mu-}$, and for the fast cooling part of the electron
population, both spectral and temporal indices are affected by
$\alpha_t$. The detailed values of $\alpha$ and $\beta$ are given in
Table~\ref{tab:fastnoIC} in Appendix~\ref{sec:appFnu}. See also
Figs.~\ref{fig:spec_ainfm42pp1} and \ref{fig:spec_asupm42pp1} for an
illustration of the spectral shapes of the synchrotron spectra.

Synchrotron self-absorption is negligible at all frequencies shown in
this figure, see Sec.~\ref{sec:synch-self} for details.

\subsection{Strong inverse Compton losses}
Accounting for inverse Compton losses modifies of course the cooling
history of the particle. The importance of inverse Compton losses is
generally quantified through the $Y$ Compton parameter, which may be
written in a first approximation (e.g. Sari \& Esin 2001, Wang et
al. 2010) as
\begin{equation}
Y(\gamma)\left[1+Y(\gamma)\right]\,\simeq\,
\frac{\epsilon_e}{\epsilon_B}\frac{\nu F_\nu\left[\nu_{\rm
      KN}(\gamma)\right]}
{\nu F_\nu\left(\nu_{\rm peak}\right)} f_{\rm cool}\ ,
\end{equation}
where the notation $Y(\gamma)$ indicates that $Y$ in general depends
on the Lorentz factor of the particle, due to Klein-Nishina
effects. These latter are quantified through the ratio of $\nu
F_\nu\left[\nu_{\rm KN}(\gamma)\right]$, i.e. the synchrotron flux at
the frequency at which Klein-Nishina suppression becomes effective, to
the peak of the synchrotron flux, $\nu F_\nu\left(\nu_{\rm
    peak}\right)$. Finally, the factor $f_{\rm cool}$ denotes the
fraction of cooling electrons, $\simeq 1$ for fast cooling, $\simeq
\left(\gamma_{\rm c}/\gamma_{\rm m}\right)^{1-s}$ for slow cooling.
The double dependence of $Y$ on Lorentz factor $\gamma$ and distance
to the shock front (through $\delta B$) that arises in the present
model renders this problem quite complex. To simplify this task, it is
assumed here that inverse Compton losses dominate everywhere
throughout the blast. This would occur for instance, if
$\epsilon_e>\epsilon_B$ in the undecayed part of the microturbulent
layer (a generic assumption), if $f_{\rm cool}$ does not lie too far
below unity, and at energies such that Klein-Nishina effects can be
neglected. Under such conditions, one may expect $Y>1$ everywhere
else, because the magnetic field decays away from the shock.
 
\begin{figure}
\includegraphics[width=0.49\textwidth]{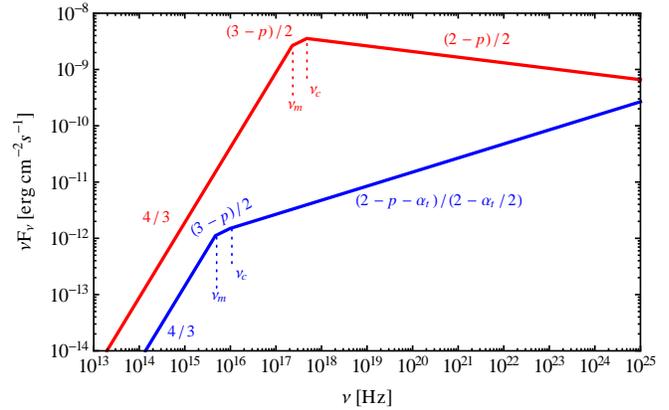}
\caption{Same as Fig.~\ref{fig:spec1}, now comparing the synchrotron
  spectrum in a homogeneous turbulence of strength
  $\epsilon_B=10^{-2}$ (top red curve) including inverse Compton
  losses with $Y=3$, with a decaying microturbulence such that
  $\alpha_t=-0.8$, $\Delta_{\mu}=10^{2}$, also including inverse Compton
  losses with $Y_\mu=3$ (bottom blue curve).  For both, the blast
  parameters are $\gamma_{\rm b}=245$, $n_{-3}=1$ and the injection
  distribution index $p=2.2$. Here $t_{\rm dyn}\,\ll\,t_{\mu-}$
  (corresponding to case 1 of Fig.~\ref{fig:spec_ICasupm42pp1}). The
  regime is slow cooling.
  \label{fig:spec3} }
\end{figure}

In the presence of inverse Compton losses, one can always define a
cooling Lorentz factor $\gamma_{\rm c}$ irrespectively of the value of
$\alpha_t$. As the cooling history of the particles is modified, so
are the spectral slopes in the fast cooling part if
$\alpha_t>-4/(p+1)$, and also at low frequencies if
$\alpha_t<-4/(p+1)$. In this latter situation indeed, the synchrotron
spectrum is dominated at high frequencies by a slow cooling spectrum
in the undecayed microturbulence, with a low energy extension from
$\nu_{\rm m}$ to $\nu_{{\rm m},\delta B_\mu}$ [see
Eq.~\ref{eq:numdBm}] with slope $1+2/\alpha_t$. At low frequencies, a
secondary component associated to cooling in the background compressed
field emerges if both $t_{\rm dyn}<t_{\mu-}$ and $\gamma_{\rm
  m}<\gamma_{\mu-}$. The synchrotron spectra for $\alpha_t>-1$ share
features similar to those discussed in Sec.~\ref{sec:gradnoIC}, up to
the modifications of the spectro-temporal indices. One noteworthy
distinction is the fact that the fast cooling spectral index has
become $-(p+\alpha_t/2)/(2-\alpha_t/2)$, which may become larger than
$-1$ if $\alpha_t<2-p$, in which case most of the power lies at
$\nu_{\mu+}$. This may be attributed to a scaling of the Compton
parameter with frequency: higher frequencies correspond to higher
Lorentz factors, hence to cooling in a region of higher magnetic
field, where the Compton parameter is smaller, which implies that a
smaller fraction of the particles energy is channeled into the inverse
Compton component.

Figure~\ref{fig:spec3} presents a concrete example of a spectral
energy distribution, comparing the standard prediction for a
homogeneous turbulence with $\epsilon_B=10^{-2}$ including inverse
Compton losses with $Y=3$ (top red line) to a time evolving
microturbulence with $\alpha_t=-0.8$, also starting at
$\epsilon_B=10^{-2}$, with $\Delta_{\mu}=10^{2}$ and $Y_\mu=3$ (bottom blue
curve). For this case, $t_{\rm dyn}\,\ll\,t_{\mu-}$, corresponding to
case 1 of Fig.~\ref{fig:spec_ICasupm42pp1}; other parameters remain
unchanged compared to previous figures.  The corresponding
characteristic frequencies for the microturbulent component are
$\nu_{\rm m}\,\simeq\, 4.7\times 10^{15}\,$Hz, $\nu_{\rm c}\,\simeq\,
1.0\times 10^{16}\,$Hz and $\nu_{\mu+}\,\simeq\, 1.3\times
10^{26}\,$Hz.

The spectro-temporal indices $\alpha$ and $\beta$ are given in
Table~\ref{tab:gradIC} for $\alpha_t>-4/(p+1)$ and \ref{tab:rapIC} in
the opposite limit $\alpha_t<-4/(p+1)$. The generic spectral shapes
are illustrated in Figs.~\ref{fig:spec_ICasupm42pp1} and
\ref{fig:spec_ICainfm42pp1} for these two cases, respectively.

Here as well, synchrotron self-absorption is negligible at all
frequencies shown in this figure, see Sec.~\ref{sec:synch-self} for
details.

\section{Discussion}\label{sec:disc}
Most of the discussion so far has ignored the characteristic frequency
associated to the maximal energy of the acceleration process. Yet, the
production of GeV photons through synchrotron radiation of electrons
already push ultra-relativistic Fermi acceleration to its limits, as
discussed e.g. in Piran \& Nakar (2010), Kirk \& Reville (2010),
Barniol-Duran \& Kumar (2010, 2011a), Lemoine \& Pelletier (2011c),
Sagi \& Nakar (2012), Bykov et al. (2012). These studies assume a
homogeneous downstream turbulence, with either microscale high power
turbulence or simply a shock compressed magnetic
field. Section~\ref{sec:accel} extends these calculations to an
evolving microturbulence, which brings in further constraints and new
phenomena. A brief comparison with the light curve of GRB090510, which
so far provides the earliest follow-up in the optical through GeV, is
then proposed.

\subsection{Acceleration to high energies}\label{sec:accel}
The maximal energy is determined through the comparison of the
acceleration timescale $t_{\rm acc}$ to other relevant timescales,
e.g. the energy loss and dynamical timescales. Given that the energy
gain per Fermi cycle is of order 2 (Gallant \& Achterberg 1999,
Achterberg et al. 2001, Lemoine \& Pelletier 2003), it suffices to
compare in each respective rest frame the downstream $t_{\rm res\vert
  d}$ and the upstream $t_{\rm res\vert u}$ residence times to the
other timescales.

\subsubsection{Upstream}
The upstream residence time $t_{\rm res\vert u}\simeq 5 t_{\rm
  L,0\vert u}/\gamma_{\rm sh}$ (Lemoine \& Pelletier 2003) -- assuming
that the particle only interacts with the background field -- where
$t_{\rm L,0}= \gamma_{\rm b}\gamma_e m_e c / (eB_{\rm u})$ and
$\gamma_e$ represents the particle Lorentz factor in the downstream
frame, $\gamma_{\rm sh}=\sqrt{2}\gamma_{\rm b}$. Considering this
residence time leads to a conservative limit on the maximal energy,
because scattering in the microturbulent field seeded in the shock
precursor should also play a role, see Plotnikov et al. (2012) for
further discussion on this issue. Comparing the above residence time
to the age of the shock wave $r/c$ then leads to a maximal Lorentz
factor (downstream frame)
\begin{equation}
\gamma_{\rm max}^{(a)}\,\simeq\,1.2\times 10^9
B_{-5}E_{53}^{1/4}n_{-3}^{-1/4}t_2^{1/4}z_{+,0.3}^{-1/4}\ .\label{eq:gmaxa}
\end{equation}
Depending on the dynamics of the microturbulence downstream, this
maximal Lorentz leads to to photons of energy
\begin{equation}
  \epsilon_{\gamma,{\rm max}}^{(a)}\,\simeq\,1\,{\rm TeV}
  B_{-5}^2E_{53}^{3/4}n_{-3}^{-1/4}t_2^{-1/4}z_{+,0.3}^{-3/4}\epsilon_{B,-2}^{1/2}\ ,\label{eq:emaxadb}
\end{equation}
through radiation in the $\delta B_\mu$ field, or
\begin{equation}
  \epsilon_{\gamma,{\rm max}}^{(a)}\,\simeq\,30\,{\rm GeV}
  B_{-5}^{3}E_{53}^{3/4}n_{-3}^{-1/4}t_2^{-1/4}z_{+,0.3}^{-3/4}\ ,\label{eq:emaxabd}
\end{equation}
if the particle radiates in the background shock compressed field
$B_{\rm d}$, as proposed in Barniol-Duran \& Kumar (2010, 2011a). This
latter case applies in particular if $\alpha_t<-1$, $\gamma_{\rm
  max}^{(a)}<\gamma_{\mu+}$ and if downstream inverse Compton losses
can be neglected. If however, $\gamma_{\rm max}^{(a)}>\gamma_{\mu+}$,
then the particle radiates in the stronger $\delta B_\mu$, while if
$\alpha_t>-1$ and $\gamma_{\rm max}^{(a)}<\gamma_{\mu+}$, the
particle radiates in a field of strength $\delta B \simeq \delta B_\mu
\left(\gamma_{\rm max}^{(a)}/\gamma_{\mu+}\right)^{-\delta_t}$, which
for these high energies lies in practice close to $\delta B_\mu$. For
instance, if $\alpha_t=-0.5$, one finds
\begin{equation}
  \epsilon_{\gamma,{\rm max}}^{(a)}\,\simeq\,0.55\,{\rm TeV}
  B_{-5}^{2.5}E_{53}n_{-3}^{-1/4}t_2^{-0.5}z_{+,0.3}^{-0.5}\Delta_{\mu,2}^{0.5}\epsilon_{B,-2}\ .
\label{eq:emaxadbi}
\end{equation}
Equation~\ref{eq:emaxabd} matches the ``confinement'' estimates of
Piran \& Nakar (2010), Barniol-Duran \& Kumar (2011a), who considered
only a shock compressed background field in the downstream, while the
other estimates Eqs.~\ref{eq:emaxadb}, \ref{eq:emaxadbi} hold in the
present more realistic setting because of the high value of magnetic
field behind the shock front.

Note that the particle is actually confined close to the shock front,
because in the ultra-relativistic regime, the shock front never lags
further than $c t_{\rm res\vert u}/\gamma_{\rm sh}^2$ behind the
particle; escape can only take place through the boundaries, but this
would lead to a less stringent condition on maximal energy than the
above age constraint, see e.g. Bykov et al. (2012). 

The above suggests that the upstream magnetic field can lie well below
$10\,\mu$G and yet lead to the production of multi-GeV photons because
the maximal Lorentz factor particles cool in a strong field, close to
the microturbulent value $\delta B_\mu$ behind the shock. Furthermore,
if one neglects altogether the upstream magnetic field, the discussion
in Plotnikov et al. (2012) suggests that scattering the microturbulent
field generated in the shock precursor may lead to Lorentz factors
sufficiently large to produce GeV photons through synchrotron
radiation in $\delta B_\mu$.

Li \& Waxman (2006) and Li \& Zhao (2011) have argued that the
comparison of the upstream residence time with the timescale for
inverse Compton losses imposes a stringent lower bound on the upstream
magnetic field, suggesting an efficient amplification of the
latter. More specifically, Li \& Zhao (2011) find that $1\,{\rm
  mG}\,n_0^{9/8}\,\lesssim \, \bar B_{\rm u}\,\lesssim\,100\,{\rm
  mG}\,n_0^{3/8}$ (with $\bar B_{\rm u}$ denoting the total magnetic
field upstream of the blast) is imposed by the synchrotron model of
$Fermi-$LAT bursts with extended GeV emission. These bounds have been
challenged in Barniol-Duran \& Kumar (2011b) and Sagi \& Nakar
(2012). Nevertheless, such bounds agree well with the general picture
that microinstabilities self-generate a microturbulent field in the
shock precursor, which is then transmitted downstream and there forms
the microturbulent layer. One expects that in the shock precursor, the
microturbulent field $\bar B_{\rm u}\sim \delta B_\mu/\gamma_{\rm b}$
(as measured in the upstream frame), in which case one finds $\bar
B_{\rm u} \sim 1\,{\rm mG}\,\epsilon_{B,-2}^{1/2}n_{-3}^{1/2}$. This
field satisfies the constraints of Li \& Zhao (2011).

\subsubsection{Downstream}
Consider now the situation downstream, which leads to much more severe
contraints on $\epsilon_{\gamma,\rm max}$ through the comparison of
$t_{\rm res\vert d}$ with the timescale of energy loss. Out of
commodity and simplicity, the limit on the maximal energy is often
quoted with a Bohm estimate $t_{\rm res\vert d}\simeq \kappa_{\rm
  B}t_{\rm L}$, with $t_{\rm L}=(\gamma_e m_e c)/(e \delta B)$ in the
downstream frame. Ignoring inverse Compton losses, this leads to
\begin{equation}
\gamma_{\rm max}^{(b)}\,\simeq\,2.1\times 10^8 \kappa_{\rm B}^{-1/2}t_2^{3/16} E_{53}^{-1/16}
n_{-3}^{-3/16} \epsilon_{B,-2}^{-1/4}\ ,\label{eq:gmaxb}
\end{equation}
assuming for the moment that $\delta B$ is homogeneous, with strength
characterized by $\epsilon_B$. This leads to
\begin{equation}
\epsilon_{\gamma,\rm max}^{(b)}\,\simeq\,30\,{\rm
  GeV}\,\kappa_{\rm B}^{-1}E_{53}^{1/8}t_2^{-3/8}n_{-3}^{-1/8}\ ,\label{eq:emaxb1}
\end{equation}
which matches the estimate of Piran \& Nakar (2010) up to a factor 2.
However, this estimate must be corrected for two effects: (1) the Bohm
approximation likely breaks down at these large Lorentz factors; (2)
the evolution of the microturbulence away from the shock front
modifies the scattering rate of the particles.

The Bohm approximation for $t_{\rm res\vert d}$ fails at the maximal
Lorentz factors because
\begin{equation}
\frac{r_{\rm L}\left(\gamma_{\rm
      max}^{(b)}\right)}{\lambda_\mu}\,\simeq\, 1.7 \times
10^2\,\lambda_{\mu,1}t_2^{9/16}E_{53}^{-3/16}n_{-3}^{-1/16}\epsilon_{B,-2}^{-3/4}\gg
  1\ ,
\end{equation}
for a Lorentz factor given by Eq.~\ref{eq:gmaxb}. This implies that
the particle only suffers a random small angle deflection of order
$\lambda_\mu/r_{\rm L}$ as it crosses a coherence cell of size
$\lambda_{\mu}=10\lambda_{\mu,1}c/\omega_{\rm pi}$, hence the
residence timescale is rather given as: $t_{\rm res\vert d}\simeq
\kappa_{\rm sc} \left(r_{\rm L}/\lambda_\mu\right)r_{\rm L}$, much
larger than the above Bohm estimate. This of course reduces
significantly the maximal energy of photons (Kirk \& Reville 2010,
Lemoine \& Pelletier 2011c, Plotnikov et al. 2012). At the present
time, one cannot exclude that the spectrum of turbulent modes extends
to wavenumbers $k\sim r_{\rm L}^{-1}$ that would allow gyroresonant
interactions in the downstream; whether or not this happens depends on
the physics of instabilities in the far upstream, which are still
debated. Were this the case, the Bohm estimate should nevertheless be
corrected by a factor that accounts for the diminished magnetic power
at those resonant scales, compared to the maximum power at
$\lambda_{\mu}^{-1}$. In order to obtain a conservative estimate of
the maximal energy, one should therefore rely on the above $t_{\rm
  res\vert d}\propto r_{\rm L}^2$. Comparing this residence timescale
with the synchrotron loss time, this would lead to a maximal Lorentz
factor
\begin{equation}
  \gamma_{\rm max}^{(c)}\,\simeq\, 3.9\times 10^7\,\kappa_{\rm
    sc}^{-1/3}\lambda_{\mu,1}n_{-3}^{-1/6}\ ,
\end{equation}
corresponding to a maximum photon energy
\begin{equation}
  \epsilon_{\gamma,\rm max}^{(c)}\,\simeq\, 0.9\,{\rm GeV}\,\kappa_{\rm sc}^{-2/3}
  E_{53}^{1/4}n_{-3}^{-1/12}\lambda_{\mu,1}^{2/3}\epsilon_{B,-2}^{1/2}t_2^{-3/4}\
  ,
\end{equation}
\emph{provided the radiation takes place in the microturbulent field
  of strength $\delta B_\mu$}. If radiation were to take place in the
background shock compressed field $B_{\rm d}$, the maximal photon
energy would rather be
\begin{equation} 
 \epsilon_{\gamma,\rm max}^{(c)}\,\simeq\, 30\,{\rm MeV}\,\kappa_{\rm sc}^{-2/3}
  E_{53}^{1/4}n_{-3}^{-7/12}\lambda_{\mu,1}^{2/3}B_{-5}t_2^{-3/4}\
  .
\end{equation}

However, once the evolution of the microturbulence is accounted for,
these estimates are modified towards more optimistic values. Consider
first the scaling of $r_{\rm L}/\lambda$, with $r_{\rm L}$ now
designing the gyroradius in the local magnetic field, and $\lambda$
the coherence length of this local magnetic field. Then
\begin{equation}
\frac{r_{\rm L}}{\lambda}\,\propto\, t^{-\alpha_t/2-1/\alpha_\lambda}\
.
\end{equation}
With $\alpha_\lambda\sim 2-3$, the following qualitative picture
emerges. If the turbulence decays gradually, meaning $\alpha_t\gtrsim
-1$, the coherence length increases more rapidly away from the shock
front than the magnetic field loses power, so that the scattering
timescale tends towards a Bohm scaling (albeit, in a weaker magnetic
field) as time increases (see also Katz et al. 2007 for a similar
picture). One then obtains (see further below) maximal energy
estimates close to the Bohm scaling. If, however, the turbulence
decays fast, meaning $\alpha_t\lesssim -1$, then the scattering
timescale increases with time, possibly faster than $t$ if
$\alpha_t\lesssim-1 -1/\alpha_\lambda$, in which case scattering
becomes impossible in the decaying part of the microturbulent
layer. Furthermore, mirror like interactions in the background shock
compressed magnetic field, which lies transverse to the shock normal,
do not allow repeated returns to the shock front (Lemoine et al. 2006,
see also Sec.~\ref{sec:fcons}). The particles must then scatter in the
undecayed microturbulent field, if $\alpha_t\lesssim -1$, in which
case the maximal energy is then determined by the comparison of
$t_{\rm res\vert d}$ to $t_{\mu+}$. These two scenarios are examined
in turn.

Assume first $\alpha_t\gtrsim -1$, meaning more specifically $\alpha_t
>-2/\alpha_\lambda$. Then the small angle scattering timescale $t_{\rm
  res\vert d}\propto t^{-\alpha_t-1/\alpha_\lambda}$, which must be
compared to the synchrotron loss time $t_{\rm syn}\propto \gamma_{\rm
  max}^{-1}\delta B_\mu \left(t_{\rm res\vert
    d}/t_{\mu+}\right)^{-\alpha_t}$, where the last factor accounts
for the evolution of the magnetic field strength with time, outside
the undecayed layer. This leads to
\begin{equation}
\gamma_{\rm max}^{(d)}\simeq \gamma_{\mu+}\left[\frac{r_{\rm
      L}(\gamma_{\mu+})^2}{\lambda_\mu\,
    ct_{\mu+}}\right]^{-(1+\alpha_t)/(3+3\alpha_t + 1/\alpha_\lambda)}\
.
\end{equation}
The term within the brackets can be rewritten in a more compact way as
$4\gamma_{\mu+}^3 r_e \lambda_\mu^{-1}/9$, with $r_e$ the classical
electron radius, so that in the limit $\alpha_\lambda\rightarrow
+\infty$, one recovers the estimate $\gamma_{\rm max}^{(c)}$ in a
homogeneous microturbulence. The fiducial values $\alpha_t=-0.5$ and
$\alpha_\lambda=2$ discussed previously lead however to
\begin{equation}
  \gamma_{\rm max}^{(d)}\,\sim\, 1.2 \times 10^8\,
  \Delta_{\mu,2}^{-0.25}E_{53}^{-0.06}
  n_{-3}^{-0.19}\epsilon_{B,-2}^{-0.25}\lambda_{\mu,1}^{0.25}t_2^{0.19}\
  ,
\end{equation}
which, through synchrotron radiation in a magnetic field of strength
$\delta B \simeq \delta B_\mu \left(\gamma_{\rm
    max}^{(d)}/\gamma_{\mu+}\right)^{-\delta_t}$ leads to photons of
typical energy
\begin{equation}
\epsilon_{\gamma,\rm max}^{(d)}\,\simeq\,2\,{\rm GeV}\,
E_{53}^{0.22}\lambda_{\mu,1}^{0.63}
\Delta_{\mu,2}^{-0.13} n_{-3}^{-0.094}\epsilon_{B,-2}^{0.38}t_2^{-0.66}\ .
\end{equation}
One should stress that this result depends on the particular value of
$\alpha_t$ and $\alpha_\lambda$, but for the above fiducial values, it
remains close to the Bohm estimate and allows the production of GeV
photons for conservative assumptions. Note finally that the stretching
of the coherence length with time does not imply that the particles of
Lorentz factor $\gamma_{\rm max}^{(d)}$ suffer gyroresonant
interactions away from the shock, nor does it require that the
turbulent power spectrum extends over many decades. For the above
fiducial values, the particles interact with modes of wavelength
$l\,\sim\, 30 \lambda_\mu\,\ll r_{\rm L}$.

In a fast decaying scenario, meaning more exactly $\alpha_t <
-1-1/\alpha_\lambda$, scattering must take place in the undecayed part
of the microturbulent layer, otherwise the particles would not see
the microturbulent layer and Fermi acceleration would not take place.
The comparison $t_{\rm res\vert d}<t_{\mu+}$ leads to
\begin{equation}
\gamma_{\rm max}^{(e)}\,\simeq\,4\times
10^6\,\kappa_{\rm sc}^{-1/2}E_{53}^{1/8}\lambda_{\mu,1}^{1/2}
\epsilon_{B,-2}^{1/2}n_{-3}^{-1/8}t_2^{-3/8}z_{+,0.3}^{3/8} \Delta_{\mu
  ,2}^{1/2}\ ,
\end{equation}
which would lead to ${\cal O}(10)\,$MeV photons only, even through
radiation in a field as strong as $\delta B_\mu$. To reach the GeV
range, ceteris paribus, one needs to increase the spatial extent of
the undecayed microturbulence, i.e. to increase $\Delta_{\mu,2}$ by
$2-3$ orders of magnitude. For $\Delta_{\mu}\gtrsim 10^{4}$, indeed,
$\gamma_{\rm max}^{(e)}\gtrsim \gamma_{\mu+}$, so that the particle
both scatters and cools in the undecayed part of the microturbulent
layer.

To summarize, the above analysis of scattering and maximal photon
energies leads to the following qualitative picture. If the turbulence
decays gradually, the particles scatter while the microturbulence
decays and the maximal energy of the photons typically falls in the
GeV range for fiducial parameters and rather conservative estimates of
the scattering properties. If, however, the turbulence decays rapidly,
GeV photons can only be produced if the undecayed part of the
turbulent layer extends far enough to accomodate both the scattering
and the cooling of the maximal energy electrons in that layer. This
latter requirement places strong constraint on the parameter $\Delta_{\mu}$
that characterizes for how long the undecayed turbulence can
survive. Requisite values can be obtained if, for instance,
$\lambda_\mu \sim 30 c/\omega_{\rm pi}$ and $\alpha_\lambda\sim3 $,
which cannot be excluded at present.  Finally, it has also been noted
that the constraints on the upstream residence time are much weaker if
the maximal energy electrons can radiate in a field close to $\delta
B_\mu$. In practice, this implies that the value of the upstream
magnetic field is not well constrained. This alleviates the apparent
strong magnetization problem discussed in He et al. (2011).

\subsection{Early GRB light curves}\label{sec:grb}
Section~\ref{sec:synchspec} together with Appendix~\ref{sec:appFnu}
have emphasized the differences between the standard synchrotron
spectra for homogeneous turbulence and those calculated with the
account of decay of the microturbulence. This Section confronts such
signatures with the observational data on GRB090510, which so far has
provided the earliest follow-up on a broad spectral range, with near
simultaneous detection in the optical, X range and GeV range (de
Pasquale et al. 2010; Ukwatta et al. 2009; Guiriec et al. 2009;
Ackermann et al. 2010; Longo et al. 2009; Hoversten et al. 2009;
Golenetskii et al. 2009; Ohmori et al. 2009). This choice is further
motivated by the analyses of Barniol-Duran \& Kumar (2009, 2010, 2011a)
and He et al. (2011), who have argued that the extended emission, from
the optical to the GeV range is compatible with an afterglow spectrum
calculated without magnetic field generation beyond direct shock
compression of the interstellar field. One may also interpret the data
with a high magnetization of the blast, closer to the traditional
estimates of $\epsilon_B$, although it then requires an
extraordinarily low upstream density $\lesssim 10^{-6}\,$cm$^{-3}$
(e.g. de Pasquale et al. 2009, Corsi et al. 2010); this possibility is
not considered here. The model of Barniol-Duran \& Kumar (2009, 2010,
2011a) then suggests that magnetic field generation processes are
ineffective, and that this burst provides the closest relative to a
``clean'' relativistic blast wave in a weakly magnetized medium

As discussed in the previous Sections, one should expect that the GeV
photons have been produced in the microturbulent layer rather than in
the background magnetic field. However, lower energy photons in the
optical and in the X range are typically produced by particles of
Lorentz $\gamma_{\rm m}$, which lose energy on a much longer timescale
than the particles of Lorentz factor $\gamma_{\rm max}$, hence in a
weaker magnetic field, possibly the background shock compressed value
$B_{\rm d}$. In any case the microturbulence is to play a key role in
shaping the spectra from the GeV down to lower energies. This
motivates further the search for a signature of this microturbulence.

Unfortunately, the comparison is not straightforward, because of the
diversity and complexity of the synchrotron spectra. The quantities
that characterize the microturbulence appear as new free parameters
from the point of view of phenomenology, hence they enlarge the
dimensionality of parameter space and add degeneracy. For this reason,
the following does not attempt to fit accurately the light curves in
the GeV, X-ray and optical, but rather accomodates the different
spectro-temporal indices in the different time domains of the
observations. The key observation of the Barniol-Duran \& Kumar (2009,
2010, 2011a) interpretation is that the spectrum is produced in a slow
cooling regime, and that in the case of GRB090510, $\nu_{\rm m}$
transits across the optical at $10^3\,$s, in order to produce the
observed break. Once this condition is satisfied, and provided
spectro-temporal indices match the observed values, one obtains a
satisfactory fit to the light curves (as it has been checked). The
following discussion reveals that the current data do not allow one to
pick out a preferred model, meaning conversely that it is possible to
reproduce the salient features of the Barniol-Duran \& Kumar (2009,
2010, 2011a) model with concrete characterizations of the
microturbulence. Each model leads to rather specific signatures that
might be identified with further higher accuracy and earlier
observations. Some of these signatures are pointed out in the
following, although more work is needed to define an observation
strategy capable of pinpointing the characteristics of the
microturbulence.

GRB090510 is a short burst at redshift $z=0.92$ whose light curve is
characterized as follows: emission is seen in the LAT range up to
$\simeq 100\,$s, with spectro-temporal indices $\alpha_{\rm GeV}\simeq
1.4$, $\beta_{\rm GeV}\simeq 1.1$; X-ray and optical follow-up start
at about this time; from $\simeq 100\,$s to about $1.5\times 10^3\,$s,
$\alpha_{\rm X}\simeq 0.74$, increasing to $2.2$ afterwards, while
$\beta_{\rm X}\simeq 0.5-0.8$; $\alpha_{\rm opt}\simeq -0.5$ before
$10^3\,$s, then $1.1$ afterwards. The late time evolution beyond
$10^3\,$s is more difficult to reproduce, in particular the steepening
of the X-ray lightcurve, which is attributed by He et al. (2011) to
sideways expansion, but which does not match the apparent shallower
decay in the optical up to $2\times 10^4\,$s (Nicuesa Guelbenzu et
al. 2012).

To compare this light curve with the afterglow computed in
App.~\ref{sec:appFnu}, consider first the possiblity of $\alpha_t>-1$
(gradual decay) without inverse Compton losses. The latter assumption
has been discussed at length in He et al. (2011). As shown in
Fig.~\ref{fig:spec_asupm1}, the spectral shape of the synchrotron
spectrum in the slow cooling limit with $\alpha_t>-1$ retains the same
form as the standard afterglow, but the values of the characteristic
frequencies and therefore the time dependences are
modified. Therefore, $\beta_{\rm X}\simeq 0.6$ if $\nu_{\rm m}<1\,{\rm
  keV}\,<\,\nu_{\rm c}$ as observed. In the model of Barniol-Duran \&
Kumar (2010), He et al. (2011), $\nu_{\rm c}$ lies below the LAT
range, so that
\begin{equation}
\beta_{\rm GeV} \simeq -\frac{p - \alpha_t(1-2p)/2}{2+3\alpha_t/2}\ ,
\end{equation}
which takes values between $-1.10$ and $-1.00$ for $p=2.2$ and
$\alpha_t\in[0,-1]$, in good agreement with the data, although the
influence of $\alpha_t$ is too weak to provide significant
constraints. The temporal slope in the GeV range also depends mildly
on $\alpha_t$, taking values from $\alpha=1.15$ at $\alpha_t=0$
($p=2.2$) to $\alpha=1.25$ at $\alpha_t=-1$ ($p=2.2$). It fits with
the observations, given that the measured slope may be contaminated by
prompt GeV emission, see the discussion in He et al. (2011). The best
probes for $\alpha_t$ remain the temporal slopes in the X and optical
domains: $\alpha$ takes values between $0.9$ and $1.4$ in the X range
(for $p=2.2$), and between $-0.50$ and $-0.30$ ($p=2.2$) in the
optical when $1\,{\rm eV\,}<\nu_{\rm m}$. Current data do not allow a
precise determination of $\alpha_t$, but it may be noted that the
fiducial value $\alpha_t=-0.5$ fits well the data. Finally, requiring
$\nu_{\rm m}\sim1\,$eV at $10^3\,$s determines $\alpha_t$ as a
function of the other parameters. For the fiducial values
characterizing the turbulence, $n\sim10^{-3}\,$cm$^{-3}$ and $B_{\rm
  d}\lesssim 1\,\mu$G, $\nu_{\rm m}=\nu_{\rm p}\left[\gamma_{\rm
    m};\delta B(t_{\rm dyn})\right]$ because the turbulence has not
had time yet to relax to the background magnetic field $B_{\rm
  d}$. Then
\begin{eqnarray}
  \nu_{\rm m}&\,\simeq\,& 2.3\times 10^{17}\,{\rm Hz}\,
  e^{4.851\alpha_t/(1+\alpha_t)} E_{53}^{1/2 + \alpha_t/16}
  n_{-3}^{3\alpha_t/16}\nonumber\\
 & &\,\,\, \times
  \Delta_{\mu,2}^{-\alpha_t/2}\epsilon_{B,-2}^{1/2}\epsilon_{e,-0.3}^2
t_2^{-3/2+5\alpha_t/16}\ ,\label{eq:atA}
\end{eqnarray}
so that $\nu_{\rm m}\sim1\,$eV at $10^3\,$s implies $\alpha_t=-0.6$ up
to small logarithmic corrections. It is somewhat remarkable that the
value of $\alpha_t$ falls close to the fiducial value discussed
previously. Also, one may note that the upstream magnetic field does
not enter Eq.~\ref{eq:atA}, because the particles actually cool in the
decaying microturbulence. One may verify that the above choice of
parameters gives a light curve in good agreement with the observed
data, except for the late time shallow decay in the optical mentioned
above.

To pursue this comparison, consider now the possibility that
$-4/(p+1)<\alpha_t<-1$.  The fast decay of the microturbulence in this
case actually mimics somewhat the scenario originally proposed by
Barniol-Duran \& Kumar (2010, 2011a): the particles get accelerated in
a microturbulent layer behind the shock front but cool where the
microturbulence has died away. Although, as discussed in some details
in Sec.~\ref{sec:accel}, it appears necessary to require that the
particles of Lorentz factor $\gamma_{\rm max}$ actually cool in the
microturbulent layer in order to produce the GeV photons, in other
words $\gamma_{\mu+}\lesssim 4\times 10^7$ at $t_{\rm obs}\sim
100\,$s, which implies
\begin{equation}
\Delta_{\mu}\gtrsim 10^{4}
E_{53}^{-1/4}n_{-3}^{-1/4}\epsilon_{B,-2}^{-1}t_2^{1/4}\ ,
\end{equation}
or to put it more simply, that the undecayed part of the
microturbulent layer with $\epsilon_{B,-2}\sim 10^{-2}$ extends for
some $10^4$ skin depths at least. 

Two different behaviors can be observed, depending on the ratio
$t_{\rm dyn}/t_{\mu-}$, i.e. whether the turbulence has relaxed down
to $B_{\rm d}$, in which case a synchrotron component associated to
$B_{\rm d}$ would emerge on top of the microturbulent synchrotron
spectrum. The maximum value of $t_{\rm dyn}/t_{\mu-}$ is obtained for
$\alpha_t\rightarrow -4/(p+1)\simeq -1.3$:
\begin{equation}
\frac{t_{\rm dyn}}{t_{\mu-}}\,\sim\, 0.3B_{-5}^{1.53}E_{53}^{1/8}n_{-3}^{-0.39}
\epsilon_{B,-2}^{-0.77}\Delta_{\mu,4}^{-1}t_2^{5/8}\ ,
\end{equation}
so that both possibilities have to be envisaged. If $t_{\rm
  dyn}<t_{\mu-}$ up to late times, and $\nu_{\rm m}<1\,{\rm
  eV}\,<\,\nu_{\mu0}$ beyond $10^3\,$s, the optical flux should decay
fast with $\alpha_{\rm opt}\simeq 1.4\rightarrow 1.6$, which disfavors
this possibility. If now $t_{\rm dyn}>t_{\mu-}$ at some point, the
synchrotron component produced by the cooling of particles in $B_{\rm
  d}$ emerges and dominates over that produced by the microturbulence,
at least at low frequencies (optical, X). This $B_{\rm d}$ component
cuts off at $\nu_{\mu0}\,\equiv\,\nu_{\rm p}\left[\gamma_{\mu+};B_{\rm
    d}\right]$, as discussed in Sec.~\ref{sec:fastnoIC};
$\nu_{\mu0}$ increases with time, and for reasonable choices of
parameters, it lies above the X range. At the time at which $t_{\rm
  dyn}=t_{\mu-}$, however, the temporal decay slope in the X range
changes from $(6p-1)/8$ (due to cooling in the microturbulence) to
$3(p-1)/4$, which implies a change from steep to shallow, which is not
observed. Therefore, this scenario is disfavored as well. It is
difficult at the present to rule it out clearly, as other effects, due
to jet breaking for instance, could complicate the temporal scaling.

Consider now the possibility $-3<\alpha_t<-4/(p+1)$, without inverse
Compton losses as above. Similar considerations regarding the maximal
energy suggests that $\gamma_{\mu+}\lesssim 4\times10^7$, implying
$\Delta_{\mu}\gtrsim 10^{4}$. The turbulence decays fast so that, for a
reasonable choice of parameters, e.g. $\alpha_t\lesssim -1.5$,
$n_{-3}\sim1$, $B_{-5}\sim 1$, the condition $t_{\rm dyn}>t_{\mu-}$
may be fulfilled at $t_{\rm obs}\gtrsim 100\,$s. This implies that the
background synchrotron component emerges on top of the microturbulent
synchrotron spectrum, and dominates at low frequencies. One then
recovers the standard afterglow spectrum proposed by Barniol-Duran \&
Kumar (2010), except that the GeV photons are produced in the
microturbulent layer. In the GeV range, the indices $(\alpha,\beta)$
match the standard predictions for the fast cooling population and
thus agree with the data; the overall afterglow provides a
satisfactory fit to the available data.  In order to discriminate this
model, one would need to access the early light curve in the frequency
range $\nu_{\rm m}<\nu<\nu_{\rm m,\delta B_\mu}$ (i.e. X-ray) in which
both the spectral and temporal indices evolves strongly with
$\alpha_t$: $(\alpha,\beta)$ go from $(1.6,0.5)$ to $(0.1, -0.4)$ as
$\alpha_t$ goes from $-1.3$ to $-3$. For reference, $\nu_{\rm m,\delta
  B_\mu}\simeq 2.3\times 10^{17}\,{\rm
  Hz}\,E_{53}^{1/2}\epsilon_{B,-2}^{1/2}\epsilon_{e,-0.3}^2
t_2^{-3/2}$.

Another interesting solution arises in this scenario when $t_{\rm
  dyn}<t_{\mu-}$ up to late times. This may be realized if $B_{-5}\ll
1$ for instance. Then the frequency $\nu_{\rm m}=\nu_{\rm
  p}\left[\gamma_{\rm m};\delta B(t_{\rm dyn})\right]$ transits
through the optical at $10^3\,$s if $\alpha_t\sim-1.5$, as
required. During the interval $10^3-10^4\,$s, the optical lies in the
range $\nu_{\rm m} - \nu_{{\rm m},\delta B_\mu}$ with
$(\alpha,\beta)=-\left[(24+7\alpha_t)/(8\alpha_t),
  1+2/\alpha_t\right]$. To reconcile $\alpha$ with $\alpha_{\rm
  opt}\sim 0.8$ (Nicuesa Gulbenzu et al. 2012), one would need to
impose $\alpha_t\sim -1.8$, while $\beta_{\rm opt}\sim 0.9$ at that
time rather requires $\alpha_t\simeq -1.1$. Thus $\alpha_t\simeq -1.5$
provides a compromise, which is marginally acceptable by the data,
that presents the clear advantage of a shallower decay in the optical
than in the X-ray at a time at which it is observed. At $10^4\,$s,
$\nu_{{\rm m},\delta B_\mu}$ transits across the optical (see its
expression above), so that the optical and the X-ray domains decline
alongside at later times.

Whichever ratio $t_{\rm dyn}/t_{\mu-}$ is considered, the model with
$\alpha_t<-4/(p+1)$ predicts a clear signature in the early optical
lightcurve, with $\alpha_{\rm opt}=-1/8$ when $1\,{\rm
  eV}\,<\,\nu_{\rm m}$, which leads to a mild growth the optical
flux. De Pasquale et al. (2010) report $\alpha_{\rm
  opt}=-0.5^{+0.11}_{-0.13}$ between $10^2\,$s and $10^3\,$s, while
Nicuesa Gulbenzu et al. (2012) rather indicate $\alpha_{\rm
  opt}=-0.2\pm0.2$, therefore it is not possible to rule out or
confirm this model at present.

Consider finally the possiblity of strong inverse Compton losses. As
before, it is assumed here that these losses dominate throughout the
blast, although a more realistic model should include the possibility
of weak inverse Compton losses at the highest energies due to the
Klein-Nishina suppression of the cross section. Nevertheless, one may
constrain this scenario with the optical and X data, as follows.

The dominance of inverse Compton losses for particles of Lorentz
factor $\gamma_{\rm m}$ implies that $\gamma_{\rm c}\sim \gamma_{\rm
  m}$, since
\begin{equation}
  \frac{\gamma_{\rm c}}{\gamma_{\rm m}}\,\simeq\, 1.5
  E_{53}^{-1/2}n_{-3}^{-1/2}\epsilon_{B,-2}^{-1}\epsilon_{e,-0.3}^{-1}\frac{4}{1+Y_\mu}t_2^{0.5}\
  ,
\end{equation}
see Eq.~\ref{eq:Ym} for the definition of $Y_\mu$.  Consider first the
possibility $\alpha_t>-4/(p+1)$, this implies that $\nu_{\rm
  m}\lesssim 1\,$keV and $\nu_{\rm c}\lesssim 1\,$keV at $t_{\rm
  obs}\gtrsim100\,$s, so the X range always fall in the fast cooling
portion. The spectral slope cannot be $-p/2$ in that range, otherwise
it could not match the observed $\beta_{\rm X}$; the slope must thus
be $-(p+\alpha_t/2)/(2-\alpha_t/2)$, which fits $\beta_X\simeq
0.5-0.8$ only for $\alpha_t\sim -0.8 \rightarrow -1.3$. In turn, this
would imply a positive value of $\alpha_{\rm opt}$ if $\nu_{\rm
  m}>\nu_{\rm c}$ (fast cooling), so the regime at time $t_{\rm
  obs}\gtrsim 100\,$s must be that of slow cooling (case 1 or 2
depicted in Fig.~\ref{fig:spec_ICasupm42pp1}). One should further
require that $\nu_{\rm m}$ goes into the optical at $10^3\,$s as
before. These features can be achieved for the typical fiducial values
considered before, $n_{-3}\sim1$, $E_{53}\sim 1$, with
$\alpha_t\sim-0.8 \rightarrow -1.3$.

An interesting feature of this scenario is that the ratio $t_{\rm
  dyn}/t_{\mu-}$ can take values smaller or larger than unity for
standard parameters of the microturbulence and $\alpha_t\sim -1$. As
it depends sensitively on the value of the upstream magnetic field, as
$B_{\rm u}^{-2/\alpha_t}$, the transition from $t_{\rm dyn}<t_{\mu-}$
(case 1 in Fig.~\ref{fig:spec_ICasupm42pp1}) to $t_{\rm dyn}>t_{\mu-}$
(case 2 in Fig.~\ref{fig:spec_ICasupm42pp1}) may well occur around
$10^2-10^3\,$s. Once $t_{\rm dyn}/t_{\mu-}>1$, the optical domain lies
in the fast cooling region corresponding to $\nu_{\rm c}<1\,{\rm
  eV}\,<\nu_{\mu-}$ and indices
$(\alpha,\beta)=\left[(3p-2)/4,p/2\right]\,\simeq\,(1.1, 1.1)$. The
optical domain then decays more slowly than the X domain, $\alpha$
being in marginal agreement with the inferred value $\alpha_{\rm
  opt}=0.8\pm0.1$ (Nicuesa Gulbenzu et al. 2012). Note that the
frequency $\nu_{\mu-}$ increases in time as $t_{\rm obs}^{3/4}$, but
remains below the X domain if $B_{-5}\lesssim 1$. Therefore the break
at $2\times 10^4\,$s in the optical cannot be attributed to another
frequency crossing the optical domain; it might however result from
jet sideways expansion in that model. In order to probe and
discriminate this model, one would need to access the very early light
curve, while it is in the fast cooling regime and measure the temporal
decay slopes of the X and optical domains. The above modelling has
implicitly assumed a Compton parameter $Y_\mu\simeq 3$; one can check
that for such a value, the inverse Compton component of optical
photons at $100\,$s remains below the synchrotron flux in the GeV
domain.

If $\alpha_t<-4/(p+1)$, the spectra depicted in
Fig.~\ref{fig:spec_ICainfm42pp1} generally resemble at high energies a
slow cooling spectrum in the microturbulent field. Since $\nu_{\rm
  m,\delta B_\mu}\lesssim 1\,$keV at $t_{\rm obs}\gtrsim 100\,$s, the
X domain always lies on the slow cooling portion of the microturbulent
synchrotron spectrum with therefore correct values for $\alpha_{\rm
  X}$, $\beta_{\rm X}$. Among the 5 cases shown in
Fig.~\ref{fig:spec_ICainfm42pp1}, the only two reasonable ones are 5
and 2. Indeed, case 1 is unlikely because it requires a small external
density to achieve slow cooling, but then $t_{\rm dyn}/t_{\mu-}$
becomes large and case 2 actually applies. Similarly, case 4 is
unlikely because $\nu_{\rm m}\sim \nu_{\rm c}$ as argued above. Case 3
leads to $\alpha_{\rm opt}>0$, which goes contrary to the
observational data. Therefore, cases 5 and/or 2 remain as the viable
alternatives. However, one must now impose $\nu_{\mu-}\lesssim 1\,$keV
at $10^2-10^4\,$s, so that the secondary synchrotron component
associated to $B_{\rm d}$ cuts off below the X range, otherwise the X
domain would fall in the fast cooling part of that synchrotron
component. This constraint is fulfilled for the previous fiducial
values $B_{-5}\lesssim 1$, $n_{-3}\sim1 $, $E_{53}\sim1 $ and for
$\alpha_t\gtrsim -1.5$. Interestingly, the optical falls on the fast
cooling part of the secondary synchrotron component at $\gtrsim
10^3\,$s, since $\nu_{\rm m}\sim\nu_{\rm c}$, therefore it decays less
fast than the X range, with $\alpha$ in marginal agreement with the
observed value, as previously.  The temporal slope in the optical at
very early times $\lesssim 10^2\,$s provides an interesting check of
this scenario: it should be flat or even weakly decaying,
corresponding to case 3 in Fig.~\ref{fig:spec_ICainfm42pp1}, which
represents the analog of case 5 when $t_{\rm dyn}<t_{\mu-}$. Finally,
as discussed before, one must require $\Delta_{\mu}\gtrsim 10^{4}$ in
order to produce GeV photons directly in the undecayed part of the
microturbulent layer. In this model, the microturbulent layer produces
both the GeV and X photons, while the optical photons are produced in
the shock compressed background field.

\subsection{Further considerations}\label{sec:fcons}
According to the modelling of Barniol-Duran \& Kumar (2009, 2010,
2011a), the magnetization of the blast wave can take a large range of
values, from $\epsilon_B\sim 10^{-9}$ to $\epsilon_B\sim 10^{-3}$,
depending on the external density. The latter value is given in de
Pasquale et al. (2009) and Corsi et al. (2010), assuming very low
upstream densities $n\lesssim 10^{-6}\,$cm$^{-3}$.  The general trend
is that, independently of $n$, the downstream magnetic field appears
to coincide with the shock compressed magnetic field $B_{\rm
  d}=4\gamma_{\rm b}B_{\rm u}$, corresponding to $B_{\rm u}\sim
1-10\,\mu$G. As noted in He et al. (2011), this generally implies large
upstream magnetizations, since $\sigma_{\rm u}\simeq \epsilon_B/2$ if
$\epsilon_B$ is calculated in terms of $B_{\rm d}$, whereas the
typical interstellar magnetization is more of the order of
$10^{-9}$. As discussed in the previous sections, accounting for a
decaying microturbulent layer alleviates the need for a large value of
$B_{\rm u}$, which allows to reconcile the apparent $\epsilon_B$ with
a more standard upstream magnetization.

Nevertheless, it is interesting to ask what would happen if the
upstream magnetization were truly high, e.g. if $B_{\rm u}\gtrsim
1\,\mu$G and $n\ll 10^{-3}\,$cm$^{-3}$. If $\sigma_{\rm u}$ is brought
upwards by several orders of magnitude, then magnetization effects are
to affect the shock physics. For instance, the PIC simulations of
Sironi \& Spitkovsky (2011) indicate that, for $\gamma_{\rm b}=15$ and
$\sigma_{\rm u}\gtrsim 10^{-4}$, the shock is no longer mediated by
the filamentation instability, but by the magnetic barrier associated
to the (transverse) background magnetic field. As discussed in Lemoine
\& Pelletier (2010, 2011a), such a transition occurs when the growth
timescale of the filamentation instability becomes larger than the
timescale on which upstream plasma elements cross the shock
precursor. This transition is predicted to occur when $\gamma_{\rm
  sh}^2\sigma_{\rm u}\xi_{\rm cr}^{-1}\sim 1$, with $\xi_{\rm cr}\sim
0.1$ the fraction of shock energy carried by suprathermal
particles. For $\gamma_{\rm b}\simeq 300$, the limiting magnetization
becomes $\sigma_{\rm c}\simeq 10^{-6}$, to be compared with
$\sigma_{\rm u}=5\times 10^{-6} B_{-5}^2n_{-3}^{-1}$ in terms of the
above fiducial values.

If $\sigma_{\rm u}\gg \sigma_{\rm c}$, microturbulence cannot be
excited at the shock front, hence the downstream plasma must be
permeated with the background shock compressed field only.  However,
one should not expect particle acceleration to proceed and indeed, the
PIC simulations of Sironi \& Spitkovsky (2011) indicate no particle
acceleration for $\sigma_{\rm u}\gtrsim 10^{-4}$, $\gamma_{\rm b}=
15$. In order to obtain particle acceleration, one needs to find a
source of turbulence in the downstream plasma, that would be able to
scatter particles back to the shock front faster than they are
advected. This possibility cannot be discarded at the present time but
it leads to some form of paradox: given that accelerated particles
probe a length scale $\simeq r_{\rm L}$ behind the shock if they
interact with a magnetic field coherent over scales $\gg r_{\rm L}$
(Lemoine \& Revenu 2006, Lemoine et al. 2006), the instability needs
to produce large scale modes on spatial scales $> r_{\rm L}$ with a
growth timescale $<r_{\rm L}/c$. The most reasonable scenario is thus
to assume that microturbulence exists behind the shock, which
restricts the possible values of the upstream magnetization. Finally,
if $\sigma_{\rm u}\gtrsim \sigma_{\rm c}$ at some early time, but
$\sigma_{\rm u}\lesssim \sigma_{\rm c}$ at some late time, because
$\sigma_{\rm c}\propto \gamma_{\rm b}^{-2}$ while $\sigma_{\rm u}$
remains constant, one should observe a very peculiar signature in the
light curve due to the initial absence of an extended shock
accelerated powerlaw (Lemoine \& Pelletier 2011b).

\section{Conclusions}\label{sec:concl}
Current understanding of the formation of weakly magnetized
collisionless shocks implies the formation of an extended
microturbulent layer, both upstream and downstream of the shock. It is
expected that the microturbulence behind the shock decays away on some
hundreds of skin depth scales, and recent high performance particle in
cell simulations have confirmed this picture. Borrowing from these
simulations, Section~\ref{sec:synchspec} has described the decay of
the microturbulence as a powerlaw in time, with an index that can take
mild or pronounced negative values, depending on the breadth of the
spectrum of magnetic perturbations seeded immediately behind the
shock. The synchrotron signal of a shock accelerated powerlaw of
particles radiating in this evolving microturbulence has been
calculated, and the results are reported in App.~\ref{sec:appFnu}.
This calculation reveals a rather large diversity of possible
signatures in the spectro-temporal evolution of the synchrotron flux
of a decelerating relativistic blast wave, commonly parameterized
under the form $F_\nu\,\propto \,t^{-\alpha}\nu^{-\beta}$. The
diversity of signals is associated with the possible values of the
ratio $t_{\rm dyn}/t_{\mu-}$, which characterizes whether the blast
has had time to relax to the background shock compressed field or not
by the comoving dynamical time $t_{\rm dyn}$, as well as the different
cooling regimes (fast cooling or slow cooling on timescale $t_{\rm
  dyn}$) and the influence or not of inverse Compton losses on the
cooling history of particles. One point on which emphasis should be
put, is that these signatures are potentially visible in early
follow-up observations of gamma-ray bursts on a wide range of
frequencies. The detailed synchrotron shapes and indices $\alpha$,
$\beta$ are discussed in App.~\ref{sec:appFnu}.

The microturbulence controls the scattering process of particles
during the acceleration stage and it therefore controls the maximal
energy of shock accelerated particles. Detailed estimates for this
maximal energy and for the corresponding energy of synchrotron photons
have been provided. It has been argued that the observation of GeV
synchrotron photons from gamma-ray burst afterglows implies either
that the microturbulence decays rather slowly (as actually observed so
far in PIC simulations), or that the spatial extent of the undecayed
layer of microturbulence is large enough to accomodate both the
scattering and the cooling of these particles.  In the former case,
the stretching of the coherence length of the microturbulence that
accompanies the erosion of magnetic power improves the acceleration
efficiency and brings it close to a Bohm scaling of the scattering
frequency.

Finally, this paper has provided a concrete basis for the
phenomenological models of Barniol-Duran \& Kumar (2009, 2010, 2011a),
which interpret the extended GeV emission of a fraction of gamma-ray
bursts as synchrotron radiation of shock accelerated particles in the
background shock compressed field. Microturbulence must play an
important role in these models for several reasons: (1) it is expected
to exist behind the shock front of a weakly magnetized relativistic
shock wave; (2) it ensures the development of Fermi acceleration,
which cannot proceed without efficient scattering; (3) it controls the
production of photons of energy as high as a few GeV. Using the
predictions of the spectro-temporal dependence of $F_\nu$ with a
dynamical microturbulence, one can reproduce satisfactorily the
phenomenology of the models of Barniol-Duran \& Kumar (2009, 2010,
2011a), as examplified here with the particular case of GRB090510. In
the present construction, the decaying microturbulent layer permits
the generation of high energy photons close to the shock front, while
the low energy photons are produced away from the shock, where the
microturbulence has relaxed to (or close to) the background shock
compressed field. In this setting, the broadband early follow-up
observations of gamma-ray bursts afterglows open an exceptional window
on the physics of collisionless shocks.  More work is certainly needed
to study the phenomenology of evolving microturbulence and its
relation with gamma-ray bursts light curves, in particular to define a
proper observational strategy capable of inferring the characteristics
of the microturbulence.

\section*{Acknowledgments}
It is a pleasure to thank P. Kumar, Z.  Li, G. Pelletier and
X.-Y. Wang for useful discussions.  This work has been supported in
part by the PEPS-PTI Program of the INP (CNRS). It has made use of the
Mathematica software and of the LevelScheme scientific figure
preparation system (Caprio, 2005).

\appendix

\section{Synchrotron spectra with time decaying
  microturbulence}\label{sec:appFnu}
To compute the synchrotron spectral power of a relativistic blast
wave, one usually solves a steady state transport equation for the
blast integrated electron distribution and folds this distribution
with the individual synchrotron power (Sari et al. 1998). This method
may be generalized to the present case by including an explicit
spatial transport term and solving the transport equation through the
method of characteristics. It appears however more convenient to
proceed with an approximation of the calculation of Gruzinov \& Waxman
(1999), which integrates over the cooling history of each electron in
the blast. The present approximation neglects the integral over the
angular coordinates (1d hypothesis), it neglects the hydrodynamical
profile of the blast (as in Sari et al. 1998) and it neglects the time
evolution of the blast physical characteristics over the cooling
history of a freshly accelerated electron. This reduces the integral
to a more tractable form, with a closed expression, at the price of
reasonable approximations. The angular part of the integral indeed
leads to a smoothing of the temporal profile of emission which may be
re-introduced at a later time, see Panaitescu \& Kumar (2000). The
blast evolves on hydrodynamical timescales that are longer, or much
longer than the radiative timescales.

The synchrotron spectral power of the blast is written as discussed in
Section~\ref{sec:radmu},
\begin{equation}
  P_{\nu} \,=\,\frac{4\gamma_{\rm b}^2}{3}\, 
  \int_{\gamma_{\rm m}}^{\gamma_{\rm max}}{\rm d}\gamma_{e,0}\,\frac{{\rm
      d}\dot N_e}{{\rm d}\gamma_{e,0}}\,\,\int_{0}^{t_{\rm dyn}} {\rm d}t\,\frac{{\rm d}E_{\rm
      syn}}{{\rm d}\nu{\rm d}t}\ ,\label{eq:appdFnu}
\end{equation}
where ${\rm d}\dot N_e/{\rm d}\gamma_{e,0}$ denotes the differential
per Lorentz factor interval of the number of electrons swept by the
shock wave by unit time. It is defined in
Eqs.~\ref{eq:dNedg},\ref{eq:dNedt}. For commodity, the spectral power
$P_\nu$ is written in the circumbust medium rest frame at redshift
$z$, whence the beaming factor $4\gamma_{\rm b}^2/3$, and all
frequencies are written in the observer rest frame. This means in
particular, that the integrands are defined in the comoving blast
frame, except ${\rm d}\nu$ of course, which is an observer frame
quantity. The received spectral flux $F_\nu$ then reads
\begin{equation}
F_\nu \,=\, \frac{P_\nu}{4\pi D_{\rm L}^2}\ ,
\end{equation}
with $D_{\rm L}$ the luminosity distance to the blast.

The time integral in Eq.~\ref{eq:appdFnu} integrates the synchrotron
power along the particle trajectory in the blast, i.e. from shock
entry until the particle reaches the back of the blast. As it ignores
the spatial profile of the blast, the upper bound $t_{\rm dyn}$ is
defined up to a factor of unity. Such prefactors of order unity are
ignored in the following calculations, which furthermore always
approximate down to broken powerlaws. A correct overall normalization
is provided at the end of the calculation.

The (comoving) dynamical timescale is defined following Panaitescu \&
Kumar (2000),
\begin{equation}
  t_{\rm dyn} \equiv \frac{1}{c}\int \frac{{\rm d}r}{\gamma_{\rm b}}\ .\label{eq:tdyn}
\end{equation}

The spectral power radiated by an electron at time $t$, of Lorentz
factor $\gamma_e$ is approximated as
\begin{equation}
  \frac{{\rm d}E_{\rm syn}}{{\rm d}\nu{\rm d}t}\,\approx\,
  \frac{1}{6\pi}\sigma_{\rm T}\delta B(t)^2\gamma_e(t)^2c
  \frac{4}{3}\frac{1}{\nu_e}
  \left(\frac{\nu}{\nu_e}\right)^{1/3}\Theta(\nu_e-\nu)\ ,
\label{eq:Esyn}
\end{equation}
with
\begin{equation}
\nu_e\,=\,\nu_{\rm p}\left[\gamma_e;\delta B(t)\right]\ .
\end{equation}
The peak frequency $\nu_{\rm p}$ for Lorentz factor $\gamma_e$ and
magnetic field $\delta B$ is defined as (e.g. Wijers \& Galama 1999)
\begin{equation}
  \nu_{\rm p}\left[\gamma_e;\delta B\right]\,\equiv\,
  \frac{3x_p}{2\pi}\frac{e\,\delta B}{m_e c} \gamma_e^2\frac{4\gamma_{\rm
      b}}{3(1+z)}\ ,
\label{eq:nusyn}
\end{equation}
with $x_p\simeq 0.29$.  For most cases, one might further approximate
Eq.~(\ref{eq:Esyn}) as a delta function peaked around $\nu_e$, but the
low energy part in $\nu^{1/3}$ does actually play a role in several
specific limits. The additional factor of $4/3$ in Eq.~(\ref{eq:Esyn})
ensures proper normalization to the synchrotron energy loss rate
integrated over frequency.

The physics of diffusive synchrotron radiation has received a lot of
attention lately, notably because it might lead to distortions of the
above spectral shape below and above $\nu_{\rm p}$ (see e.g. Medvedev
2000, Fleishman \& Urtiev 2010, Kirk \& Reville 2010, Reville \& Kirk
2010, Mao \& Wang 2011, Medvedev et al. 2011). The magnitude of such
distortions is characterized by the wiggler parameter $a$:
\begin{equation}
a \,\equiv\, \frac{e \delta B \lambda_{\delta B}}{m_e c^2}\ .
\end{equation}
If $a>\gamma_e$, the particle suffers a deflection of order unity in
each cell of coherence of the turbulent field, hence the standard
synchrotron regime applies. At the opposite extreme, if $a<1$, the
deflection per coherence cell does not exceed the emission angle of
$1/\gamma_e$. This leads to diffusive synchrotron radiation, with
different slopes at low and high frequencies. In the intermediate
regime, $1<a<\gamma_e$, the spectrum remains synchrotron like with
some departures at $\nu\gg\nu_{\rm p}$ and $\nu\lesssim a^{-3}\nu_{\rm
  p}$ (Medvedev et al. 2011).

In the present scenario, one can define a quantity $a_\mu$ immediately
behind the shock front, in terms of $\delta B_\mu$ and
$\lambda_{\mu}$,
\begin{equation}
  a_\mu\,\simeq\, \sqrt{8}\epsilon_B^{1/2}\frac{\lambda_{\mu}}{c/\omega_{\rm pi}}\gamma_{\rm
    b}\frac{m_p}{m_e}\,\approx\, 20 \gamma_{\rm
    m}\,\epsilon_{B,-2}^{1/2}\epsilon_{e,-0.3}^{-1}\lambda_{\mu,1}\
  ,
\end{equation}
with $\lambda_{\mu,1}\equiv \lambda_{\mu}/(10c/\omega_{\rm pi})$ and
$\gamma_{\rm m}\simeq 6.4\times 10^4
E_{53}^{1/8}n_{-3}^{-1/8}t_{2}^{-3/8}\epsilon_{e,-0.3}z_{+,0.3}^{3/8}$
the minimum Lorentz factor of the electron population. For generic
parameters, one thus finds $a_\mu>\gamma_e$ or $1\ll a_\mu<\gamma_e$,
but in any case, diffusive synchrotron effects can be neglected close
to a highly relativistic shock front.

As the turbulence decays, its coherence length evolves, hence $a$
evolves in a non-trivial way, starting from $a_\mu$. By the time
$t_{\mu-}$, i.e. when the turbulence has relaxed to $B_{\rm d}$, one
finds
\begin{equation}
  a(t_{\mu-})\,\simeq\, a_\mu \left(\frac{B_{\rm d}}{\delta
      B_\mu}\right)^{1+2/(\alpha_t\alpha_\lambda)}\ .
\end{equation}
With $\alpha_t<0$, one can check that this value remains much larger
than unity for generic GRB blast wave characteristics.  Considering
for instance the example $\alpha_t=-3$ (fast decay of the turbulence),
$\alpha_\lambda=2$, one finds $ a(t_{\mu-}) \sim 1.3\times 10^5\,
B_{-5}^{2/3} E_{53}^{1/8} t_{2}^{-3/8} n_{-3}^{-11/24}
\epsilon_{B,-2}^{1/6}\lambda_{\mu,1} z_{+,0.3}^{3/8}$. Diffusive
synchrotron effects become important at late times when the blast wave
has become moderately relativistic, but for the present purposes they
can be neglected; this justifies the above choice Eq.~\ref{eq:Esyn}
for ${\rm d}E/{\rm d}\nu{\rm d}t$.

The influence of synchrotron self-absorption is discussed briefly in
Sec.~\ref{sec:synch-self}. It does not affect the synchrotron spectra
around and above the optical range but it may lead to interesting time
signatures at frequencies below a GHz.

\subsection{Gradual decay: $-1<\alpha_t<0$; no Inverse Compton
  cooling}\label{sec:appnoICasupm1}
If $-1<\alpha_t<0$, the particle cools gradually in the decaying
microturbulent field, see Eq.~\ref{eq:cool2}. This Section ignores the
possibility of inverse Compton losses, the effects of which are
discussed in Sec.~\ref{sec:appICgrad}. One then defines the critical
frequencies
\begin{equation}
\nu_{\mu+}\,\equiv\,\nu_{\rm p}\left[\gamma_{\mu+};\delta B_\mu\right]\ ,
\end{equation}
which is associated to particles of Lorentz factor $\gamma_{\mu+}$ and
magnetic strength $\delta B_\mu$, and its counterpart
\begin{equation}
\nu_{\mu-}\,\equiv\,\nu_{\rm p}\left[\gamma_{\mu-};B_{\rm d}\right]\ .
\end{equation}

The cooling Lorentz factor enters the calculation through the upper
bound on the integral defined in Eq.~(\ref{eq:appdFnu}): $\gamma_{\rm
  c}$ denotes as usual that which allows cooling on a timescale
$t_{\rm dyn}$.  To calculate $\gamma_{\rm c}$, one first define the
usual $\gamma_{{\rm c},\delta B_\mu}=\gamma_e\,t_{\rm
  syn}\left[\gamma_e;\delta B_\mu\right]/t_{\rm dyn}$, which corresponds
to the cooling Lorentz factor in a homogeneous turbulence of strength
$\delta B_\mu$. The generic notation $t_{\rm syn}\left[\gamma_e;\delta
  B\right]$ refers to the synchrotron loss time for a particle of
Lorentz factor $\gamma_e$ in a magnetic field of strength $\delta B$.

Then, if $t_{\rm dyn}<t_{\mu+}$, cooling indeed occurs inside the
undecayed part of the microturbulence so that $\gamma_{\rm
  c}=\gamma_{{\rm c},\delta B_\mu}$. If $t_{\rm dyn}>t_{\mu-}$,
cooling instead occurs in the background field $B_{\rm d}$, meaning
$\gamma_{\rm c}=\gamma_{{\rm c},\delta B_\mu}\delta B_\mu^2/B_{\rm
  d}^2$. In the intermediate regime $t_{\mu+}<t_{\rm dyn}<t_{\mu-}$,
cooling takes place in the decaying part of the turbulence. The
cooling Lorentz factor is then defined as the solution of
\begin{equation} 
  t_{\rm
    syn}\left[\gamma_{\rm c};\delta B\left(t_{\rm
        dyn}\right)\right]=t_{\rm dyn}\ ,
\end{equation}
 which leads to
\begin{equation}
\gamma_{\rm c}\,=\, \gamma_{{\rm c},\delta
  B_\mu}^{1+\alpha_t}\gamma_{\mu+}^{-\alpha_t}\ .\label{eq:gammac}
\end{equation}
To summarize,
\begin{equation}
\gamma_{\rm c} \,=\,\begin{cases}
\displaystyle{\gamma_{{\rm c},\delta B_\mu}} & \text{if}\,\, t_{\rm dyn}<t_{\mu+} \\
\displaystyle{\gamma_{{\rm c},\delta
  B_\mu}^{1+\alpha_t}\gamma_{\mu+}^{-\alpha_t}} &
\text{if}\,\, t_{\mu+}<t_{\rm dyn}<t_{\mu-}\ ,\\
\displaystyle{\gamma_{{\rm c},\delta B_\mu}\frac{\delta B_\mu^2}{B_{\rm d}^2}} & \text{if}\,\, t_{\mu-}<t_{\rm dyn}
\end{cases}\label{eq:gammac2}
\end{equation}

The characteristic cooling frequency
\begin{equation}
\nu_{\rm c}\,=\,\nu_{\rm p}\left[\gamma_{\rm c};\delta B_{\gamma_{\rm
      c}}\right]\ ,\label{eq:nuc}
\end{equation}
with 
\begin{equation}
  \delta B_{\gamma_{\rm c}}\,=\,\max\left\{B_{\rm d},\,\delta B_\mu
    \left(\frac{t_{\rm dyn}}{t_{\mu+}}\right)^{\alpha_t/2}\right\}\ .\label{eq:dBgc}
\end{equation}
The field $\delta B_{\gamma_{\rm c}}$ corresponds to the field
strength at the point at which particles of Lorentz factor
$\gamma_{\rm c}$ cool through synchrotron radiation. The latter
expression means that if $t_{\rm dyn}<t_{\mu-}$, $\delta
B_{\gamma_{\rm c}}$ is given by the value of $\delta B$ at time
$t_{\rm dyn}$, while if $t_{\rm dyn}<t_{\mu-}$, $\delta B_{\gamma_{\rm
    c}}=B_{\rm d}$.  The factor $(t_{\rm dyn}/t_{\mu+})^{\alpha_t/2}$
can also be written as $(\gamma_{\rm c}/\gamma_{\mu+})^{-\delta_t}$,
with
\begin{equation}
\delta_t\,\equiv\, \frac{\alpha_t}{2(1+\alpha_t)}\ .\label{eq:delta}
\end{equation}
Note that the value $\delta B_{\gamma_{\rm c}}$ can also be understood
as the smallest value of the magnetic field in the blast.

The characteristic frequency $\nu_{\rm m}$ associated to particles of
Lorentz factor $\gamma_{\rm m}$ can also be derived similarly:
\begin{equation}
\nu_{\rm m}\,=\,\nu_{\rm p}\left[\gamma_{\rm m}; \delta B_{\gamma_{\rm
      m}}\right]\ ,\label{eq:num}
\end{equation}
with 
\begin{equation}
\delta B_{\gamma_{\rm m}}\,=\,\max\left\{
\delta B_{\gamma_{\rm c}}, \delta B_\mu
\left(\frac{\gamma_{\rm m}}{\gamma_{\mu+}}\right)^{-\delta_t}\right\}\
.\label{eq:dBgm}
\end{equation}
In short, this implies that $\delta B_{\gamma_{\rm m}}$ is determined
by $\delta B_{\gamma_{\rm c}}$ if $\gamma_{\rm m}<\gamma_{\rm c}$, and
by $\delta B_\mu \left(\gamma_{\rm
    m}/\gamma_{\mu+}\right)^{-\delta_t}$ otherwise, if the
$\gamma_{\rm m}$ particles can cool in the decaying layer. To
understand the former value, one should note that for $\alpha_t>-1$,
most of the power of non-cooling particles ($\gamma_{\rm
  m}<\gamma_{\rm c}$) is generated at the back of the blast, since the
time integrated power $\propto \int^{t_{\rm dyn}} {\rm d} t\, \delta
B^2(t)\propto \delta B_{\gamma_{\rm c}}^2$.

\begin{figure}
\includegraphics[width=0.49\textwidth]{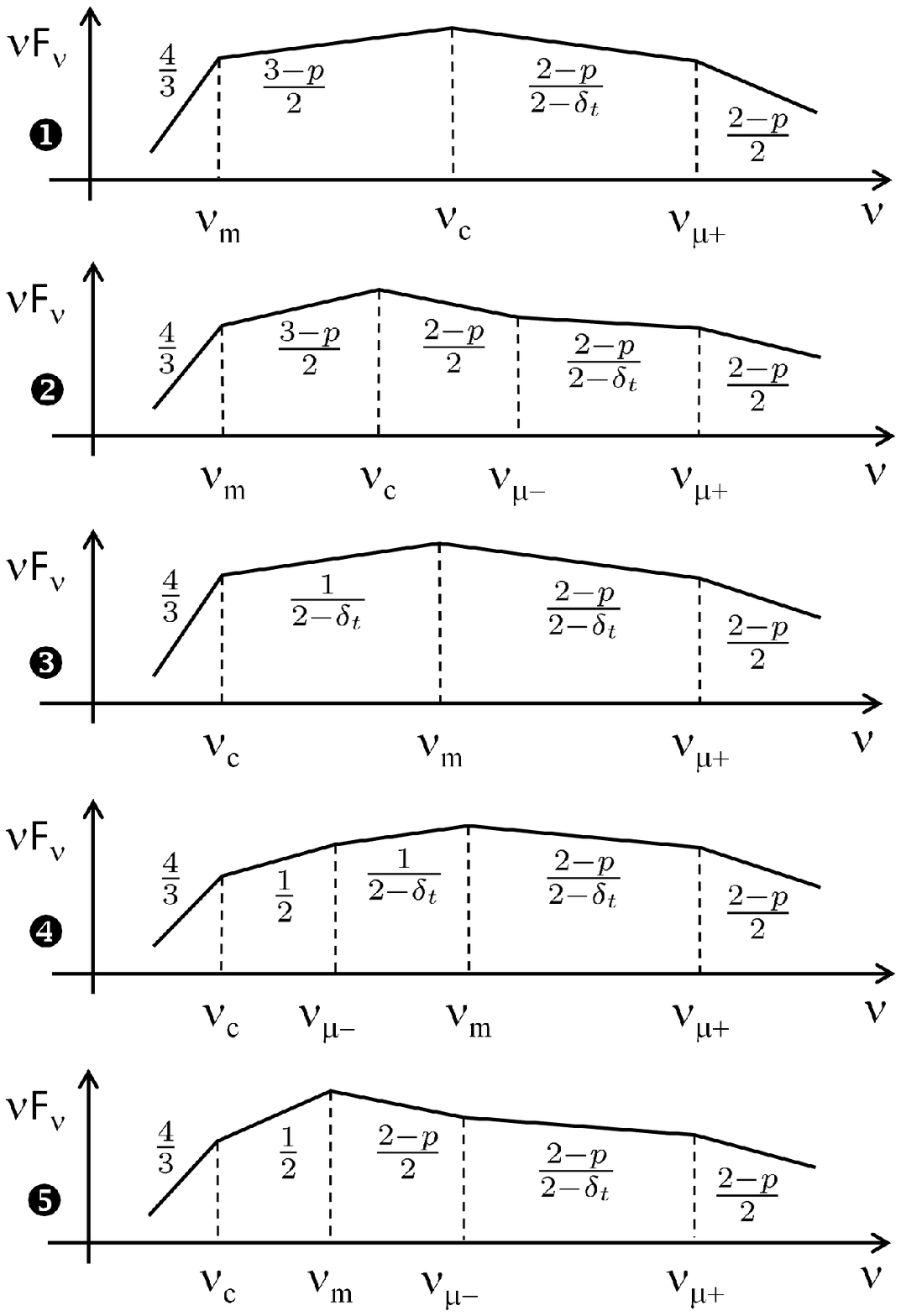}
\caption{Generic synchrotron spectra in time decaying microturbulence
  with $-1<\alpha_t<0$, neglecting inverse Compton losses, with the
  spectral indices as indicated.  Case 1:
  slow cooling scenario ($\gamma_{\rm c}>\gamma_{\rm m}$) with
  $t_{\rm dyn}<t_{\mu-}$, in which case the turbulence has not had
  time to relax to the background value $B_{\rm d}$; case 2: slow
  cooling scenario with $t_{\rm dyn}>t_{\mu-}$; case 3: fast cooling
  scenario with $\nu_{\mu-}<\nu_{\rm c}<\nu_{\rm m}$; case 4: fast
  cooling scenario with $\nu_{\rm c}<\nu_{\mu-}<\nu_{\rm m}$; case
  5: fast cooling scenario with $\nu_{\rm c}<\nu_{\rm
    m}<\nu_{\mu-}$.  See the discussion in Sec.~\ref{sec:appnoICasupm1}
  for the precise definitions of the various frequencies and for
  modifications to these spectra if $\nu_{\rm c}$ or $\nu_{\rm m}$
  exceeds $\nu_{\mu+}$, or if $\nu_{\rm max}$ is taken into account.
  \label{fig:spec_asupm1} }
\end{figure}

Depending on how $\nu_{\rm c}$, $\nu_{\rm m}$, $\nu_{\mu-}$ and
$\nu_{\mu+}$ are ordered one relatively to each other, one obtains
different synchrotron spectra. The various possible shapes of these
spectra with their characteristic indices are summarized in the five
cases depicted in Fig.~\ref{fig:spec_asupm1} and the spectro-temporal
indices of $F_\nu\propto t^{-\alpha}\nu^{-\beta}$ are provided in
Table~\ref{tab:gradnoIC}. This table assumes a decelerating blast wave
in an external density profile $\propto r^{-k}$.

\begin{table*}
\centering
\begin{minipage}{110mm}
  \caption{Spectral ($\beta$) and temporal ($\alpha$) indices of
    $F_\nu\propto t^{-\alpha}\nu^{-\beta}$ for various orderings of
    the characteristic frequencies, assuming a decaying
    microturbulence with $-1<\alpha_t<0$, neglecting inverse Compton
    losses. The different spectra match those depicted in
    Fig.~\ref{fig:spec_asupm1}. Case 1: slow cooling, $t_{\rm
      dyn}<t_{\mu-}$; case 2: slow cooling, $t_{\rm dyn}>t_{\mu-}$;
    case 3: fast cooling scenario with $\nu_{\mu-}<\nu_{\rm
      c}<\nu_{\rm m}$; case 4: fast cooling scenario with $\nu_{\rm
      c}<\nu_{\mu-}<\nu_{\rm m}$; case 5: fast cooling scenario with
    $\nu_{\rm c}<\nu_{\rm m}<\nu_{\mu-}$. Note that the exact values
    of the characteristic frequencies vary from case to case, see the
    accompanying text for details. For all cases, $-\alpha=(2-3p)/4$
    and $-\beta=-p/2$ if $\nu>\nu_{\mu+}$. The quantity $k$ refers to
    the external density profile $n\propto r^{-k}$.}
\begin{tabular}{@{}llrr@{}}
  \hline
  Case & Frequency range & $-\beta$ & $-\alpha$ \\
  \hline
  Case 1 &  $\nu<\nu_{\rm m}$ &
  $\frac{1}{3}$ & $\frac{6 (-2+k)+(-5+k) \alpha_t}{6 (-4+k)}$ \\
  &  $\nu_{\rm m}<\nu<\nu_{\rm c}$ & $\frac{1-p}{2}$  & $\frac{k (10-6
    p)+24 (-1+p)+(-5+k) (1+p) \alpha_t}{8 (-4+k)}$ \\
  & $\nu_{\rm c}<\nu<\nu_{\mu+}$ & $\frac{-2 p+(1-2 p) \alpha_t}{4+3
    \alpha_t}$ & $\frac{-2 (-4+k) (-2+3 p)+(-6+k (4-5 p)+15 p)
    \alpha_t}{2 (-4+k) \left(4+3 \alpha_t\right)}$ \\
  \hline
  Case 2 & $\nu<\nu_{\rm m}$ & $\frac{1}{3}$ & 
$\frac{2 (-3+k)}{3 (-4+k)}$ \\
 & $\nu_{\rm m}<\nu<\nu_{\rm c}$ & $\frac{1-p}{2}$ & 
$-\frac{(-3+k) (-1+p)}{-4+k}$ \\
 & $\nu_{\rm c}<\nu<\nu_{\mu-}$ & $-\frac{p}{2}$ & 
$\frac{-2+k+3 p-k p}{-4+k}$ \\
 & $\nu_{\mu-}<\nu<\nu_{\mu+}$ & 
$\frac{-2 p+(1-2 p) \alpha_t}{4+3 \alpha_t}$ &
$\frac{-2 (-4+k) (-2+3 p)+(-6+k (4-5 p)+15 p) \alpha_t}{2
  (-4+k) \left(4+3 \alpha_t\right)}$ \\
\hline
 Case 3 & $\nu<\nu_{\rm c}$ & $\frac{1}{3}$ & 
$\frac{-4+6 k+3 (-5+k) \alpha_t}{6 (-4+k)}$ \\
 & $\nu_{\rm c}<\nu<\nu_{\rm m}$ & 
$-\frac{2+\alpha_t}{4+3 \alpha_t}$ & 
$-\frac{-8+2 k+(-9+k) \alpha_t}{2 (-4+k) \left(4+3 
\alpha_t\right)}$ \\
& $\nu_{\rm m}<\nu<\nu_{\mu+}$ & 
$\frac{-2 p+(1-2 p) \alpha_t}{4+3 \alpha_t}$ & 
$\frac{-2 (-4+k) (-2+3 p)+(-6+k (4-5 p)+15 p) \alpha_t}{2
  (-4+k) \left(4+3 \alpha_t\right)}$ \\
\hline
Case 4 & $\nu<\nu_{\rm c}$ & $\frac{1}{3}$ & $\frac{2}{12-3 k}$\\
 & $\nu_{\rm c}<\nu<\nu_{\mu-}$ & $-\frac{1}{2}$ & $\frac{1}{-4+k}$ \\
 & $\nu_{\mu-}<\nu<\nu_{\rm m}$ & 
$-\frac{2+\alpha_t}{4+3 \alpha_t}$ & 
$-\frac{-8+2 k+(-9+k) \alpha_t}{2 (-4+k) \left(4+3 
\alpha_t\right)}$ \\
 & $\nu_{\rm m}<\nu<\nu_{\mu+}$ & 
$\frac{-2 p+(1-2 p) \alpha_t}{4+3 \alpha_t}$ & 
$\frac{-2 (-4+k) (-2+3 p)+(-6+k (4-5 p)+15 p) \alpha_t}{2
  (-4+k) \left(4+3 \alpha_t\right)}$ \\
\hline
Case 5 & $\nu<\nu_{\rm c}$ & $\frac{1}{3}$ & $\frac{2}{12-3 k}$ \\
 & $\nu_{\rm c}<\nu<\nu_{\rm m}$ & $-\frac{1}{2}$ & $\frac{1}{-4+k}$
 \\
 & $\nu_{\rm m}<\nu<\nu_{\mu-}$ & $-\frac{p}{2}$ & $\frac{-2+k+3 p-k
   p}{-4+k}$ \\
 & $\nu_{\mu-}<\nu<\nu_{\mu+}$ & 
$\frac{-2 p+(1-2 p) \alpha_t}{4+3 \alpha_t}$ & 
$\frac{-2 (-4+k) (-2+3 p)+(-6+k (4-5 p)+15 p) \alpha_t}{2
  (-4+k) \left(4+3 \alpha_t\right)}$ \\
\hline
\end{tabular}
\end{minipage}
\label{tab:gradnoIC}
\end{table*}

The following describes in some detail how the spectra are obtained,
starting from the single particle spectra. Section~\ref{sec:appICgrad}
discusses similar cases under the assumption of strong inverse Compton
losses.

\subsubsection{Fast cooling}
Consider first the simplest case of fast cooling, for which $\nu_{\rm
  c}$ is smaller than all other critical frequencies. The quantity
${\rm d}P_{\nu}/{\rm d}\dot N_e$, see Eq.~\ref{eq:appdFnu}, represents
the integral over the history of one cooling particle of initial
Lorentz factor $\gamma_{\rm e,0}$. 

If $\gamma_{e,0}>\gamma_{\mu+}$, the particle starts to cool in the
undecayed part of the microturbulent layer where $\mu(t)<1$ and
continues cooling in the decaying part, see
Eq.~\ref{eq:cool2}. Straightforward integration of
Eq.~\ref{eq:appdFnu} then leads to $\nu\,{\rm d}P_{\nu}/{\rm d}\dot
N_e \propto \nu^{1-b_\nu}$ with
\begin{equation}
1-b_\nu\,=\,\begin{cases} +1/2  & \text{if}\,\, \nu_{\mu+}<\nu<\nu_{e,0,\delta B_\mu} \\
(1+\delta_t/2)/(2-\delta_t) & \text{if}\,\,{\rm
  max}\left(\nu_{\mu-},\nu_{\rm c}\right)<\nu<\nu_{\mu+}\\
+1/2 & \text{if}\,\,\nu_{\rm c}<\nu<\nu_{\mu-}\\
+4/3 &  \text{if}\,\,\nu<\nu_{\rm c}\\
\end{cases}
\end{equation}
The frequency 
\begin{equation}
  \nu_{e,0,\delta B_\mu}\,\equiv\,\nu_{\rm p}\left[\gamma_{e,0};\delta
    B_\mu\right]\ .
\end{equation}  
Of course, the spectral power vanishes above $\nu_{e,0,\delta B_\mu}$.

If $\gamma_{e,0}<\gamma_{\mu+}$, which appears much more likely as
discussed in Sec.~\ref{sec:radmu}, but $\gamma_{e,0}>\gamma_{\mu-}$
(and $\gamma_{e,0}>\gamma_{\rm c}$), then $\nu_{e,0}$ is obtained by
solving first for the time $t_{e,0}$ at which the particle actually
cools, as for $\nu_{\rm c}$ and $\nu_{\rm m}$:
\begin{equation}
t_{\rm syn}\left[\gamma_{e,0};\delta B\left(t_{e,0}\right)\right]=t_{e,0}\ ,
\end{equation}
which leads to a shifted peak frequency
\begin{equation}
  \nu_{e,0} = \nu_{e,0,\delta B_\mu}\left(\frac{\gamma_{e,0}}{\gamma_{\mu+}}\right)^{-\delta_t}\ .\label{eq:newnu}
\end{equation}
Then
\begin{equation}
1-b_\nu\,=\,\begin{cases} 2+2/\alpha_t  & \text{if}\,\,
  \nu_{e,0}<\nu<\nu_{e,0,\delta B_\mu} \\
(1+\delta_t/2)(2-\delta_t) & \text{if}\,\,{\rm
  max}\left(\nu_{\mu-},\nu_{\rm c}\right)<\nu<\nu_{e,0}\\
+1/2 & \text{if}\,\,\nu_{\rm c}<\nu<\nu_{\mu-}\\
+4/3 &  \text{if}\,\,\nu<\nu_{\rm c}\\
\end{cases}
\end{equation}
For $\nu_{e,0}<\nu<\nu_{e,0,\delta B_\mu}$, the standard index $+1/2$
has become $2 + 2/\alpha_t$, which may take large or small negative
values depending on whether $\alpha_t$ lies close to 0 or to
$-1$. This index corresponds to the radiation of a ``non cooling''
particle in a changing microturbulent magnetic field. 

Finally, if $\gamma_{e,0}\ll\gamma_{\mu-}$ (but
$\gamma_{e,0}\gg\gamma_{\rm c}$), the particle does not cool in the
decaying microturbulence. The peak frequency for $\nu\,{\rm
  d}P_{\nu}/{\rm d}\dot N_e $ has moved to
\begin{equation}
  \nu_{e,0} = \nu_{e,0,\delta B_\mu}\frac{B_{\rm d}}{\delta B_\mu}\ ,
\end{equation}
as expected, and for $\nu_{e,0}<\nu<\nu_{e,0,\delta B_\mu}$, one finds
the slope $2+2/\alpha_t$. Of course, for $\nu_{\rm c}<\nu<\nu_{e,0}$,
one obtains $+1/2$.

Folding the previous results over the particle distribution of initial
Lorentz factors is straightforward, albeit somewhat tedious. For the
fast cooling scenario considered here, meaning $\nu_{\rm c}<\nu_{\rm
  m}$, this leads to three generic spectra depicted as cases $3$, $4$
and $5$ in Fig.~\ref{fig:spec_asupm1}, depending on the ordering of
$\nu_{\rm c}$ and $\nu_{\rm m}$ relatively to $\nu_{\mu-}$:
$\nu_{\mu-}<\nu_{\rm c}<\nu_{\rm m}$ (case 3), $\nu_{\rm
  c}<\nu_{\mu-}<\nu_{\rm m}$ (case 4), $\nu_{\rm c}<\nu_{\rm
  m}<\nu_{\mu-}$ (case 5). This figure does not consider the unlikely
cases associated to the possibility $\nu_{\rm m}>\nu_{\mu+}$; these
are briefly addressed further below.  The distinctive features of
these spectra can be summarized as follows.

If $\nu_{\mu-}<\nu_{\rm c}$, the slope of $\nu F_\nu$ becomes
$+1/(2-\delta_t)$ for $\nu_{\rm c}<\nu<\nu_{\rm m}$; if $\nu_{\rm
  c}<\nu_{\mu-}$ however, the slope is $+1/2$ for $\nu_{\rm
  c}<\nu<\nu_{\mu-}$ (fast cooling in $B_{\rm d}$) and
$+1/(2-\delta_t)$ for $\nu_{\mu-}<\nu<\nu_{\rm m}$ (fast cooling in
decaying turbulence). For $\nu_{\rm m}<\nu<\nu_{\mu+}$, the slope is
$(2-p)/(2-\delta_t)$. The general trend of decaying microturbulence is
to produce flatter synchrotron spectra than in a homogeneous magnetic
field, due to the stretch in frequency associated to cooling in
regions of different magnetic field strengths.

The maximal frequency $\nu_{\rm max}$ associated to $\gamma_{\rm max}$
is calculated in a similar fashion to $\nu_{\rm m}$. This frequency
does not appear in Fig.~\ref{fig:spec_asupm1} for the sake of clarity
but its impact can be described as follows. If $\nu_{\rm
  max}>\nu_{\mu+}$, spectral power vanishes above $\nu_{\rm
  max}$. Otherwise, $\nu F_\nu$ has spectral index $2+2/\alpha_t$ for
$\nu_{\rm max}<\nu<\nu_{\rm p}\left[\gamma_{\rm max};\delta
  B_\mu\right]$, which as before may take values close to zero if
$\alpha_t$ is close to $-1$, and it vanishes beyond $\nu_{\rm
  p}\left[\gamma_{\rm max};\delta B_\mu\right]$.

The peak power $\nu F_\nu$ is radiated at $\nu_{\rm m}$ as usual,
with
\begin{equation}
\left .\nu F_{\nu}\right\vert_{\nu=\nu_{\rm m}}\,\approx\,
\frac{0.28}{4\pi D_L^2}\frac{4}{3}\gamma_{\rm b}^2 \dot N_e
\gamma_{\rm m} m_e c^2\ .\label{eq:nFnfast}
\end{equation}
The above relates the maximum power to the incoming electron energy
per unit time, which avoids specifying the value of the magnetic field
in which cooling takes place. The numerical prefactor $0.28$ matches
the prefactors derived in Panaitescu \& Kumar (2000) for non-decaying
turbulence, Lorentz beaming is included through the factor
$4\gamma_{\rm b}^2/3$ and $\dot N_e$ has been defined in
Eq.~(\ref{eq:dNedg}).

\subsubsection{Slow cooling}
For high energy particles with $\gamma_{e,0}>\gamma_{\rm c}$, the
individual spectra of the quantity $\nu\,{\rm d}P_{\nu}/{\rm d}\dot
N_e$ mimic those discussed in the fast cooling section before and this
discussion thus concentrates on the bulk of electrons for which
$\gamma_{e,0}<\gamma_{\rm c}$. Such electrons do not cool
substantially anywhere in this slow cooling limit, therefore one finds
either a slope $+4/3$ at low frequencies, or a slope $2+2/\alpha_t$ at
high frequencies corresponding to the changing magnetic field. There
are two frequencies associated to $\gamma_{e,0}$: $\nu_{e,0,\delta
  B_\mu}$ as before, corresponding to the peak frequency of emission
when the particle experiences the undecayed $\delta B_\mu$, and
$\nu_{e,0}$ the frequency to be calculated in the lowest magnetic
field found downstream, i.e. at the back of the blast. Then
\begin{equation}
  \nu_{e,0}\,=\,  \nu_{\rm  p}\left[\gamma_{e,0};\delta B_{\gamma_{\rm
      c}}\right]\ ,
\end{equation}
provided $\gamma_{e,0}<\gamma_{\mu+}$. 

As before, folding over the particle initial Lorentz distribution
leads to the all-particle spectra. The possible spectra are displayed
as cases $1$ and $2$ in Fig.~\ref{fig:spec_asupm1}, depending on the
ordering of $t_{\rm dyn}$ vs $t_{\mu-}$: case $1$ if $t_{\rm
  dyn}<t_{\mu-}$, case 2 if $t_{\rm dyn}>t_{\mu-}$, which implies
$\nu_{\rm c}<\nu_{\mu-}$ and $\delta B_{\gamma_{\rm c}}=B_{\rm d}$. If
$\nu_{\rm c}>\nu_{\mu+}$, one would of course recover the spectrum of
a standard slow cooling scenario in a homogeneous turbulence of
strength $\delta B_\mu$.

The distinctive features of the $\nu F_{\nu}$ spectra can be
summarized as follows: the slope $1-p/2$ of the fast cooling part of
the particle population has been turned into $(2-p)/(2-\delta_t)$ for
${\rm max}\left(\nu_{\rm c},\nu_{\mu-}\right)<\nu<\nu_{\mu+}$. The
$+4/3$ slope for $\nu<\nu_{\rm m}$ remains unchanged, just as
$(3-p)/2$ for $\nu_{\rm m}<\nu<\nu_{\rm c}$, or $1-p/2$ for
$\nu_{\mu+}<\nu$ (assuming $\nu_{\mu+}<\nu_{\rm max}$ of course).

The peak power is now radiated at $\nu_{\rm c}$, with
\begin{equation}
  \left .\nu F_{\nu}\right\vert_{\nu=\nu_{\rm c}}\,\approx\,
  \frac{0.28}{4\pi D_L^2}\frac{4}{3}\gamma_{\rm b}^2 \dot N_e
  \gamma_{\rm c} m_e c^2\left(\frac{\gamma_{\rm c}}{\gamma_{\rm
        m}}\right)^{1-p}\ .
  \label{eq:nFnslow}
\end{equation}

\subsection{Rapid decay: $\alpha_t<-1$; no inverse Compton cooling}\label{sec:appnoICainfm1}
This Section now considers the limit in which the energy density
stored in the microturbulence decreases faster than $t^{-1}$. Inverse
Compton losses are neglected here; their impact is discussed in
Sec.~\ref{sec:appICrapid}. The crucial difference between the limit
$\alpha_t<-1$ and that discussed in the previous Section has to do
with the cooling history of a particle. In the present case, either
the initial Lorentz factor $\gamma_{e,0}>\gamma_{\mu+}$, in which case
the particle cools down to $\gamma_{\mu+}$ after crossing the
undecayed part of the microturbulent layer, or $\gamma_{e,0}\leq
\gamma_{\mu+}$, in which case it does not cool anywhere in the
decaying microturbulent layer, see Eq.~(\ref{eq:cool2}). In this
latter case, the particle eventually cools in the background magnetic
field, provided $t_{\rm dyn}>t_{\mu-}$.

The cooling Lorentz factor and its corresponding frequency should
therefore be defined as follows. If $t_{\rm dyn}<t_{\mu+}$, one
recovers a trivial case as it means that the turbulence has not
relaxed beyond $\delta B_\mu$, hence the turbulence is homogeneous
downstream. This case is not discussed further here. If
$t_{\mu-}<t_{\rm dyn}$, the turbulence has relaxed to $B_{\rm d}$ by
$t_{\rm dyn}$; in this case, $\gamma_{\rm c}$ is defined as usual in
terms of $t_{\rm dyn}$ and $B_{\rm d}$. Indeed, cooling cannot take
place in the decaying part of the microturbulent layer, but cooling is
possible in the background shock compressed field $B_{\rm d}$.  The
cooling frequency then reads $\nu_{\rm c}=\nu_{\rm p}\left[\gamma_{\rm
    c};B_{\rm d}\right]$. In the intermediate limit, $t_{\mu+}<t_{\rm
  dyn}<t_{\mu-}$, the turbulence has not had time to relax down to
$B_{\rm d}$. The cooling Lorentz factor remains undefined; however,
one can understand the spectra obtained with the correspondence
$\gamma_{\rm c}\rightarrow\gamma_{\mu+}$ and $\nu_{\rm
  c}\rightarrow\nu_{\mu+}$, since particles above $\gamma_{\mu+}$ do
cool down to $\gamma_{\mu+}$ in $\delta B_\mu$.

For a particle of initial Lorentz factor $\gamma_{e,0}$, one should
define two critical frequencies: $\nu_{e,0,\delta B_\mu}\equiv\nu_{\rm
  p}\left[\gamma_{e,0};\delta B_\mu\right]$ and $\nu_{e,0,B_{\rm
    d}}\equiv\nu_{\rm p}\left[\gamma_{e,0}; B_{\rm d}\right]$. Note
that the Lorentz factors $\gamma_{\mu+}$ and $\gamma_{\mu-}$ and their
corresponding frequencies $\nu_{\mu+}$ and $\nu_{\mu-}$ remain
unchanged here. It proves necessary to define a new frequency
associated to particles of Lorentz factor $\gamma_{\mu+}$ radiating in
the lowest magnetic field $\delta B_{\gamma_{\rm c}}$, with as before
$\delta B_{\gamma_{\rm c}}\rightarrow B_{\rm d}$ if $t_{\rm
  dyn}>t_{\mu-}$:
\begin{equation}
  \nu_{\mu0} \,\equiv\,\nu_{\rm p}\left[\gamma_{\mu+};\,
    \delta B_{\gamma_{\rm c}}\right]
  \ .
\end{equation}
One also define 
\begin{equation}
\nu_{{\rm m},\delta B_\mu}\,=\,\nu_{\rm p}\left[\gamma_{\rm m};\,
\delta B_\mu\right]\ .\label{eq:numdBm}
\end{equation}
The expression of the characteristic frequency $\nu_{\rm m}$ is given
further below, case by case.

Spectra $\nu\,{\rm d}P_{\nu}/{\rm d}\dot N_e\propto \nu^{1-b_\nu}$
integrated over the cooling history of a particle of initial Lorentz
factor $\gamma_{e,0}$ show:
\begin{equation}
  1-b_\nu\,=\,\begin{cases}
    +1/2 &\text{if}\,\, \nu_{\mu+}<\nu<\nu_{e,0,\delta
      B_\mu}\\
    +1/2 &\text{if}\,\,\nu_{\rm
      c}<\nu< \min\left(\nu_{\mu-},\nu_{e,0,B_{\rm d}}\right)\\
    \min\left(2 + 2/\alpha_t,\,4/3\right) & \text{if}\,\,
    \min\left(\nu_{\mu-},\nu_{e,0,B_{\rm d}}\right)<\nu \\ 
    & \text{and}\,\,\nu<\min\left(\nu_{\mu+},\nu_{e,0,\delta B_\mu}\right)\\
\end{cases}
\end{equation}
To understand the latter slope, one may recall that in a decaying
microturbulence, radiation in a region of small extent but high
magnetic power competes with radiation in a region of large extent at
small magnetic power. If $\alpha_t<-3$, decay is so fast that most of
the radiation is produced in $\delta B_\mu$ and one collects at low
frequencies the $+4/3$ tail. If however, $-3<\alpha_t<-1$, the
radiation produced by the particle as it crosses the decaying part of
the turbulence dominates this tail and the slope becomes
$2+2/\alpha_t$. This latter can be much flatter, possibly giving rise
to a flat energy spectrum in the limit $\alpha_t\rightarrow-1$.

\begin{figure}
\includegraphics[width=0.49\textwidth]{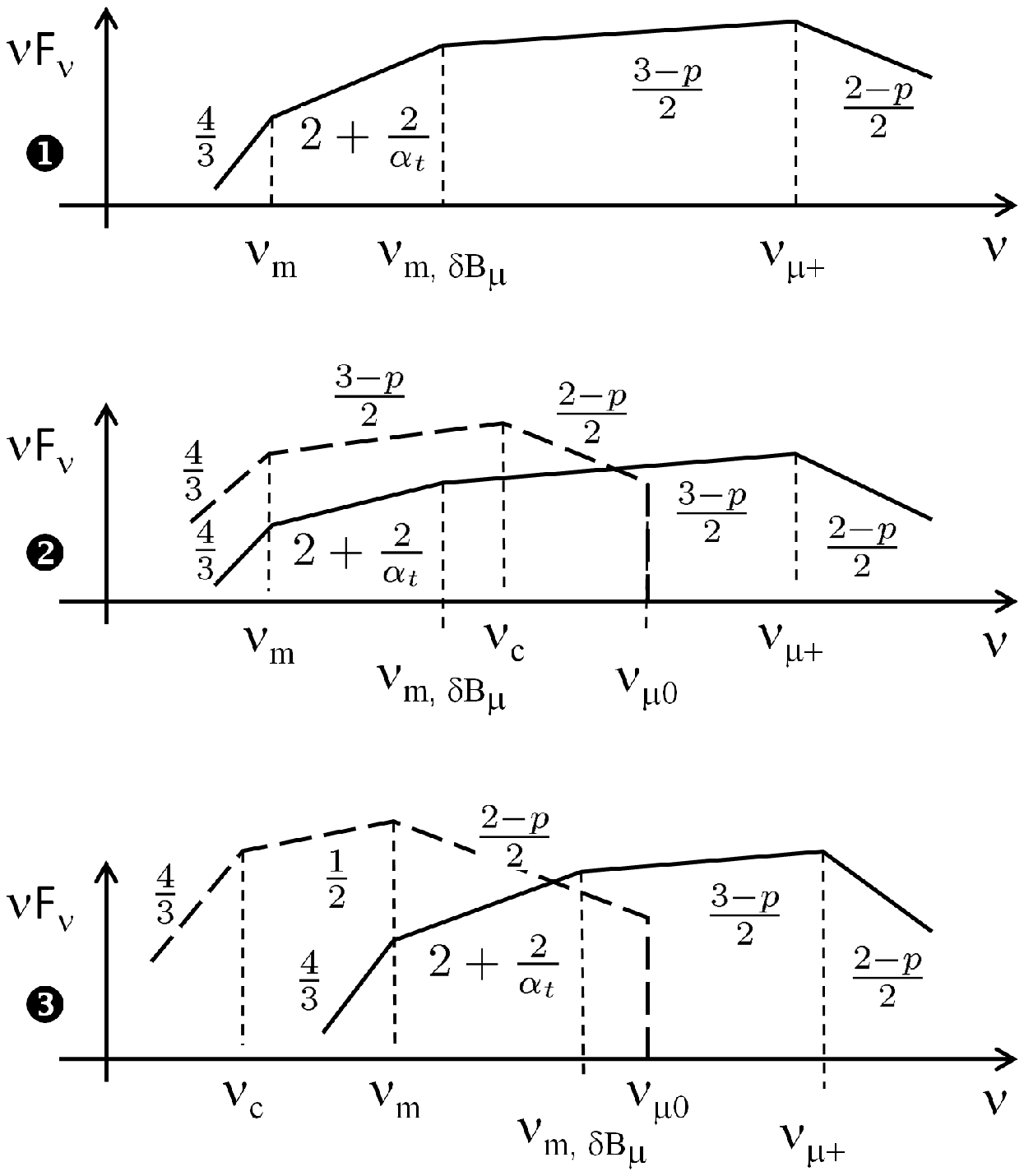}
\caption{Generic synchrotron spectra in time decaying microturbulence
  with $-3<\alpha_t<-4/(p+1)$, neglecting inverse Compton losses, with
  the spectral indices as indicated.  Case 1: scenario with $t_{\rm
    dyn}<t_{\mu-}$, in which case the turbulence has not had time to
  relax to the background value $B_{\rm d}$; case 2: slow cooling
  scenario with $t_{\rm dyn}>t_{\mu-}$; case 3: fast cooling scenario
  with $t_{\rm dyn}>t_{\mu-}$.  See the discussion in
  Sec.~\ref{sec:appnoICainfm1} for the precise definitions of the
  various frequencies. To obtain the spectra for $\alpha_t<-3$, it
  suffices to carry out the replacement $2+2/\alpha_t \rightarrow
  4/3$. The dashed line represents the secondary synchrotron component
  associated to cooling in the background shock compressed field
  $B_{\rm d}$, whenever $t_{\rm
    dyn}>t_{\mu-}$.\label{fig:spec_ainfm42pp1} }
\end{figure}

\begin{figure}
\includegraphics[width=0.49\textwidth]{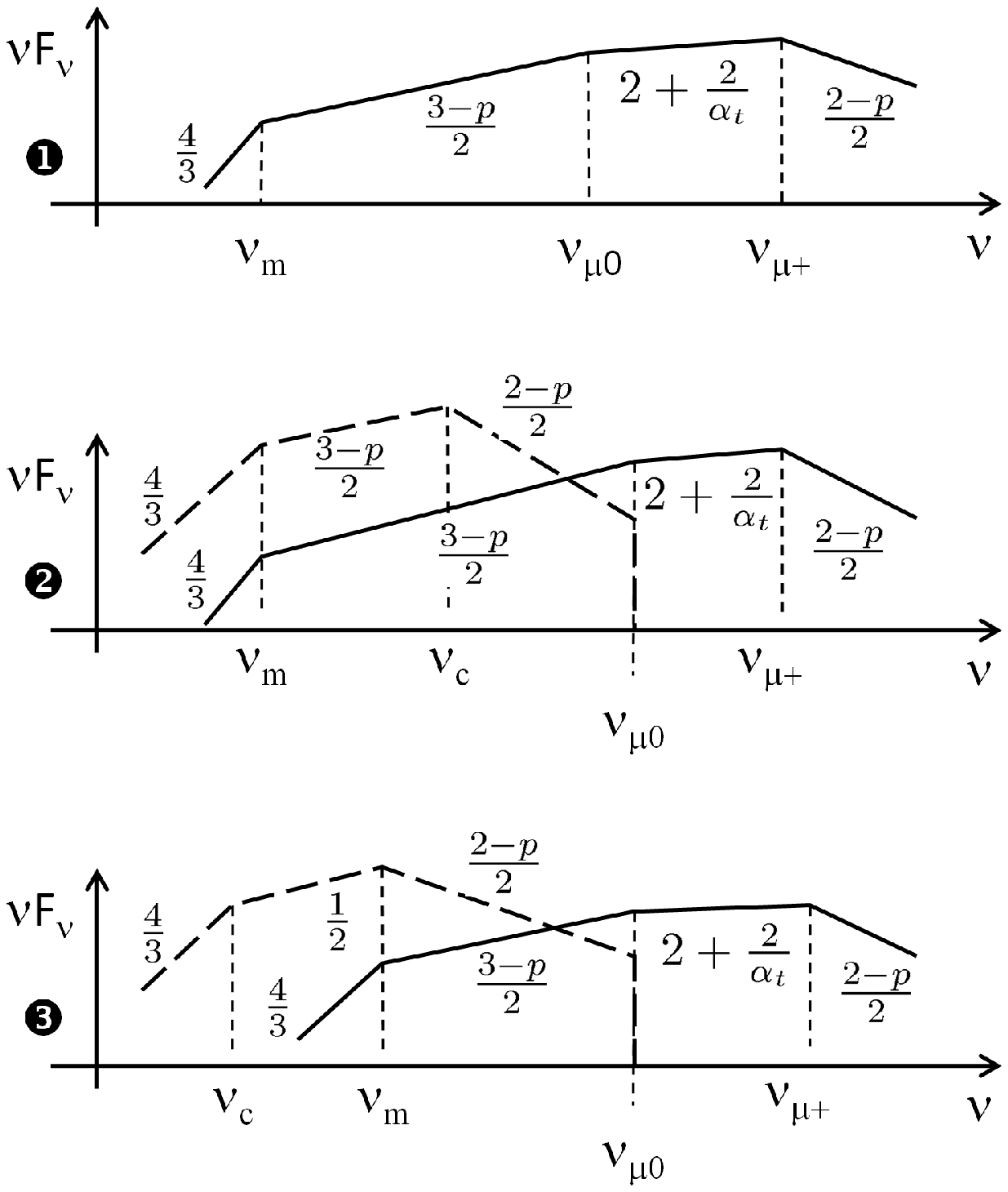}
\caption{Same as Fig.~\ref{fig:spec_ainfm42pp1} for
  $-4/(p+1)<\alpha_t<-1$.\label{fig:spec_asupm42pp1} }
\end{figure}

After folding over the particle population, one obtains the generic
spectra depicted in Fig.~\ref{fig:spec_ainfm42pp1} for
$-3<\alpha_t<-4/(p+1)$ and in Fig.~\ref{fig:spec_asupm42pp1} for
$-4/(p+1)<\alpha_t<-1$. The salient features of these spectra can be
summarized as follows.

Consider for simplicity the limit $-3<\alpha_t<-4/(p+1)$. The other
limit $-4/(p+1)<\alpha_t<-1$ can be understood in a similar way, while
the limit $\alpha_t<-3$ follows from the former after replacing
$2+2/\alpha_t$ with $4/3$. If $t_{\rm dyn}<t_{\mu-}$, the turbulence
has not had time to relax down to $B_{\rm d}$. At high frequencies
$>\nu_{{\rm m},\delta B_\mu}$, the spectrum then takes the form of a
slow cooling scenario in a homogeneous turbulence of strength $\delta
B_\mu$, with $\nu_{\rm c}\rightarrow \nu_{\mu+}$. As discussed above,
the turbulence decays so fast that the early emission in the region of
high magnetic power dominates that further away from the shock. At
frequencies $\nu_{\rm m}<\nu<\nu_{{\rm m},\delta B_\mu}$, one collects
the low energy extension with slope $2+2/\alpha_t$, instead of $4/3$
as discussed before. Note that the frequency $\nu_{\rm m}=\nu_{\rm
  p}\left[\gamma_{\rm m};\delta B_{\gamma_{\rm c}}\right]$. For
$\nu<\nu_{\rm m}$, one recovers the slope $+4/3$ as expected. This
case is denoted case 1 in Fig.~\ref{fig:spec_ainfm42pp1}.

In case 2, one now assumes $t_{\rm dyn}>t_{\mu-}$ (relaxed turbulence)
and $\nu_{\rm c}>\nu_{\rm m}$ (slow cooling). The afterglow comprises
two contributions: one associated to slow cooling in $\delta B_\mu$,
as before; plus a second one associated to slow cooling in the
background $B_{\rm d}$. This latter is indicated in dashed lines in
Fig.~\ref{fig:spec_ainfm42pp1}; it exhibits a cut-off above
$\nu_{\mu0}$, since there are no particles with Lorentz factor above
$\gamma_{\mu+}$ beyond $t_{\mu+}$. If $\gamma_{\rm
  max}<\gamma_{\mu+}$, the cut-off would of course take place at
$\nu_{\rm p}\left[\gamma_{\rm max};B_{\rm d}\right]$. The
characteristic frequency $\nu_{\rm m}= \nu_{\rm p}\left[\gamma_{\rm
    m};\,B_{\rm d}\right]$.

The peak power for the $\nu F_{\nu,B_{\rm d}}$ component associated to
cooling in $B_{\rm d}$ is standard, while the power $\nu F_{\nu,\delta
  B_\mu}$ related to the decaying microturbulence can be written at
$\nu_{\mu+}$ as
\begin{equation}
  \left .\nu F_{\nu,\delta B_\mu}\right\vert_{\nu=\nu_{\mu+}}\,\approx\,
  \frac{0.28}{4\pi D_L^2}\frac{4}{3}\gamma_{\rm b}^2 \dot N_e
  \gamma_{\mu+} m_e c^2\left(\frac{\gamma_{\mu+}}{\gamma_{\rm
        m}}\right)^{1-p}\ .
\label{eq:nFnainf}
\end{equation}
The ratio of energy fluxes at their respective peaks reads
\begin{equation}
\frac{\nu F_{\nu,B_{\rm d}}}{\nu F_{\nu,\delta B_\mu}}\,=\,
\left(\frac{\gamma_{\rm c}}{\gamma_{\mu+}}\right)^{2-p}\,=\,
\left(\frac{\delta B_\mu^2\,t_{\mu+}}{B_{\rm d}^2\, t_{\rm
      dyn}}\right)^{2-p}\ ,
\label{eq:nFnBd}
\end{equation}
and it scales as one would expect with the ratio of the product of
synchrotron power times the exposure to the magnetic field.

One may also calculate the energy flux ratio at $\nu_{\rm m}$:
\begin{equation}
  \frac{\nu_{\rm m} F_{\nu_{\rm m},B_{\rm d}}}{\nu_{\rm m} F_{\nu_{\rm m},\delta B_\mu}}\,=\,
  \frac{t_{\rm dyn}}{t_{\mu-}}\ .
\end{equation}
 
In case 3, one now assumes a similar configuration with $t_{\rm
  dyn}>t_{\mu-}$ (relaxed turbulence), but $\nu_{\rm c}<\nu_{\rm
  m}$, meaning fast cooling in $B_{\rm d}$. The only difference with
the previous spectrum corresponds to that change of cooling regime for
the component $\nu F_{\nu,B_{\rm d}}$. As discussed before, the bulk
of electrons cannot cool in the microturbulent field (as long as
$\gamma_{\rm m}<\gamma_{\mu+}$).

Figure~\ref{fig:spec_asupm42pp1} shows the corresponding spectra for
the case $-4/(p+1)<\alpha_t<-1$. The discussion is very similar to the
previous one for $\alpha_t<-4/(p+1)$ and the differences are as
follows. For $-4/(p+1)<\alpha_t<-1$, the spectral index $2+2/\alpha_t$
is softer than $(3-p)/2$; the low energy tail of index $2+2/\alpha_t$
of the high energy population thus extends from $\nu_{\mu+}$ down to
$\nu_{\mu0}$ as it dominates the slow cooling contribution of the bulk
of electrons.

The spectro-temporal indices of $F_\nu\propto t^{-\alpha}\nu^{-\beta}$
for these two values of $\alpha_t$ are provided in
Table~\ref{tab:fastnoIC}.

\begin{table*}
\centering
\begin{minipage}{120mm}
  \caption{Spectral ($\beta$) and temporal ($\alpha$) indices of
    $F_\nu\propto t^{-\alpha}\nu^{-\beta}$ assuming a decaying
    microturbulence with $-3<\alpha_t<-1$, with negligible inverse
    Compton losses. In cases 2 and 3, one must superimpose a
    synchrotron component associated to cooling in the background
    shock compressed magnetic field; the spectro-temporal slopes given
    here concern only the synchrotron component associated to the
    decaying microturbulent layer, not the latter. Case 1: $t_{\rm
      dyn}<t_{\mu-}$; case 2: $t_{\rm dyn}>t_{\mu-}$ with slow cooling
    in the background shock compressed field; case 3: $t_{\rm
      dyn}>t_{\mu-}$ with fast cooling in the background shock
    compressed field. The corresponding synchrotron spectra are shown
    in Fig.~\ref{fig:spec_ainfm42pp1} for $-3<\alpha_t<\-4/(p+1)$ and
    in Fig.~\ref{fig:spec_asupm42pp1} for $-4/(p+1)<\alpha_t<-1$. For
    all cases, $-\alpha=(2-3p)/4$ and $-\beta=-p/2$ if
    $\nu>\nu_{\mu+}$. The quantity $k$ refers to the external
    density profile $n\propto r^{-k}$.}
\begin{tabular}{@{}llrr@{}}
  \hline
  Case & Frequency range & $-\beta$ & $-\alpha$ \\
  \hline
  Case 1 & $\nu<\nu_{\rm m}$ & $\frac{1}{3}$ & 
  $\frac{6 (-2+k)+(-5+k) \alpha_t}{6 (-4+k)}$ \\
  $\left[-4/(p+1)<\alpha_t<-1\right]$   & $\nu_{\rm m}<\nu<\nu_{\mu0}$ & $\frac{1-p}{2}$ &
  $\frac{k (10-6 p)+24 (-1+p)+(-5+k) (1+p) \alpha_t}{8 (-4+k)}$
  \\
  & $\nu_{\mu0}<\nu<\nu_{\mu+}$ & $1+\frac{2}{\alpha_t}$ & 
  $\frac{2 (6+k)+(2-k (-2+p)+9 p) \alpha_t}{2 (-4+k) 
    \alpha_t}$ \\
  \hline
  Cases 2 and 3  & $\nu<\nu_{\rm m}$ & 
  $\frac{1}{3}$ & 
  $ \frac{-6 k+(3+k) \alpha_t}{6 (-4+k) \alpha_t}$ \\
  $\left[-4/(p+1)<\alpha_t<-1\right]$  & $\nu_{\rm m}<\nu<\nu_{\mu0}$ & $\frac{1-p}{2}$ &
  $\frac{-2 k+(-1+k+6 p-2 k p) \alpha_t}{2 (-4+k) \alpha_t} $
  \\
  & $\nu_{\mu0}<\nu<\nu_{\mu+}$ & $1+\frac{2}{\alpha_t}$ & 
  $\frac{2 (6+k)+(2-k (-2+p)+9 p) \alpha_t}{2 (-4+k) \alpha_t} $ \\
  \hline
  Case 1 & $\nu<\nu_{\rm m}$ & $\frac{1}{3}$ & 
  $\frac{6 (-2+k)+(-5+k) \alpha_t}{6 (-4+k)}$ \\
  $\left[-3<\alpha_t<-4/(p+1)\right]$   & $\nu_{\rm m}<\nu<\nu_{{\rm m},\delta B_\mu}$ & 
  $1+\frac{2}{\alpha_t}$ & 
  $\frac{6 (-4+k)+(-7+3 k) \alpha_t}{2 (-4+k) \alpha_t} $ \\
  & $\nu_{{\rm m},\delta B_\mu}<\nu<\nu_{\mu+}$ & $\frac{1-p}{2}$ & 
  $\frac{2-3 k-12 p+3 k p}{16-4 k} $ \\
  \hline
  Cases 2 and 3 & $\nu<\nu_{\rm m}$ & $\frac{1}{3}$ & 
  $\frac{-6 k+(3+k) \alpha_t}{6 (-4+k) \alpha_t} $ \\
  $\left[-3<\alpha_t<-4/(p+1)\right]$  & $\nu_{\rm m}<\nu<\nu_{{\rm m},\delta B_\mu}$ & 
  $1+\frac{2}{\alpha_t}$ & 
  $\frac{6 (-4+k)+(-7+3 k) \alpha_t}{2 (-4+k) \alpha_t} $ \\
  & $\nu_{{\rm m},\delta B_\mu}<\nu<\nu_{\mu+}$ & $\frac{1-p}{2}$ & 
  $\frac{2-3 k-12 p+3 k p}{16-4 k} $ \\
  \hline
\end{tabular}
\end{minipage}
\label{tab:fastnoIC}
\end{table*}

\subsection{Gradual decay $\alpha_t>-4/(p+1)$, with dominant inverse
  Compton cooling}\label{sec:appICgrad}
This Section assumes $-4/(p+1)<\alpha_t<0$ and it assumes that the
Compton parameter $Y\gg1$ for all Lorentz factors, everywhere in the
blast. One defines
\begin{equation}
  Y_\mu\equiv \frac{U_{\rm rad}}{\delta B_\mu^2/(8\pi)}\ ,\label{eq:Ym}
\end{equation}
The cooling time of a particle is then defined as
\begin{equation}
t_{\rm cool}(\gamma_{e,0}) \,\equiv\, \frac{1}{1+Y_\mu}t_{\rm
    syn}\left[\gamma_{e,0};\delta B_\mu\right]\ ,
\end{equation}
and it is homogeneous throughout the blast. This simplifies the
cooling history of a particle:
\begin{equation}
  \gamma_e \,\simeq\,\begin{cases}
    \gamma_{e,0} & \text{if}\,\, t<t_{\rm cool}\left(\gamma_{e,0}\right) \\
    \gamma_{e,0} \displaystyle{\frac{t_{\rm cool}\left(\gamma_{e,0}\right)}{t}} &
    \text{if}\,\, t>t_{\rm cool}\left(\gamma_{e,0}\right)\ .
\end{cases}
\end{equation}

The definitions of the critical Lorentz factors must be adapted to
this case. One defines a Lorentz factor
\begin{equation}
\tilde\gamma_{\mu+}\,\equiv\,\frac{\gamma_{\mu+}}{1+Y_\mu}\ ,
\end{equation}
such that inverse Compton cooling takes place at the end of the
undecayed microturbulent layer. Similarly, one defines
$\tilde\gamma_{\mu-}\equiv\tilde\gamma_{\mu+} t_{\mu+}/t_{\mu-}$. The
associated frequency $\nu_{\mu+}$ (resp. $\nu_{\mu-}$) is defined as
before in terms of $\tilde\gamma_{\mu+}$
(resp. $\tilde\gamma_{\mu-}$).

The cooling Lorentz factor is defined as
\begin{equation}
\tilde \gamma_{\rm c}=\tilde \gamma_{\mu+}\frac{t_{\mu+}}{t_{\rm
    dyn}}\ .
\end{equation} 
Eqs.~\ref{eq:nuc},\ref{eq:dBgc} for the cooling frequency $\nu_{\rm
  c}$ and the definition of $\delta B_{\gamma_{\rm c}}$ remain valid.
Regarding $\nu_{\rm m}$, Eq.~\ref{eq:num} remains valid but the
definition of $\delta B_{\gamma_{\rm m}}$ must be modified to account
for inverse Compton losses:
\begin{equation}
  \delta B_{\gamma_{\rm m}}\,=\,\max\left\{ \delta B_{\gamma_{\rm c}},
    \, \delta B_\mu\left(\frac{\gamma_{\rm
          m}}{\tilde\gamma_{\mu+}}\right)^{-\alpha_t/2}\right\}\ .
\label{eq:dBgm2}
\end{equation}

The index of $\nu\,{\rm d}P_{\nu}/{\rm d}\dot N_e$ differs from the
standard case of homogeneous turbulence only if the particle radiates
in the changing magnetic field while it is cooling in the radiation
field. This applies to the spectral domain ${\rm
  max}\left(\nu_{\mu-},\nu_{\rm c}\right)<\nu<{\rm
  min}\left(\nu_{e,0},\nu_{\mu+}\right)$, in which the index of
$\nu\,{\rm d}P_{\nu}/{\rm d}\dot N_e$ becomes
$(1-\alpha_t)/(2-\alpha_t/2)$. The frequency $\nu_{e,0}$ is defined in
terms of $\gamma_{e,0}$ as $\nu_{\rm m}$ in terms of $\gamma_{\rm m}$.

\begin{figure}
\includegraphics[width=0.49\textwidth]{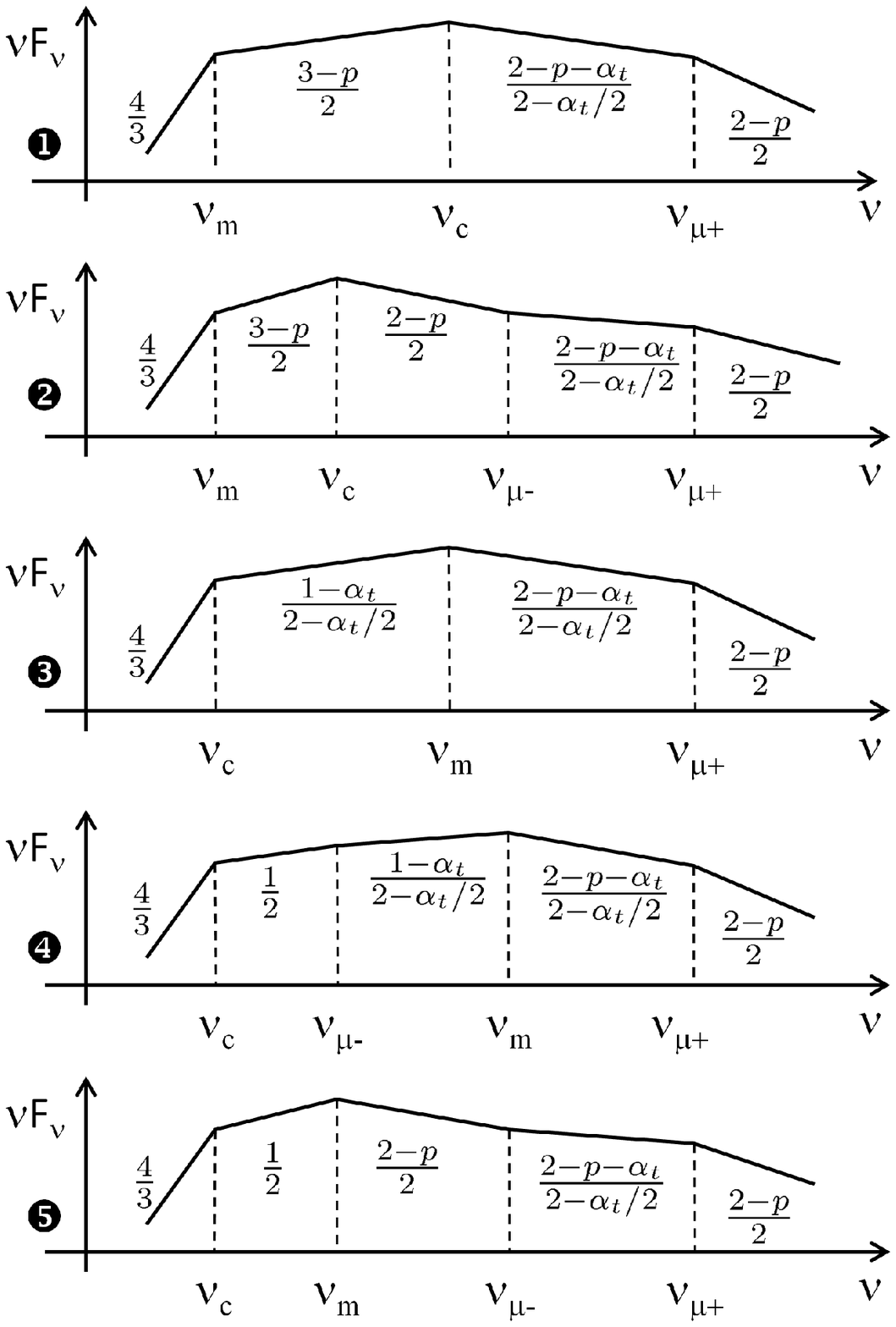}
\caption{Generic synchrotron spectra in time decaying microturbulence
  with $-4/(p+1)<\alpha_t<0$, assuming that inverse Compton losses
  dominate over synchrotron losses everywhere in the blast.  Case 1:
  slow cooling scenario ($\gamma_{\rm c}>\gamma_{\rm m}$) with $t_{\rm
    dyn}<t_{\mu-}$, in which case the turbulence has not had time to
  relax to the background value $B_{\rm d}$; case 2: slow cooling
  scenario with $t_{\rm dyn}>t_{\mu-}$; case 3: fast cooling scenario
  with $\nu_{\mu-}<\nu_{\rm c}<\nu_{\rm m}$; case 4: fast cooling
  scenario with $\nu_{\rm c}<\nu_{\mu-}<\nu_{\rm m}$; case 5: fast
  cooling scenario with $\nu_{\rm c}<\nu_{\rm min}<\nu_{\mu-}$.  See
  the discussion in Sec.~\ref{sec:appICgrad} for the precise
  definitions of the various frequencies and for modifications to
  these spectra if $\nu_{\rm c}$ or $\nu_{\rm m}$ exceeds
  $\nu_{\mu+}$, or if $\nu_{\rm max}$ is taken into account. Note that
  the spectral slope $(2-p-\alpha_t)/(2-\alpha_t/2)$ takes positive
  values whenever $\alpha_t<2-p$. \label{fig:spec_ICasupm42pp1} }
\end{figure}

After folding over the particle Lorentz distribution, one obtains the
generic full all-particle spectra represented in
Fig.~\ref{fig:spec_ICasupm42pp1}. As before, this figure ignores a
possible maximal frequency $\nu_{\rm max}$ and assumes that $\nu_{\rm
  c}$ and $\nu_{\rm m}$ are smaller than $\nu_{\mu+}$. The discussion
is very similar to that given in Sec.~\ref{sec:appnoICasupm1} and is
not be repeated here. The significant differences lie in the spectral
indices: $(2-p)/(2-\delta_t)$ has become $(2-p-\alpha_t)/(2-\alpha_t/2)$,
while $1/(2-\delta_t)$ (otherwise $1/2$ for standard fast cooling) has
become $(1-\alpha_t)/(2-\alpha_t/2)$. Inverse Compton losses take away
a large fraction of the dissipated energy, so that the synchrotron
peak power is modified as follows, for slow cooling (cases 1 and 2):
\begin{eqnarray}
  \left .\nu F_{\nu}\right\vert_{\nu=\nu_{\rm c}}&\,\approx\,&
  \frac{0.28}{4\pi D_L^2}\frac{4}{3}\gamma_{\rm b}^2 \dot N_e
  \gamma_{\rm c}m_e c^2
  \left(\frac{\tilde\gamma_{\rm c}}{\gamma_{\rm
        m}}\right)^{1-p}\times\nonumber\\
  & &\quad\quad\frac{1}{1+Y_\mu}\frac{\delta B_{\gamma_{\rm c}}^2}{\delta B_\mu^2}\ .\label{eq:nFnslowIC}
\end{eqnarray}
and for fast cooling (cases 3, 4 and 5):
\begin{eqnarray}
  \left .\nu F_{\nu}\right\vert_{\nu=\nu_{\rm m}}&\,\approx\,&
  \frac{0.28}{4\pi D_L^2}\frac{4}{3}\gamma_{\rm b}^2 \dot N_e
  \gamma_{\rm m}m_e c^2
  \times\nonumber\\
  & &\quad\quad\frac{1}{1+Y_\mu}\frac{\delta B_{\gamma_{\rm m}}^2}{\delta B_\mu^2}\ .\label{eq:nFnfastIC}
\end{eqnarray}
The ratios of magnetic energy densities $\delta B_{\gamma_{\rm
    c}}^2/\delta B_\mu^2$ (resp. $\delta B_{\gamma_{\rm m}}^2/\delta
B_\mu^2$) that appear in the both expressions, yield the proper $Y$
Compton parameter at the location at which most of the cooling of
particles of Lorentz factor $\gamma_{\rm c}$ (resp. $\gamma_{\rm m}$)
occurs.

The corresponding spectro-temporal indices for $F_\nu\propto
t^{-\alpha}\nu^{-\beta}$ are given in Table~\ref{tab:gradIC}.

\begin{table*}
\centering
\begin{minipage}{100mm}
  \caption{Spectral ($\beta$) and temporal ($\alpha$) indices of
    $F_\nu\propto t^{-\alpha}\nu^{-\beta}$ for various orderings of
    the characteristic frequencies, assuming a decaying
    microturbulence with $-4/(p+1)<\alpha_t$, with dominant inverse
    Compton losses everywhere in the blast. Case 1: slow cooling with
    $t_{\rm dyn}<t_{\mu-}$; case 2: slow cooling with $t_{\rm
      dyn}>t_{\mu-}$; case 3: fast cooling scenario with
    $\nu_{\mu-}<\nu_{\rm c}<\nu_{\rm m}$; case 4: fast cooling
    scenario with $\nu_{\rm c}<\nu_{\mu-}<\nu_{\rm m}$; case 5: fast
    cooling scenario with $\nu_{\rm c}<\nu_{\rm min}<\nu_{\mu-}$. The
    corresponding synchrotron spectra are shown in
    Fig.~\ref{fig:spec_ICasupm42pp1}. For all cases,
    $-\alpha=(2-3p)/4$ and $-\beta=-p/2$ if $\nu>\nu_{\mu+}$. The
    quantity $k$ refers to the external density profile $n\propto
    r^{-k}$.}
\begin{tabular}{@{}llrr@{}}
  \hline
  Case & Frequency range & $-\beta$ & $-\alpha$ \\
  \hline
  Case 1 &  $\nu<\nu_{\rm m}$ & $\frac{1}{3}$ & 
$\frac{6 (-2+k)+(-5+k) \alpha_t}{6 (-4+k)}$ \\
 & $\nu_{\rm m}<\nu<\nu_{\rm c}$ & $\frac{1-p}{2}$ &
$ \frac{k (10-6 p)+24 (-1+p)+(-5+k) (1+p) \alpha_t}{8
  (-4+k)}$ \\
 & $\nu_{\rm c}<\nu<\nu_{\mu+}$ & 
$\frac{2 p+\alpha_t}{-4+\alpha_t}$ & 
$\frac{2 (-4+k) (-2+3 p)+(2-k (-2+p)+9 p) \alpha_t}{2 (-4+k)
  \left(-4+\alpha_t\right)}$ \\
\hline
 Case 2 & $\nu<\nu_{\rm m}$ & $\frac{1}{3}$ & $\frac{2 (-3+k)}{3
   (-4+k)}$ \\
 & $\nu_{\rm m}<\nu<\nu_{\rm c}$ & $\frac{1-p}{2}$ & 
$-\frac{(-3+k) (-1+p)}{-4+k}$\\
 & $\nu_{\rm c}<\nu<\nu_{\mu-}$ & $-\frac{p}{2}$ & $-\frac{2+(-3+k)
   p}{-4+k}$\\ 
 & $\nu_{\mu-}<\nu<\nu_{\mu+}$ & 
$\frac{2 p+\alpha_t}{-4+\alpha_t}$ &
$\frac{2 (-4+k) (-2+3 p)+(2-k (-2+p)+9 p) \alpha_t}{2 (-4+k)
  \left(-4+\alpha_t\right)}$ \\
\hline
 Case 3 & $\nu<\nu_{\rm c}$ & $\frac{1}{3}$ & 
$\frac{-4+6 k+(-5+k) \alpha_t}{6 (-4+k)}$ \\
 & $\nu_{\rm c}<\nu<\nu_{\rm m}$ & 
$\frac{2+\alpha_t}{-4+\alpha_t}$ & 
$\frac{2 (-4+k)+(11+k) \alpha_t}{2 (-4+k) \left(-4+\alpha_t\right)}$ \\
 &  $\nu_{\rm m}<\nu<\nu_{\mu+}$ & 
$\frac{2 p+\alpha_t}{-4+\alpha_t}$ & 
$\frac{2 (-4+k) (-2+3 p)+(2-k (-2+p)+9 p) \alpha_t}{2 (-4+k)
  \left(-4+\alpha_t\right)}$ \\
\hline
Case 4 & $\nu<\nu_{\rm c}$ & $\frac{1}{3}$ & $\frac{2 (-1+k)}{3
  (-4+k)}$\\
 & $\nu_{\rm c}<\nu<\nu_{\mu-}$ & $-\frac{1}{2}$ & $\frac{1-k}{-4+k}$
 \\
 & $\nu_{\mu-}<\nu<\nu_{\rm m}$ & 
$\frac{2+\alpha_t}{-4+\alpha_t}$ & 
$\frac{2 (-4+k)+(11+k) \alpha_t}{2 (-4+k) \left(-4+\alpha_t\right)}$ \\
 & $\nu_{\rm m}<\nu<\nu_{\mu+}$ & 
$\frac{2 p+\alpha_t}{-4+\alpha_t}$ & 
$\frac{2 (-4+k) (-2+3 p)+(2-k (-2+p)+9 p) \alpha_t}{2 (-4+k)
  \left(-4+\alpha_t\right)}$ \\
\hline
Case 5 & $\nu<\nu_{\rm c}$ & $\frac{1}{3}$ & $\frac{2 (-1+k)}{3
  (-4+k)}$ \\
 & $\nu_{\rm c}<\nu<\nu_{\rm m}$ & $-\frac{1}{2}$ & 
$\frac{1-k}{-4+k}$ \\
 & $\nu_{\rm m}<\nu<\nu_{\mu-}$ & $-\frac{p}{2}$ & 
$-\frac{2+(-3+k) p}{-4+k}$ \\
 & $\nu_{\mu-}<\nu<\nu_{\mu+}$ & 
$\frac{2 p+\alpha_t}{-4+\alpha_t}$ & 
$\frac{2 (-4+k) (-2+3 p)+(2-k (-2+p)+9 p) \alpha_t}{2 (-4+k)
  \left(-4+\alpha_t\right)}$ \\
\hline
\end{tabular}
\end{minipage}
\label{tab:gradIC}
\end{table*}

\subsection{Rapid decay $-3<\alpha_t<-4/(p+1)$, with dominant inverse
  Compton cooling}\label{sec:appICrapid}
\begin{figure}
\includegraphics[width=0.49\textwidth]{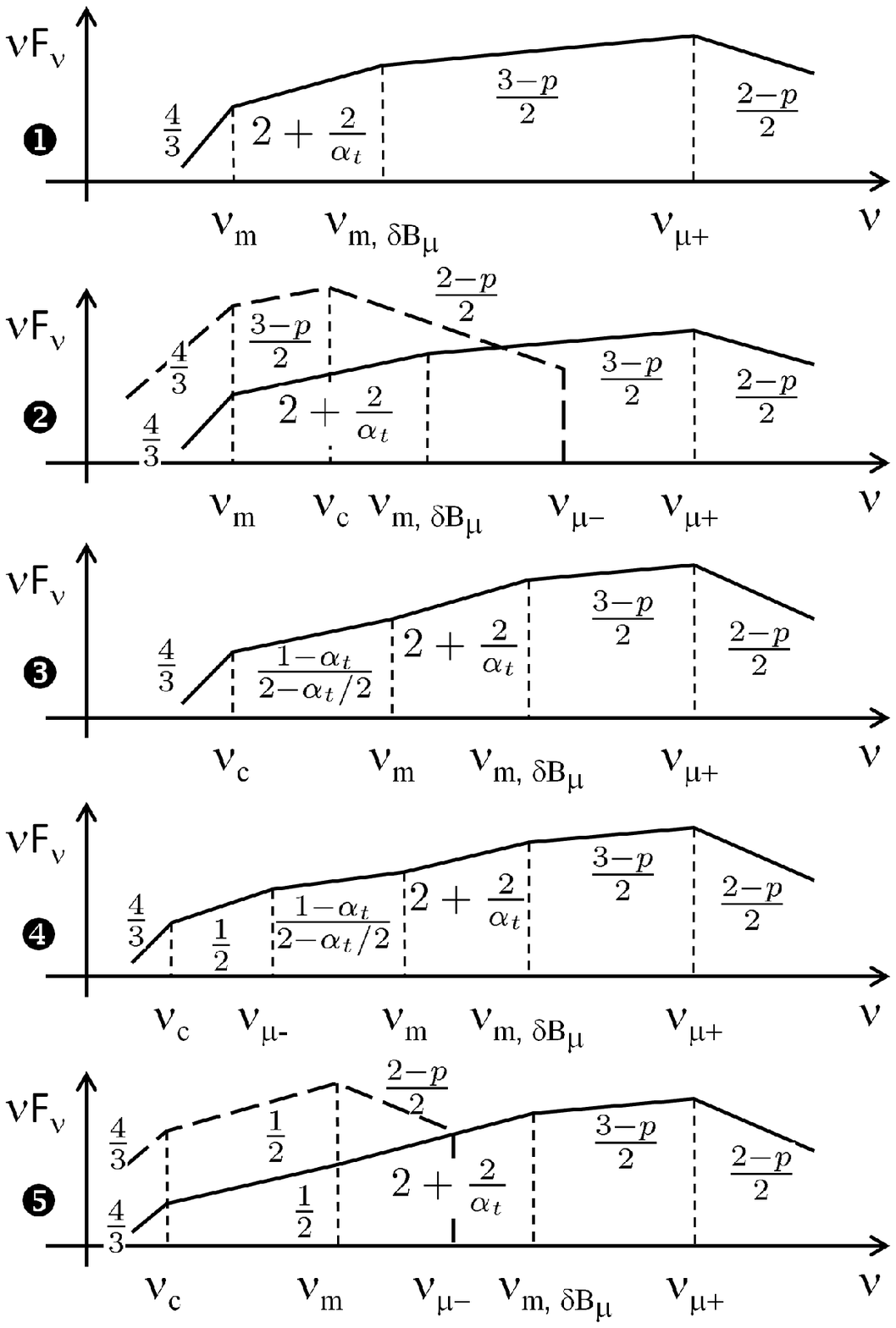}
\caption{Generic synchrotron spectra in time decaying microturbulence
  with $-3<\alpha_t<-4/(p+1)$, assuming that inverse Compton losses
  dominate throughout the blast. The spectral indices as
  indicated. Case 1: slow cooling scenario with $t_{\rm
    dyn}<t_{\mu-}$, in which case the turbulence has not had time to
  relax to the background value $B_{\rm d}$; case 2: slow cooling
  scenario with $t_{\rm dyn}>t_{\mu-}$; case 3: fast cooling scenario
  with $\nu_{\mu-}<\nu_{\rm c}<\nu_{\rm m}$; case 4: fast cooling
  scenario with $\nu_{\rm c}<\nu_{\mu-}<\nu_{\rm m}$; case 5: fast
  cooling scenario with $\nu_{\rm c}<\nu_{\rm m}<\nu_{\mu-}$.  See the
  discussion in Sec.~\ref{sec:appICrapid} for the precise definitions
  of the various frequencies and for modifications to these spectra if
  $\nu_{\rm c}$ or $\nu_{\rm m}$ exceeds $\nu_{\mu+}$, or if $\nu_{\rm
    max}$ is taken into account. The dashed line represents the
  secondary synchrotron component associated to cooling in the
  background shock compressed field $B_{\rm d}$ as in
  Fig.~\ref{fig:spec_ainfm42pp1}.\label{fig:spec_ICainfm42pp1} }
\end{figure}

Finally, one must consider the possibility that $-3<\alpha_t<-4/(p+1)$
with dominant inverse Compton losses, as in the previous
Section~\ref{sec:appICgrad}. The generic spectra obtained after
folding over the particle population are depicted in
Fig.~\ref{fig:spec_ICainfm42pp1}. The critical frequencies are defined
as in the previous Section~\ref{sec:appICgrad}.

The characteristic features of these spectra mix those of
Sec.~\ref{sec:appICgrad} [gradual decay, $\alpha_t>-4/(p+1)$ with
inverse Compton losses] with those of Sec.~\ref{sec:appnoICainfm1}
(rapid decay, $\alpha_t<-1$, no inverse Compton losses). They can be
summarized and understood as follows.

Cases 1 and 2 depict a slow cooling scenario $\gamma_{\rm
  m}<\tilde\gamma_{\rm c}$ with either $t_{\rm dyn}<t_{\mu-}$ (case 1)
or $t_{\rm dyn}>t_{\mu-}$ (case 2). Because the bulk of electrons do
not actually cool, the spectra obtained match those of cases 1 and 2
in Fig.~\ref{fig:spec_ainfm42pp1}, for which inverse Compton losses
are absent.

Cases 3 and 4 depict fast cooling regimes with different orderings of
$\nu_{\rm c}$ relatively to $\nu_{\mu-}$, but $\nu_{\rm
  m}>\nu_{\mu-}$. The peak power still occurs at $\nu_{\mu+}$ due to
the rapid decay of the microturbulence and the spectra keep a shape
similar to those of cases 1 and 2, up to the fast decaying index at
low frequencies $\nu_{\rm c}<\nu<\nu_{\rm m}$. Case 4 differs somewhat
from its counterpart when inverse Compton losses are neglected, which
is represented as case 3 in Fig.~\ref{fig:spec_ainfm42pp1}, because it
does not reveal an additional synchrotron fast cooling component in
the background shock compressed magnetic field. This absence is
directly related to the presence of inverse Compton losses, which cut
the electron distribution down to $\gamma_{\mu-}<\gamma_{\rm m}$ by
the end of the microturbulent layer, so that the electron
distribution is mostly monoenergetic at that point and cooling in the
background magnetic field ensues with generic slope $1/2$.

In contrast, $\gamma_{\rm m}<\gamma_{\mu-}$ in case 5. Then the
electron distribution maintains a powerlaw shape between $\gamma_{\rm
  m}$ and $\gamma_{\mu-}$ at the end of the microturbulence layer, so
that an additional component associated to cooling in the background
field emerges, as shown in Fig.~\ref{fig:spec_ICainfm42pp1}.

The corresponding spectro-temporal indices for $F_\nu\propto
t^{-\alpha}\nu^{\beta}$ are given in Table~\ref{tab:rapIC}.

The peak power for the synchrotron component associated to the
microturbulent layer occurs at $\nu_{\mu+}$ in all cases with
\begin{equation}
  \left .\nu F_{\nu,\delta B_\mu}\right\vert_{\nu=\nu_{\mu+}}\,\approx\,
  \frac{0.28}{4\pi D_L^2}\frac{4}{3}\gamma_{\rm b}^2 \dot N_e
  \frac{\tilde\gamma_{\mu+}}{1+Y_\mu} m_e c^2
  \left(\frac{\tilde\gamma_{\mu+}}{\gamma_{\rm
        m}}\right)^{1-p}\ .
\label{eq:nFnainfIC}
\end{equation}
The ratio between the peak powers of each synchrotron component for
case 5 reads
\begin{equation}
  \frac{\nu F_{\nu,B_{\rm d}}}{\nu F_{\nu,\delta B_\mu}}\,=\,
  \left(\frac{\gamma_{\rm m}}{\tilde\gamma_{\mu+}}\right)^{2-p}\frac{B_{\rm d}^2}{\delta B_\mu^2}\ ,
\label{eq:nFnBdIC}
\end{equation}
which resembles Eq.~\ref{eq:nFnBdIC} up to an additional factor
$B_{\rm d}^2/\delta B_\mu^2$ at the benefit of the high energy
microturbulent component, due to the competition between synchrotron
and inverse Compton losses.

\begin{table*}
\centering
\begin{minipage}{100mm}
  \caption{Spectral ($\beta$) and temporal ($\alpha$) indices of
    $F_\nu\propto t^{-\alpha}\nu^{-\beta}$ for various orderings of
    the characteristic frequencies, assuming a decaying
    microturbulence with $-3<\alpha_t<-4/(p+1)$, with dominant inverse
    Compton losses everywhere in the blast. In case 2 and 5, one must
    superimpose a synchrotron component associated to cooling in the
    background shock compressed magnetic field; the spectro-temporal
    slopes given here concern only the synchrotron component
    associated to the decaying microturbulent layer, not the latter.
    Case 1: slow cooling with $t_{\rm dyn}<t_{\mu-}$; case 2: slow
    cooling scenario with $t_{\rm dyn}>t_{\mu-}$; case 3: fast cooling
    scenario with $\nu_{\mu-}<\nu_{\rm c}<\nu_{\rm m}$; case 4: fast
    cooling scenario with $\nu_{\rm c}<\nu_{\mu-}<\nu_{\rm m}$; case
    5: fast cooling scenario with $\nu_{\rm c}<\nu_{\rm
      min}<\nu_{\mu-}$. The corresponding synchrotron spectra are
    shown in Fig.~\ref{fig:spec_ICainfm42pp1}. For all cases,
    $-\alpha=(2-3p)/4$ and $-\beta=-p/2$ if $\nu>\nu_{\mu+}$.  The
    quantity $k$ refers to the external density profile $n\propto
    r^{-k}$.}
\begin{tabular}{@{}llrr@{}}
  \hline
  Case & Frequency range & $-\beta$ & $-\alpha$ \\
  \hline
  Case 1 & $\nu<\nu_{\rm m}$ & $\frac{1}{3}$ & 
$\frac{6 (-2+k)+(-5+k) \alpha_t}{6 (-4+k)}$ \\ 
 & $\nu_{\rm m}<\nu<\nu_{{\rm m},\delta B_\mu}$ & 
$1+\frac{2}{\alpha_t}$ & 
$\frac{6 (-4+k)+(-7+3 k) \alpha_t}{2 (-4+k) \alpha_t}$ \\
 & $\nu_{{\rm m},\delta B_\mu}< \nu<\nu_{\mu+}$ & $\frac{1-p}{2}$ &
$\frac{2-3 k-12 p+3 k p}{16-4 k}$ \\
\hline
Case 2 &  $\nu<\nu_{\rm m}$ & $\frac{1}{3}$ &
$\frac{-6 k+(3+k) \alpha_t}{6 (-4+k) \alpha_t}$ \\
 & $\nu_{\rm m}<\nu<\nu_{{\rm m},\delta B_\mu}$ & 
$1+\frac{2}{\alpha_t}$ & 
$\frac{6 (-4+k)+(-7+3 k) \alpha_t}{2 (-4+k) \alpha_t}$ \\
 & $\nu_{{\rm m},\delta B_\mu}< \nu<\nu_{\mu+}$ & $\frac{1-p}{2}$
 & $\frac{2-3 k-12 p+3 k p}{16-4 k}$ \\
\hline
Case 3 & $\nu<\nu_{\rm c}$ & $\frac{1}{3}$ & 
$\frac{-4+6 k+(-5+k) \alpha_t}{6 (-4+k)}$ \\
 & $\nu_{\rm c}<\nu<\nu_{\rm m}$ & 
$\frac{2+\alpha_t}{-4+\alpha_t}$ & 
$\frac{2 (-4+k)+(11+k) \alpha_t}{2 (-4+k) \left(-4+\alpha_t\right)}$ \\
 & $\nu_{\rm m}<\nu<\nu_{{\rm m},\delta B_\mu}$ & 
$1+\frac{2}{\alpha_t}$ & 
$\frac{6 (-4+k)+(-7+3 k) \alpha_t}{2 (-4+k) \alpha_t}$ \\
 & $\nu_{{\rm m},\delta B_\mu}<\nu<\nu_{\mu+}$ & $\frac{1-p}{2}$ &
$\frac{2-3 k-12 p+3 k p}{16-4 k}$ \\
\hline
Case 4 & $\nu<\nu_{\rm c}$ & $\frac{1}{3}$ & 
$\frac{-16+6 k-5 (-5+k) \alpha_t}{24 (-4+k)}$ \\
 & $\nu_{\rm c}<\nu<\nu_{\mu-}$ & $-\frac{1}{2}$ & $\frac{1-k}{-4+k}$
 \\
 & $\nu_{\mu-}<\nu<\nu_{\rm m}$ & 
$\frac{2+\alpha_t}{-4+\alpha_t}$ & 
$\frac{2 (-4+k)+(11+k) \alpha_t}{2 (-4+k) \left(-4+\alpha_t\right)}$ \\
 & $\nu_{\rm m}<\nu<\nu_{{\rm m},\delta B_\mu}$ & 
$1+\frac{2}{\alpha_t}$ & 
$\frac{6 (-4+k)+(-7+3 k) \alpha_t}{2 (-4+k) \alpha_t}$ \\
 & $\nu_{{\rm m},\delta B_\mu}<\nu<\nu_{\mu+}$ & $\frac{1-p}{2}$ &
$\frac{2-3 k-12 p+3 k p}{16-4 k}$ \\
\hline
Case 5 & $\nu<\nu_{\rm c}$ & $\frac{1}{3}$ & 
$\frac{-6 k+(23+k) \alpha_t}{6 (-4+k) \alpha_t}$\\
 & $\nu_{\rm c}<\nu<\nu_{\rm m}$  & $-\frac{1}{2}$ & 
$\frac{-2 k+(11-3 k) \alpha_t}{2 (-4+k) \alpha_t}$\\
 & $\nu_{\rm m}<\nu<\nu_{{\rm m},\delta B_\mu}$ & 
$1+\frac{2}{\alpha_t}$ & 
$\frac{6 (-4+k)+(-7+3 k) \alpha_t}{2 (-4+k) \alpha_t}$\\
 & $\nu_{{\rm m},\delta B_\mu}<\nu<\nu_{\mu+}$ & $\frac{1-p}{2}$ &
$\frac{2-3 k-12 p+3 k p}{16-4 k}$\\
\hline
\end{tabular}
\end{minipage}
\label{tab:rapIC}
\end{table*}

\subsection{Synchrotron self-absorption} \label{sec:synch-self}

At very low frequencies, the synchrotron spectrum may be modified by
opacity effects. The absorption break frequency generally lies well
below the optical domain, e.g. for a homogeneous turbulence, $k=0$
(constant density profile) and a slow cooling regime, one finds
$\nu_{\rm abs}\sim 3\times 10^7\,{\rm Hz}\,
E_{53}^{1/5}n_{-3}^{3/5}\epsilon_{B,-2}^{1/5}\epsilon_{e,-0.3}^{-1}z_{+,0.3}^{-1}$
(see Panaitescu \& Kumar 2000). In this standard case, the frequency
index of $F_\nu$ is $2$ below the absorption frequency, e.g. Granot et
al. (1999) and references therein.

Accounting for the decaying microturbulence modifies the situation as
follows. All calculations are performed in the comoving downstream
frame, as indicated by the primes. Moreover, the self-similar profile
of the blast and the secular evolution of the blast characteristics
are neglected, as in the rest of App.~\ref{sec:appFnu}.  One then
defines the synchrotron self-absorption coefficient (Rybicki \&
Lightman, 1979):
\begin{equation}
  \alpha'_{\nu'}\,=\,-\frac{1}{8\pi m_e \nu'^2}\,\int{\rm d}\gamma_e\,
  P'_{\nu'/e} \gamma_e^2\frac{\partial}{\partial
    \gamma_e}\left(\frac{1}{\gamma_e^2}\frac{{\rm d}n'}{{\rm
        d}\gamma_e}\right)\ , \label{eq:ap}
\end{equation}
with $P'_{\nu'/e}\,\equiv\,{\rm d}E/{\rm d}\nu'{\rm d}t'$ the spectral
synchrotron power emitted per electron per frequency interval, as
defined in Eq.~\ref{eq:Esyn} up to the change of frame for the
frequency. Here however, one considers only the low frequency part
$\propto \nu^{1/3}$ since one is interested in frequencies well below
$\nu_{\rm m}$.  The electron distribution function ${\rm d}n'/{\rm
  d}\gamma_e$ depends on the distance to the shock front when cooling
is efficient. Since only the minimum Lorentz factor is of importance
in the calculations of synchrotron self-absorption, one can
approximate the distribution as a unique powerlaw,
\begin{equation}
  \frac{{\rm d}n'}{{\rm
      d}\gamma_e}\,=\,
  \frac{1}{\gamma_i}\left(\frac{\gamma_e}{\gamma_i}\right)^{-q}n'(q-1)
\end{equation}
with $\gamma_i=\gamma_{\rm m}$ at distances such that cooling is
inefficient for particles of Lorentz factor $\gamma_{\rm m}$, and
$\gamma_i$ position dependent at larger distances from the shock
front. Of course, in a slow cooling regime, $\gamma_i=\gamma_{\rm m}$
everywhere in the blast. The index $q$ departs from $s$ only when
cooling becomes efficient; its exact value does not affect the
following up to factors of order unity. The above integral leads to
\begin{equation}
\alpha'_{\nu'}\,\simeq\, 5 \frac{e n'}{\delta B}
\gamma_i^{-5}\left(\frac{\nu'}{\nu'_i}\right)^{-5/3}\ ,\label{eq:ap2}
\end{equation}
with $\nu'_i\equiv \nu_{\rm p}'\left[\delta B;\gamma_i\right]$. This
local absorption coefficient depends on position both through
$\gamma_i$ (when cooling is efficient) and through $\delta B$, in the
presence of decaying turbulence. One may thus rewrite the absorption
coefficient as
\begin{equation}
  \alpha'_{\nu'}\,\simeq\, \alpha'_{\mu,\nu'}\left(\frac{\delta B}{\delta
      B_\mu}\right)^{2/3}\left(\frac{\gamma_i}{\gamma_{\rm
        m}}\right)^{-5/3}\ ,\label{eq:ap3}
\end{equation}
with $\alpha'_{\mu,\nu'}$ the absorption coefficient defined according to
Eq.~\ref{eq:ap2} with $\delta B \rightarrow \delta B_\mu$ and
$\gamma_i\rightarrow \gamma_{\rm m}$. This formulation allows to write
$\alpha'_{\nu'}$ as a broken power law function of the distance to the
shock front.

The emission coefficient $j'_{\nu'}$ is defined as
\begin{equation}
j'_{\nu'}=\frac{1}{4\pi}\int {\rm d}\gamma_e\,
  P'_{\nu'/e} \frac{{\rm d}n'}{{\rm
        d}\gamma_e}\ , \label{eq:jnu}
\end{equation}
assuming isotropic emission, and it can be recast similarly to
$\alpha'_{\nu'}$ into
\begin{equation}
j'_{\nu'}\,=\, j'_{\mu,\nu'}\left(\frac{\delta B}{\delta
      B_\mu}\right)^{2/3}\left(\frac{\gamma_i}{\gamma_{\rm
        m}}\right)^{-2/3}\ ,\label{eq:jnu2}
\end{equation}
with $j'_{\mu,\nu'}\,\simeq\,0.4 e^3 \delta B_\mu (m_e c^2)^{-1}n'
\left(\nu'/\nu'_{\mu,\rm m}\right)^{1/3}$ and $\nu'_{\mu,\rm m}\equiv
\nu_{\rm p}'\left[\delta B_\mu;\gamma_{\rm m}\right]$.

The specific intensity along a ray path obeys the radiative transfer
equation
\begin{equation}
  \frac{{\rm d}I'_{\nu'}}{{\rm d}s}\,=\,-\alpha'_{\nu'}I'_{\nu'} +
  j'_{\nu'}\ ,
\end{equation}
with the usual formal solution
\begin{equation}
I'_{\nu'}(x=0)\,=\, \int_{0}^{\tau_{\nu',\rm b}}
\frac{j'_{\nu'}}{\alpha'_{\nu'}}\,e^{-\tau_{\nu'}}\,{\rm d}\tau_{\nu'}\ .
\end{equation}
Here ${\rm d}\tau_{\nu'} = \alpha'_{\nu'}{\rm d} x$, $x>0$ denotes the
distance from the shock front, and $\tau_{\nu'}(x=0)\equiv0$ by
convention. The parameter $\tau_{\nu',\rm b}$ consequently represents
the total optical depth of the blast which can be derived by
integrating the broken power law form of Eq.~\ref{eq:ap3} over $x$,
from $x=0$ up to $x_{\rm b}\sim r/\gamma_{\rm b}\sim t_{\rm
  dyn}/c$. The above formulation of the solution of the equation of
radiative transfer is particularly appealing because it leads to a
simple evaluation of $I'_{\nu'}(0)$ in various cases.

In particular, if the regime is slow cooling, $\gamma_i=\gamma_{\rm
  m}$ everywhere and $j'_{\nu'}/\alpha'_{\nu'}$ becomes uniform in the
blast, as in the homogeneous (slow cooling) case. Then one finds
\begin{equation}
  I'_{\nu'}(0)=\frac{j'_{\mu,\nu'}}{\alpha'_{\mu,\nu'}}\left(1-e^{-\tau_{\nu',\rm
        b}}\right)\ .\label{eq:inu}
\end{equation}
In the optically thick limit $\tau_{\nu',\rm b}\gg1$, the ratio
$j'_{\mu,\nu'}/\alpha'_{\mu,\nu'}$ leads to the standard slope $2$ for
$F_\nu$ below the absorption frequency $\nu'_{\rm abs,b}$, which is
defined by the condition $\tau_{\nu',\rm b}=1$ at $\nu'_{\rm
  abs,b}$. Note that $\tau_{\nu',\rm b}$ shares with the absorption
coefficient $\alpha'_{\nu'}$ the scaling $\propto \nu'^{-5/3}$. In the
optically thin limit $\tau_{\nu',\rm b}\ll1$, meaning $\nu'\gg
\nu'_{\rm abs,b}$, one recovers the slope $1/3$ since $\tau_{\nu',\rm
  b}\propto \alpha'_{\mu,\nu'}$. The non trivial dependence of
$\nu'_{\rm abs,b}$ on time leads to a non trivial time dependence of
the flux below the break frequency. This dependence is not discussed
here, but can be calculated through $\tau_{\nu',\rm b}$.

The fast cooling regime leads to a more complicated shape of the
spectrum in the transition region between the asymptotic thin and
thick regimes (for a similar discussion with a homogeneous turbulence,
see Granot et al. 2000). One needs to define here an intermediate
optical depth,
\begin{equation}
\tau_{\nu',\rm m} = \int_{0}^{x_{\rm cool,m}} \alpha'_{\nu'}\,{\rm d}x\ ,
\end{equation}
with $x_{\rm cool,m}$ the location at which particles of Lorentz
factor $\gamma_{\rm m}$ start to cool. The fast cooling regime implies
that by the back of the blast, such particles have cooled down to
$\gamma_{\rm c}$. One now defines an absorption frequency $\nu'_{\rm
  abs,m}$ such that $\tau_{\nu',\rm m}=1$ at $\nu'_{\rm abs,
  m}$. Because the optical depth is an increasing function of distance
to the shock front, $\tau_{\nu',\rm m}<\tau_{\nu',\rm b}$ (possibly
$\tau_{\nu',\rm m}\ll\tau_{\nu',\rm b}$) and $\nu'_{\rm
  abs,m}<\nu'_{\rm abs, b}$. Therefore, at frequencies $\nu'<\nu'_{\rm
  abs,m}$, the shell is optically thick for $x\,\geq\,x_{\rm cool,m}$
and the standard synchrotron self-absorption spectrum emerges. At
frequencies $\nu'>\nu'_{\rm abs,b}$, the whole shell is optically thin
to synchrotron radiation, so that the standard $\propto \nu^{1/3}$
spectrum calculated above applies. However, for $\nu'_{\rm
  abs,m}<\nu'<\nu'_{\rm abs,b}$, opacity competes with local emission
at distances $>x_{\rm cool,m}$, and the overall spectrum departs from
either asymptote. The exact spectral shape can be derived from
Eq.~\ref{eq:inu}, using the calculated value of $\tau_{\nu',\rm b}$.

\end{document}